\documentclass{emulateapj}

\usepackage{graphicx}
\usepackage[]{natbib}
\usepackage{placeins}
\usepackage{lineno}
\usepackage{ulem}

\pdfoutput=1

\usepackage{color}

\newcommand{\degree}{\ensuremath{^\circ}}

\newcommand{\ltsima} {$\; \buildrel < \over \sim \;$}
\newcommand{\gtsima} {$\; \buildrel > \over \sim \;$}
\newcommand{\lta} {\lower.5ex\hbox{\ltsima}}
\newcommand{\gta} {\lower.5ex\hbox{\gtsima}}
\newcommand{\grbnos} {GRB~$120323$A}
\newcommand{\grb} {GRB~$120323$A~}

\newcommand{\fdaigne}[1]{\textcolor{black}{#1}}

\shorttitle{Photospheric component in short \grb spectra and its effects on the E$_{peak}$-L relation}
\shortauthors{Guiriec et al.}

\begin{document}

\title{ Evidence for a photospheric component in the prompt emission of the short \grb and its effects on the GRB hardness - luminosity relation}

\author{S. Guiriec\altaffilmark{1,2}, F. Daigne\altaffilmark{3}, R. Hasco\"et\altaffilmark{3}, G. Vianello\altaffilmark{4}, F. Ryde\altaffilmark{5,6}, R. Mochkovitch\altaffilmark{3}, C. Kouveliotou\altaffilmark{7}, S. Xiong\altaffilmark{8}, P.N. Bhat\altaffilmark{8}, S. Foley\altaffilmark{9}, D. Gruber\altaffilmark{10}, J. M. Burgess\altaffilmark{8}, S. McGlynn\altaffilmark{9}, J. McEnery\altaffilmark{1,11}, N. Gehrels\altaffilmark{1}}

\altaffiltext{1}{NASA Goddard Space Flight Center, Greenbelt, MD 20771, USA}
\altaffiltext{2}{NASA Postdoctoral Program Fellow}
\altaffiltext{3}{Institut d'Astrophysique de Paris â UMR 7095 Universit\'e Pierre et Marie Curie-Paris 06; CNRS 98 bis bd Arago, 75014 Paris, France}
\altaffiltext{4}{W. W. Hansen Experimental Physics Laboratory, Kavli Institute for Particle Astrophysics and Cosmology, Department of Physics and SLAC National Accelerator Laboratory, Stanford University, Stanford, CA 94305, USA}
\altaffiltext{5}{Department of Physics, Royal Institute of Technology, AlbaNova, SE-106 91 Stockholm, Sweden}
\altaffiltext{6}{The Oskar Klein Centre for Cosmo Particle Physics, AlbaNova, SE-106 91 Stockholm, Sweden}
\altaffiltext{7}{Office of Science and Technology, ZP12, NASA/Marshall Space Flight Center, Huntsville, AL 35812, USA}
\altaffiltext{8}{University of Alabama in Huntsville, NSSTC, 320 Sparkman Drive, Huntsville, AL 35805, USA}
\altaffiltext{9}{UCD School of Physics, University College Dublin, Dublin 4, Ireland}
\altaffiltext{10}{Max-Planck-Institut f\"ur extraterrestrische Physik (Giessenbachstrasse 1, 85748 Garching, Germany)}
\altaffiltext{11}{Department of Physics and Department of Astronomy, University of Maryland, College Park, MD 20742, USA}

\email{sylvain.guiriec@nasa.gov}

\begin{abstract}

The short \grb had the highest flux ever detected with the Gamma-ray Burst Monitor (GBM) on board the Fermi Gamma-ray Space Telescope. Here we study its remarkable spectral properties and their evolution using two spectral models: (i) a single emission component scenario, where the spectrum is modeled by the empirical Band function (a broken power law), and (ii) a two component scenario, where thermal (a Planck-like function) emission is observed simultaneously with a non-thermal component (a Band function). We find that the latter model fits the integrated burst spectrum significantly better than the former, and that their respective spectral parameters are dramatically different: when fit with a Band function only, the E$_\mathrm{peak}$ of the event is unusually soft for a short GRB (70 keV compared to an average of 300 keV), while adding a thermal component leads to more typical short GRB values (Epeak $\sim$300 keV). Our time-resolved spectral analysis produces similar results. We argue here that the two-component model is the preferred interpretation for \grb, based on: (i) the values and evolution of the Band function parameters of the two component scenario, which are more typical for a short GRB, and (ii) the appearance in the data of a significant hardness-intensity correlation, commonly found in GRBs, when we employee two-component model fits; the correlation is non-existent in the Band-only fits. GRB~$110721$A, a long burst with an intense photospheric emission, exhibits the exact same behavior. We conclude that \grb has a strong photospheric emission contribution, first time observed in a short GRB. Magnetic dissipation models are difficult to reconcile with these results, which instead favor photospheric thermal emission and fast cooling synchrotron radiation from internal shocks. Finally, we derive a possibly universal hardness-luminosity relation in the source frame using a larger set of GRBs ($\mathrm{L_i^{Band}}=\mathrm{(1.59\pm0.84).10^{50}~(E_{peak,i}^{rest})^{1.33\pm0.07} erg.s^{-1}}$), which could be used as a possible redshift estimator for cosmology.

\end{abstract}


\section{Introduction}

Gamma-Ray Bursts (GRBs) are extremely energetic explosions at cosmological distances~\citep{Meegan:1992,vanParadijs:1997,Bhat:2011} resulting most likely from the formation of stellar mass black holes, either through the collapse of massive stars~\citep{Woosley:1993,MacFadyen:1999,Woosley:2006} or via the merger of two compact objects~\citep{Paczynski:1986,Fryer:1999,Rosswog:2003}. Regardless of the nature and formation mechanism of the GRB central engine, the fireball model, initially proposed by~\citet{Cavallo:1978}, best explains the GRB emission from radio up to GeV gamma-rays. During the GRB explosion, the central engine produces a collimated bipolar jet, mainly composed of electrons, positrons, photons and a small amount of baryons. The central engine ejecta are accelerated to relativistic velocities forming layers of high density regions, which propagate at various speeds. When the faster layers catch up with the slowest, charged particles are accelerated through mildly relativistic collisionless shocks: this is the so-called internal shock phase~\citep{Rees:1994,Kobayashi:1997,Daigne:1998}. As a result they then produce non thermal radiation such as synchrotron emission, observed as GRB prompt emission in gamma-rays (keV$-$MeV) ~\citep[see e.g., the spectral catalogs by][]{Preece:2000,Goldstein:2012a}, and even up to several tens of GeVs \citep[see e.g.,][]{Abdo:2009:GRB080916C,Abdo:2009:GRB090902B,Ackermann:2010:GRB090510,Ackermann:2011:GRB090926A}. As the jet expands, it slows down as it interacts with the interstellar medium in a relativistic shock; the charged particles involved in this collision produce synchrotron radiation visible from radio wavelengths to X-ray energies. This is the external shock phase~\citep{Rees:1992,Meszaros:1993}, which is responsible for the afterglow emission observed during few hours to several days and even years following the prompt emission. Alternative models for the GRB prompt phase are magnetically driven, involving mechanisms such as magnetic field line reconnection.

Besides the non-thermal (synchrotron) radiation, the fireball model also predicts strong thermal emission emanating from the jet's photosphere, which would be observable when the ejecta layers become optically thin to Thompson scattering~\citep{Goodman:1986,Meszaros:2002,Rees:2005}. In the most standard version of the internal shock scenario within a thermally accelerated outflow (fireball), this thermal emission would be very intense and would overpower the non-thermal component~\citep{Daigne:2002}. Prompt GRB spectra have been traditionally adequately fitted with the empirical Band function~\citep{Band:1993}, which is a smoothly (with curvature) broken power law, with indices $\alpha$ and $\beta$ for the low and the high energy part, respectively. The break energy of the Band function, parameterized as  E$_\mathrm{peak}$, corresponds to the maximum of the $\nu$F$_{\nu}$ spectrum~\citep{Gehrels:1997}, when $\alpha>$-2 and $\beta<$-2. The Band parameters are usually compatible with non-thermal emission. The synchrotron mechanism thus remains the preferred model to explain most of the prompt emission; note, however, that the $\alpha$ parameter values are often incompatible with slow and fast electron cooling scenarios~\citep{Crider:1997,Preece:1998}.

However, a few bursts observed with the Burst And Transient Source Experiment (BATSE) onboard the {\it Compton Gamma Ray Observatory (CGRO)} were found to be well fitted by a single blackbody function, the tell-tale signature of emission from the photosphere~\citep{Ghirlanda:2003, Ryde:2004}. Furthermore, time resolved analysis of the strongest GRB pulses observed with BATSE were shown to,  in some cases, be well fitted by a combination of a blackbody and a power law function~\citep{Ryde:2005}. The pulse temperatures were found to lie within $20-100$ keV and were observed to evolve in a characteristic way, decaying as a broken power law in time~\citep{Ryde:2009}.  However, the BATSE energy range ($25-1800$ keV) made it difficult to fully assess these models~\citep[see e.g.,][]{Ghirlanda:2007}. The broader energy range (8 keV -- 40 MeV) of the Gamma-ray Burst Detector (GBM) on the {\it Fermi} gamma ray space telescope alleviates this shortcoming. Using {\it Fermi} data, \citet{Ryde:2010} proposed that the  dominant contribution to the spectrum of GRB~090902B is a modified blackbody.  Furthermore, \citet{Guiriec:2011a} reported for the first time the simultaneous fit to the data of long GRB~100724B using a blackbody (BB) and a Band function corresponding to the thermal and non-thermal emissions, respectively. In this GRB, the thermal emission identified in the time-integrated spectrum and followed in the time-resolved spectroscopy analysis was a subdominant contribution, corresponding to only few percent of the total energy. Interpreted as a photospheric emission, this low intensity BB did not support the standard fireball scenario, where the energy initially released by the central engine is only thermal; instead, it suggested an initially magnetically dominated outflow (note that this conclusion is independent of the nature of the mechanism responsible for the non-thermal emission, internal shocks or magnetic reconnection). \fdaigne{The lack of GRBs whose thermal contribution to the prompt emission overpowers the non-thermal one points towards a similar conclusion for the majority of bursts~\citep{Daigne:2002,Zhang:2009}. When a thermal component is detected, the variable ratios between the energy contained in the thermal and non-thermal components from burst to burst may indicate that there is a range of initial magnetization in GRB outflows~\citep{Hascoet:2013}.}
%\fdaigne{The lack of any bright thermal component in most GRBs, including very bright ones like GRB 080916C, points towards a similar conclusion for the majority of bursts \citep{Daigne:2002,Zhang:2009}.}

Another example is given by the strong emission pulse in GRB~$110721$A in which the presence of two components is yet again highly significant~\citep{Axelsson:2012:GRB110721A}. The temperature of the BB component is observed to have the same characteristic temporal evolution as seen in some of the BATSE bursts~\citep{Ryde:2005}. On the other hand, the peak energy of the Band component decreases  more rapidly and as a single power law in time. In the case of GRB~$110721$A the flux contribution of the BB is at most 10 \%.

%See also~\citet{Bhat:2011} for an comprehensive review of recent GRB results.

Here we discuss the possible identification of such a subdominant but intense BB component in the short \grbnos. In section~\ref{section:observation}~and~\ref{section:Spectral analysis} we describe the {\it Fermi} observations of \grb and the analysis procedure, respectively. We present the results of our time-integrated and time-resolved spectral analyses in section~\ref{section:Time-integrated spectral analysis}~and~\ref{section:Time-resolved spectral analysis}, respectively. In section~\ref{Model description}, we examine the hypothesis that the BB component might be present together with a Band component during the entire burst and compare these results against Band fits only. In Section~\ref{section:Flux-Epeak correlation} we present an intriguing result on the relation between E$_\mathrm{peak,i}^\mathrm{rest}$ and the Band function luminosity, which supports and extends the often discussed E$_\mathrm{peak}$-L correlation as well as the scenario of the simultaneous existence of the thermal and non-thermal components. We present a discussion and interpretation of our results in section~\ref{section:interpretation1} and conclude in Section~\ref{section:conclusion}. Appendix~\ref{section:Simulations} describes the simulation procedure used to assess and validate our results.

\begin{figure*}[t!]
\begin{center}

\includegraphics[totalheight=0.54\textheight, clip, viewport=0 0 487 684]{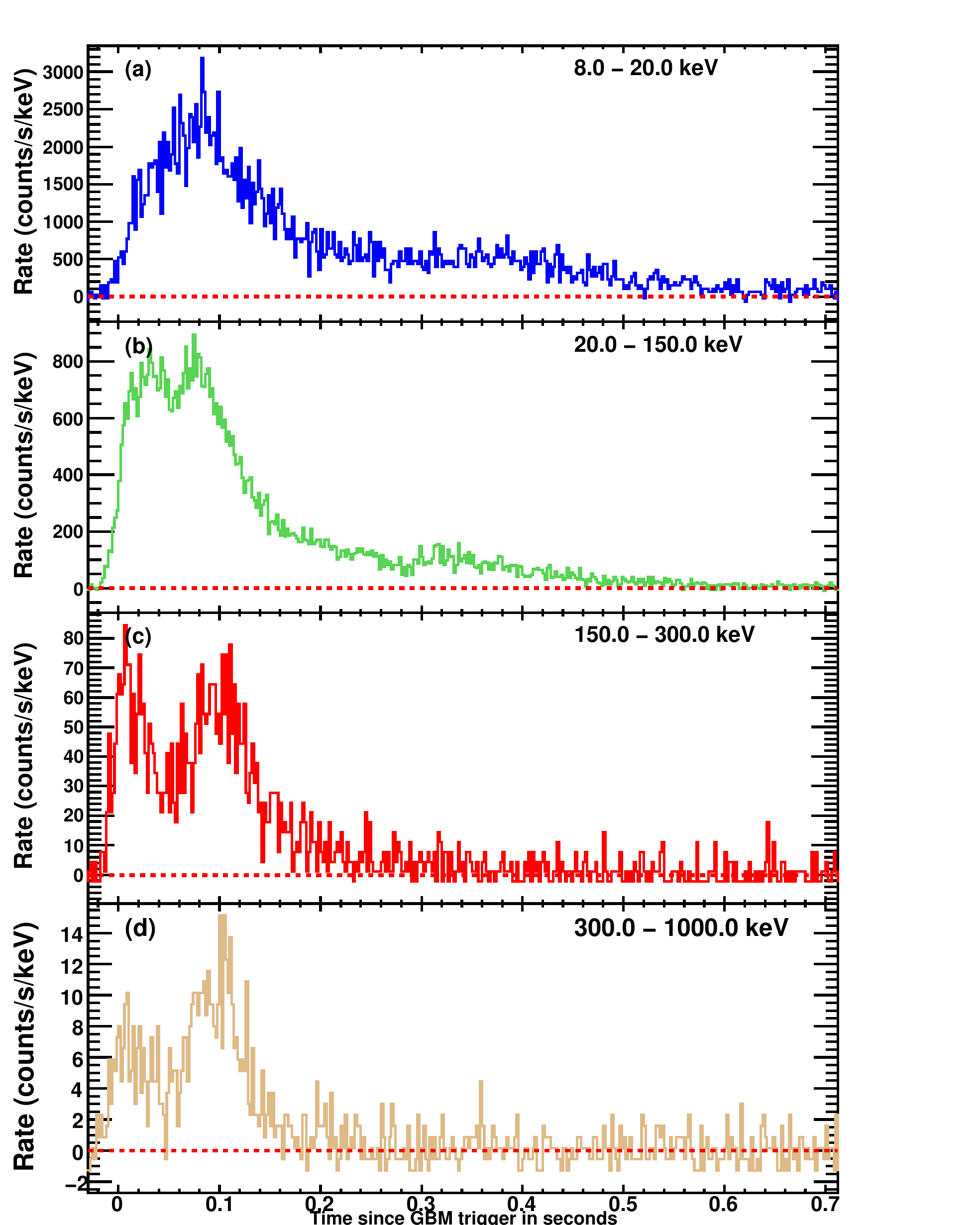}
\includegraphics[totalheight=0.54\textheight, clip, viewport=24 0 487 684]{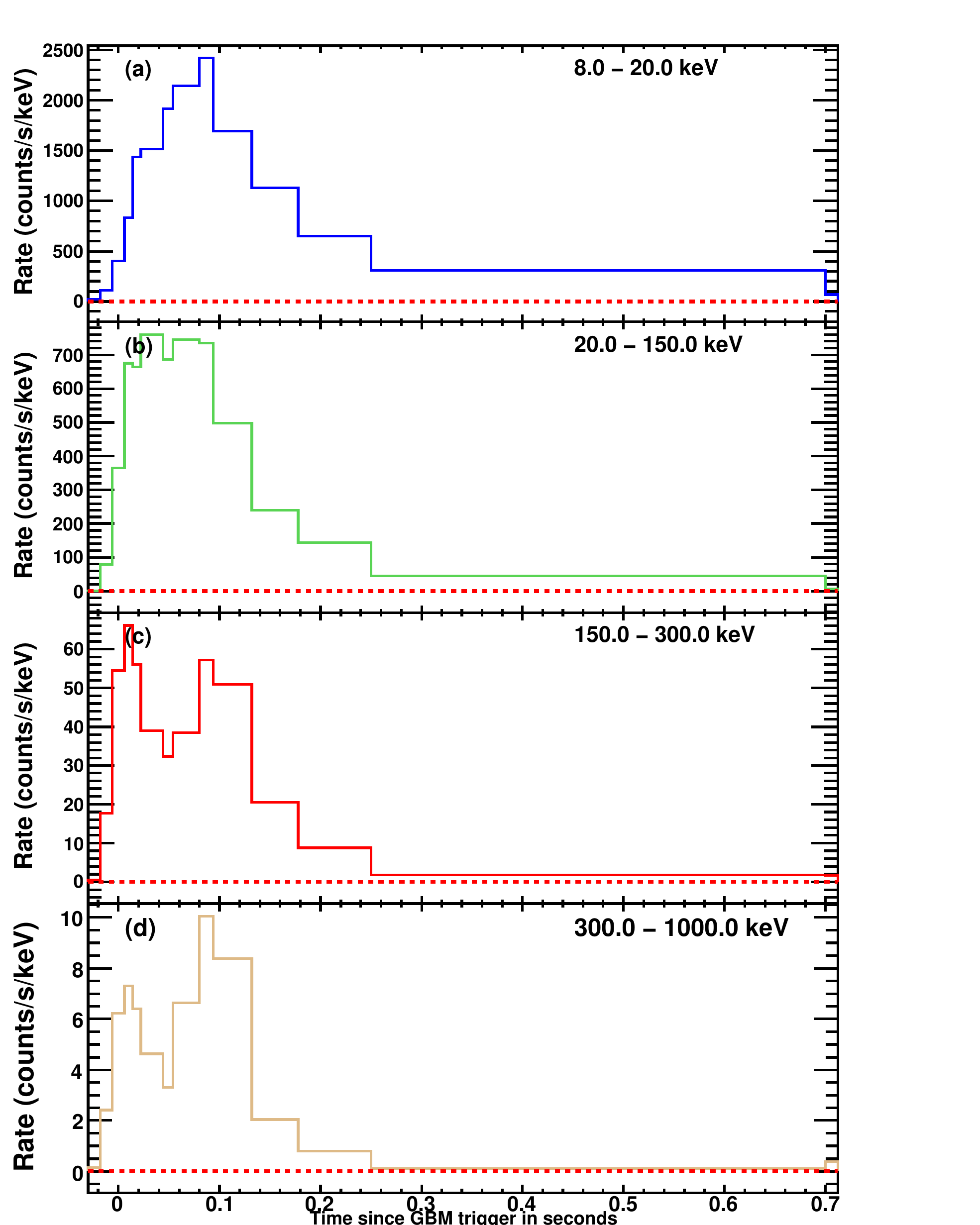}

\caption{\label{fig:GRB120323A_LCs}Background substracted count light curves of \grb in four energy bands ranging from 8 keV to 1 MeV as observed with GBM. The light curves from 8 to 300 keV are obtained after summing the signal detected in NaI detectors N0, N1 and N3, and the light curve from 300 keV to 1 MeV corresponds to the BGO detector B0. The figures on the left show the 2 ms light curves while the right ones correspond to the light curves rebinned for the fine time-resolved spectroscopy describe in section~\ref{sec:Fine time resolved}.}
\end{center}
\end{figure*}

\label{section:Band vs Band+BB}
\label{section:BandvsBand+BB}

\section{Observations}
\label{section:observation}

\begin{figure}
\begin{center}
\includegraphics[totalheight=0.25\textheight, clip]{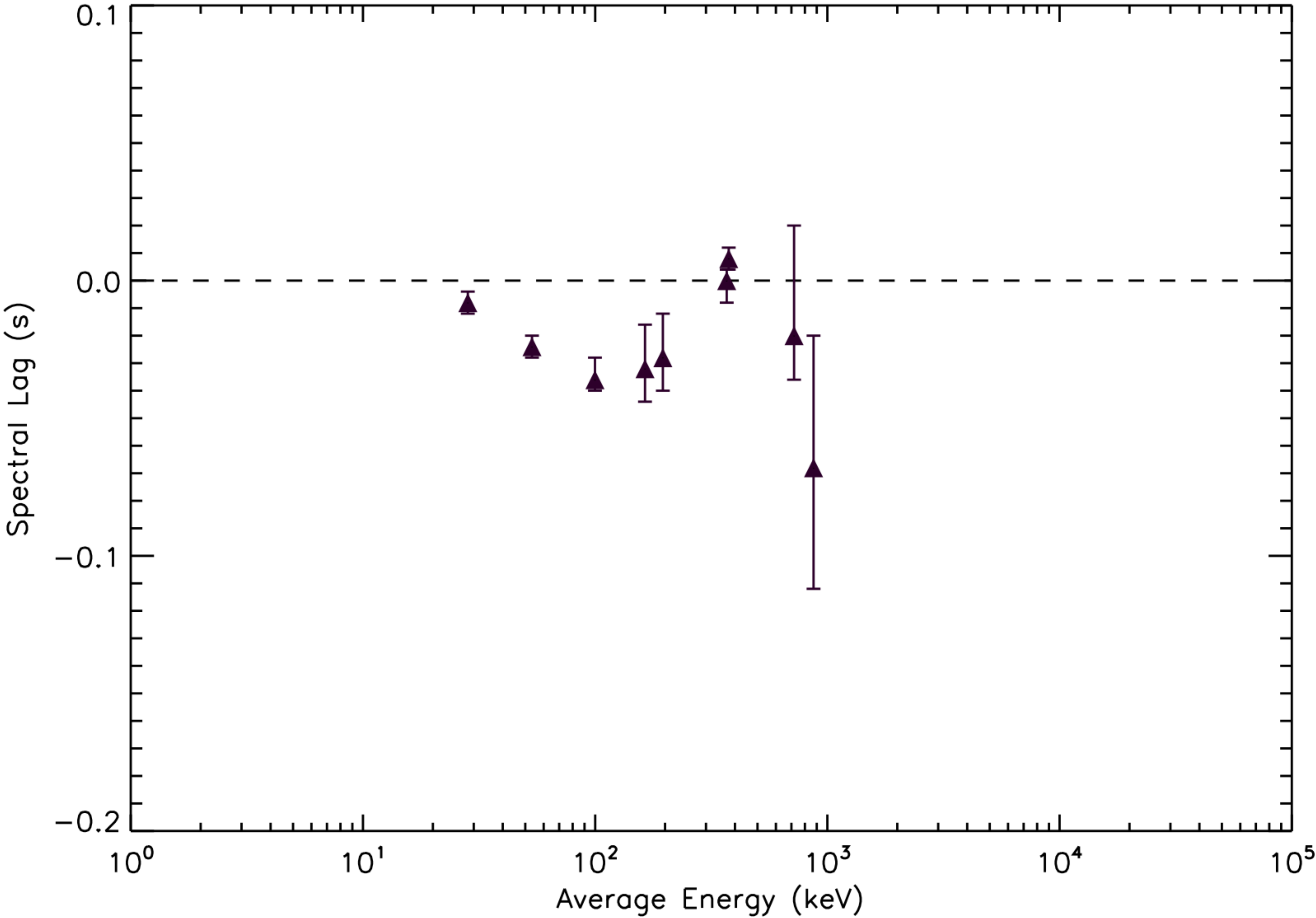}
\caption{\label{fig:lags}Spectral lags measured between the lowest energy band 8-20 keV and higher energy bands in the NaI and BGO detectors for GRB 120323A plotted as a function of mean energy of the higher energy band. The higher energy bands are: for NaI 20-40 keV, 40-70 keV, 70-140 keV, 140-270 keV, 270-525 keV, 525-985 keV; for BGO, 110-250 keV, 250-550 keV and 550-1380 keV.}
\end{center}
\end{figure}

The GBM onboard {\it Fermi} detected a very intense short burst, \grbnos, on 2012 March 23 at 12:10:19.72 UT~\citep{Gruber:2012}. \grb has the highest peak flux among all events observed with GBM thus far; the intensity of the burst fullfiled the criterion for an Autonomous Repointing Request (ARR) of the {\it Fermi} spacecraft to place the source in the field of view of the Large Area Telescope (LAT). However, the burst was unusually soft and was not detected at high energies by the LAT~\citep{Tam:2012} in the standard LAT data (100 MeV to $>$300 GeV), nor in the Low LAT Energy (LLE) data (designed to increase the LAT event acceptance at low energies and enable spectral analysis below 100 MeV). Figure~\ref{fig:GRB120323A_LCs} shows the GBM light curves of \grb in four energy bands ranging from 8 keV to 1 MeV. The T$_\mathrm{90}$ duration~\citep{kouveliotou:1993} of the event was T$_\mathrm{90}=0.448\pm0.090$ s~ between 50 and 300 keV \citep{Gruber:2012}. The evolution of the spectral lag as a function of energy is shown in Figure~\ref{fig:lags}. The spectral lags are calculated over the full duration of the GRB (T$_\mathrm{0}$-0.02s to T$_\mathrm{0}$+0.68s) using the cross-correlation method described in~\citet{Norris:2000}. The maximum spectral lag for \grb is small as expected for short bursts~\citep{Norris:2006,Zhang:2006}. While the energy dependence of the lags is also consistent with that for other intense short bursts~\citep{Guiriec:2010}, it is important to note that this may be due to pulse confusion between the two main emission peaks, with the second peak being spectrally harder than the first and so contributing more to the cross-correlation function peak at higher energies.
The evolution of the spectral lag as a function of energy is shown in Figure~\ref{fig:lags}. The spectral lags are calculated over the full duration of the GRB (T$_\mathrm{0}$-0.02s to T$_\mathrm{0}$+0.68s) using the cross-correlation method described in~\citet{Norris:2000}. The maximum spectral lag for \grb is small as expected for short bursts~\citep{Norris:2006,Zhang:2006}. While the energy dependence of the lags is also consistent with that for other intense short bursts~\citep{Guiriec:2010}, it is important to note that this may be due to pulse confusion between the two main emission peaks, with the second peak being spectrally harder than the first and so contributing more to the cross-correlation function peak at higher energies.

\grb was also detected with {\it Konus}-WIND~\citep{Golenetskii:2012B} and {\it MESSENGER}. The best location for this event was estimated with the Inter-Planetary Network (IPN) using all three satellites, to be centered at RA=340.4\degree and Dec=29.7\degree, inside an irregular error box with a maximal dimension of 0.75\degree and a minimal dimension of 0.25\degree~\citep{Golenetskii:2012A}.

\section{Data analysis procedure}
\label{section:Spectral analysis}

We performed a spectral analysis of \grb using only GBM Time Tag Event (TTE) data (8 keV - 40 MeV). In the TTE data, each event detected with a GBM detector is recorded with its trigger time and the detector energy channel. TTE data have the finest time and energy resolution and are then ideal to perform spectral analysis of such short GRB. For more information about GBM, see~\citet{Meegan:2009}. We selected the three NaI detectors, N0, N1 and N3, with angles to the source below 50\degree as these are not affected by blockage from other parts of the spacecraft nor shadowed by other detectors. We also used one BGO detector, B0, with a direct view to the source. Since there is no detection of this GRB in the regular or LLE LAT data, we did not include these datasets in our spectral analysis; however, we note here that the extrapolation of our GBM spectral analysis is consistent with the LAT upper limits.

We selected the NaI energy channels from 8 keV to the overflow channels starting at $\sim$900 keV, and the BGO data from 200 keV to the overflow channels starting at $\sim$40 MeV. We then generated the response matrices for each detector using the best known location for the event, which is the IPN location reported in Section~\ref{section:observation}. For each detector, the background was estimated by fitting a polynomial function to time intervals pre and post burst. The background during the GRB was then estimated by extrapolating the polynomial function over the source time interval. Data were fit using the spectral analysis package Rmfit provided by the GBM instrument team; the spectral fit was performed using the full spectral resolution of the instruments. We determined the best spectral parameters by optimizing the Castor C statistic value. Castor Cstat (henceforth Cstat) is a likelihood technique modified for a particular data set to converge to a ${\chi}^\mathrm{2}$ with an increase of the signal.

\section{Time-integrated spectral analysis}
\label{section:Time-integrated spectral analysis}

\begin{figure}
\begin{center}
\includegraphics[totalheight=0.175\textheight, clip]{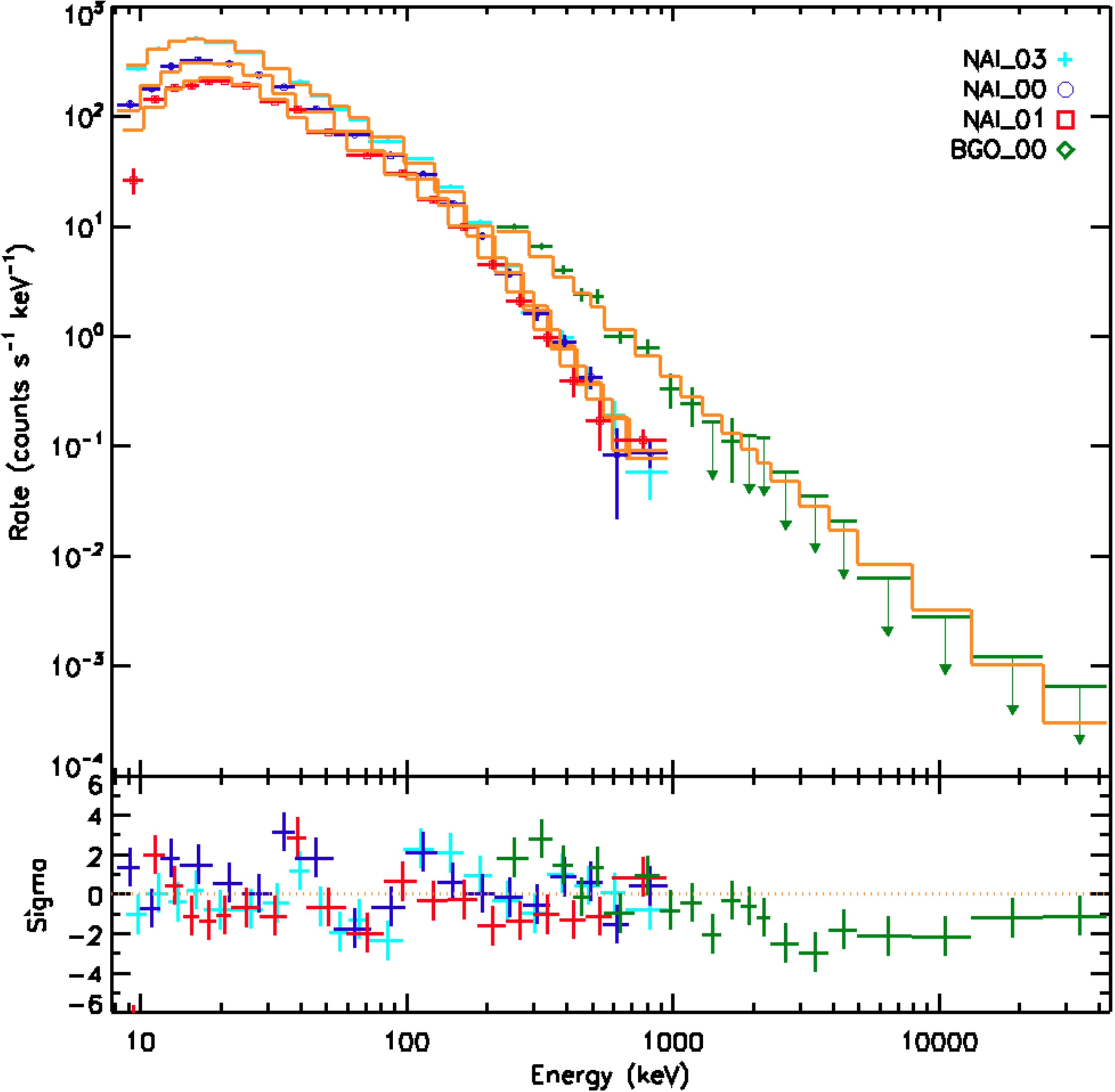}
\includegraphics[totalheight=0.175\textheight, clip]{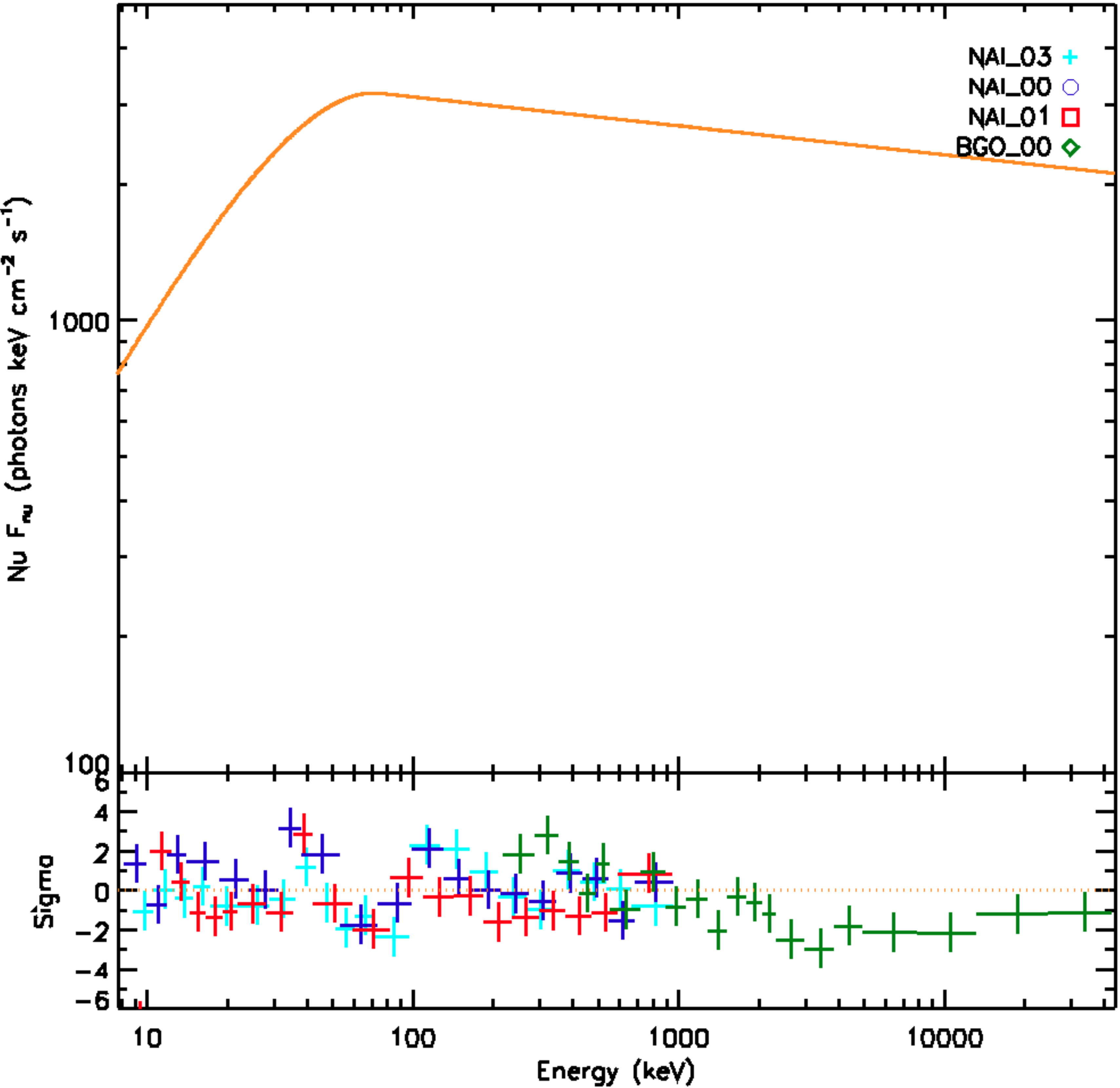}

\includegraphics[totalheight=0.172\textheight, clip]{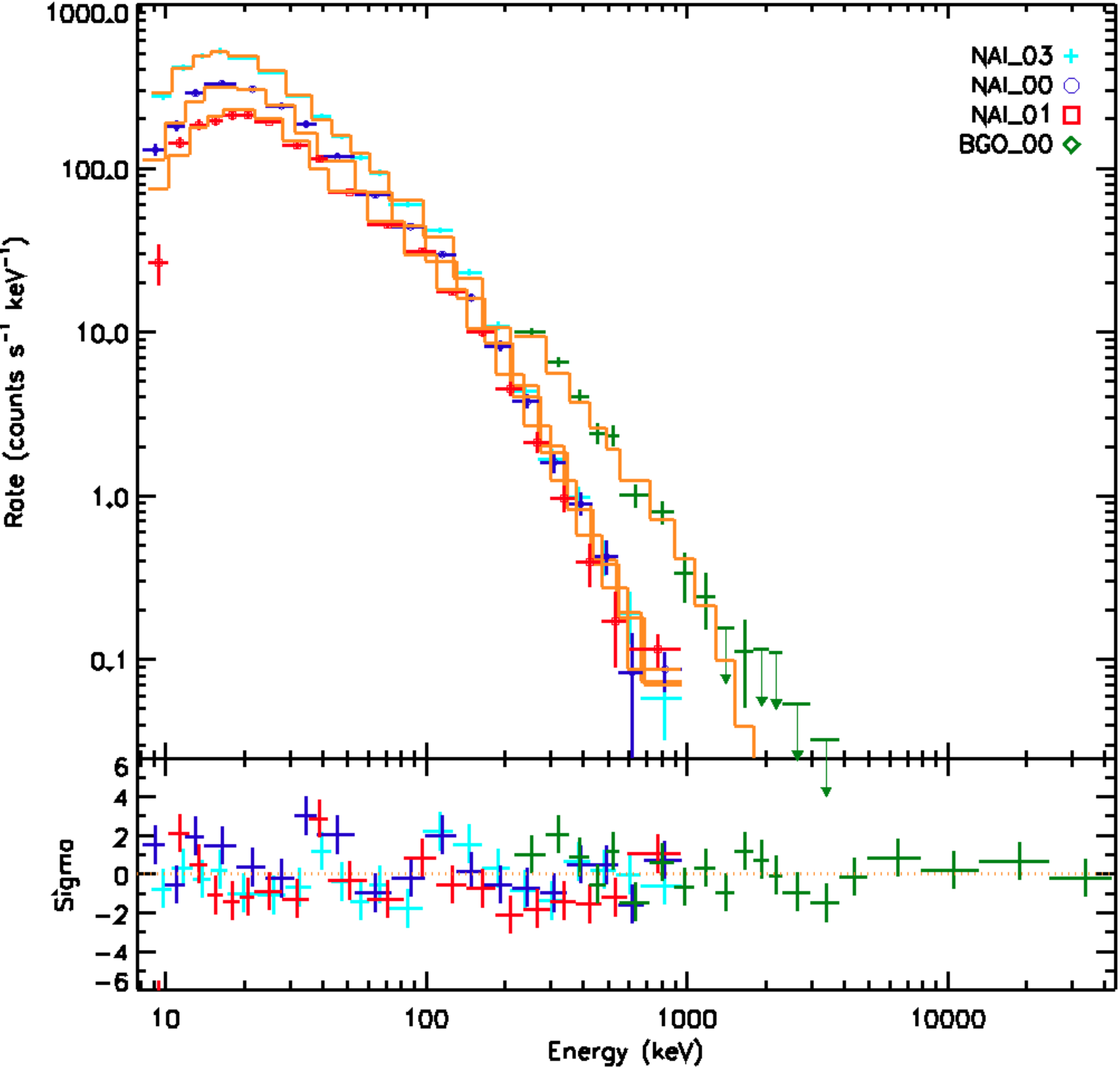}
\includegraphics[totalheight=0.172\textheight, clip]{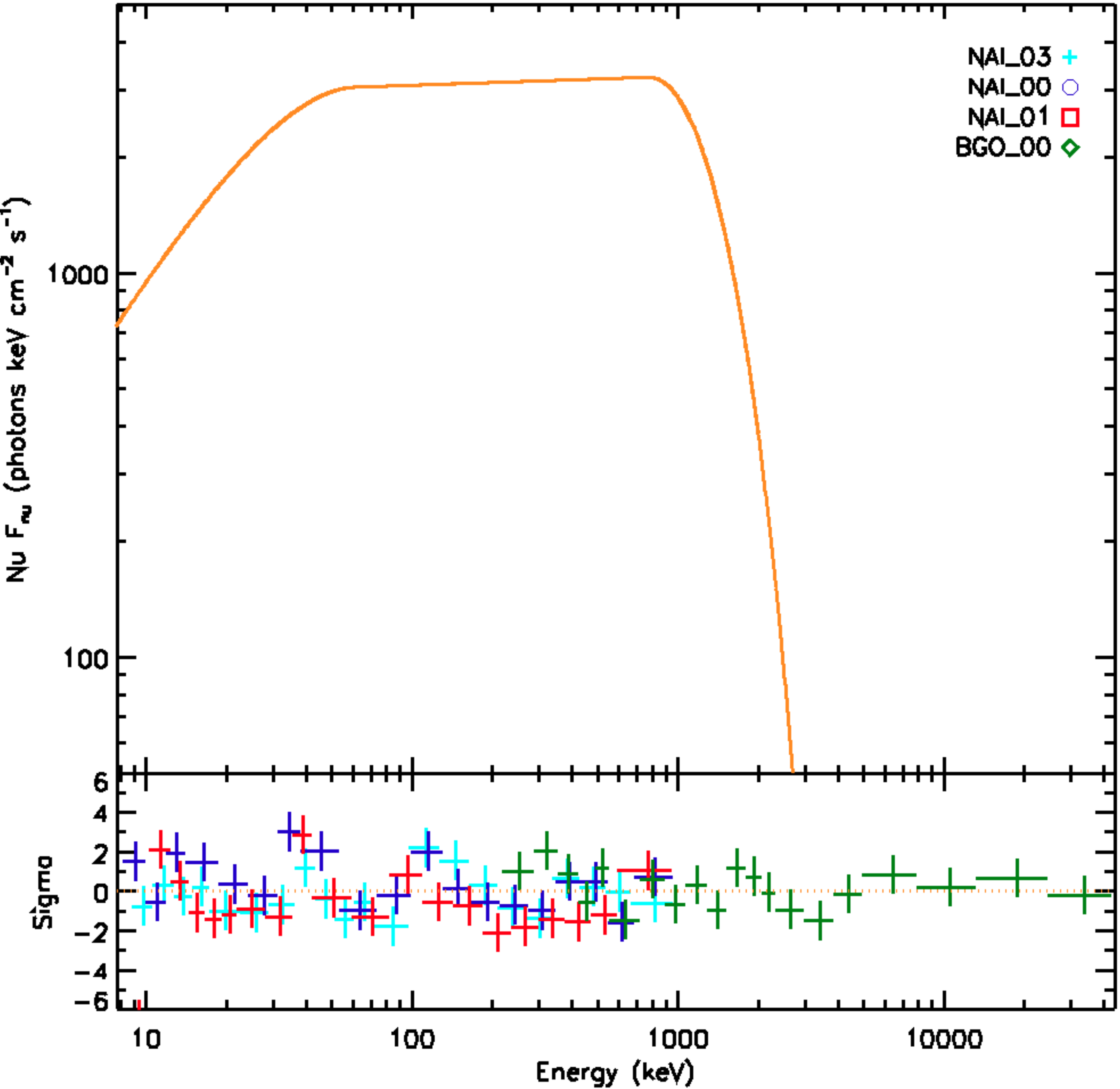}

\includegraphics[totalheight=0.173\textheight, clip]{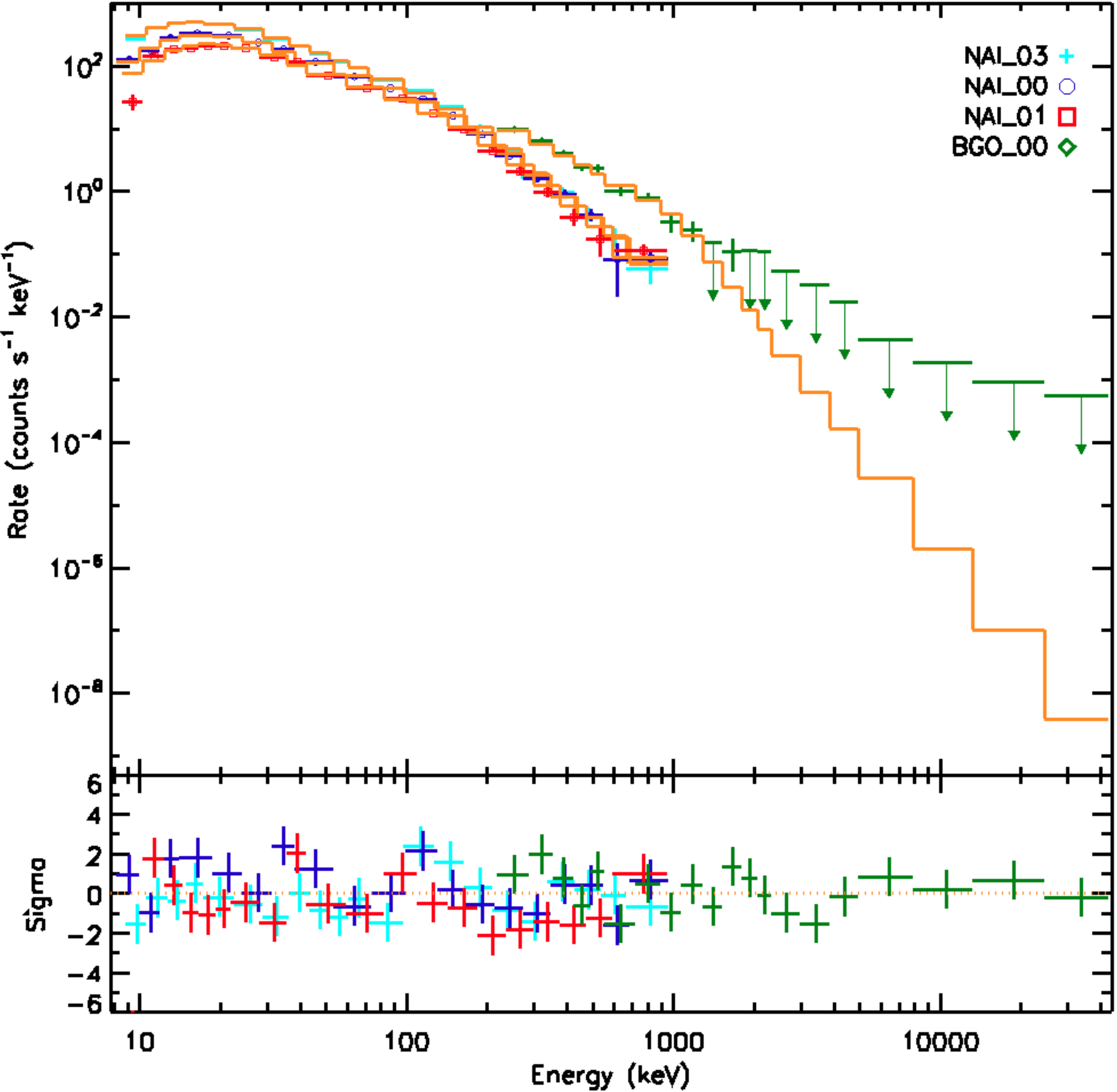}
\includegraphics[totalheight=0.173\textheight, clip]{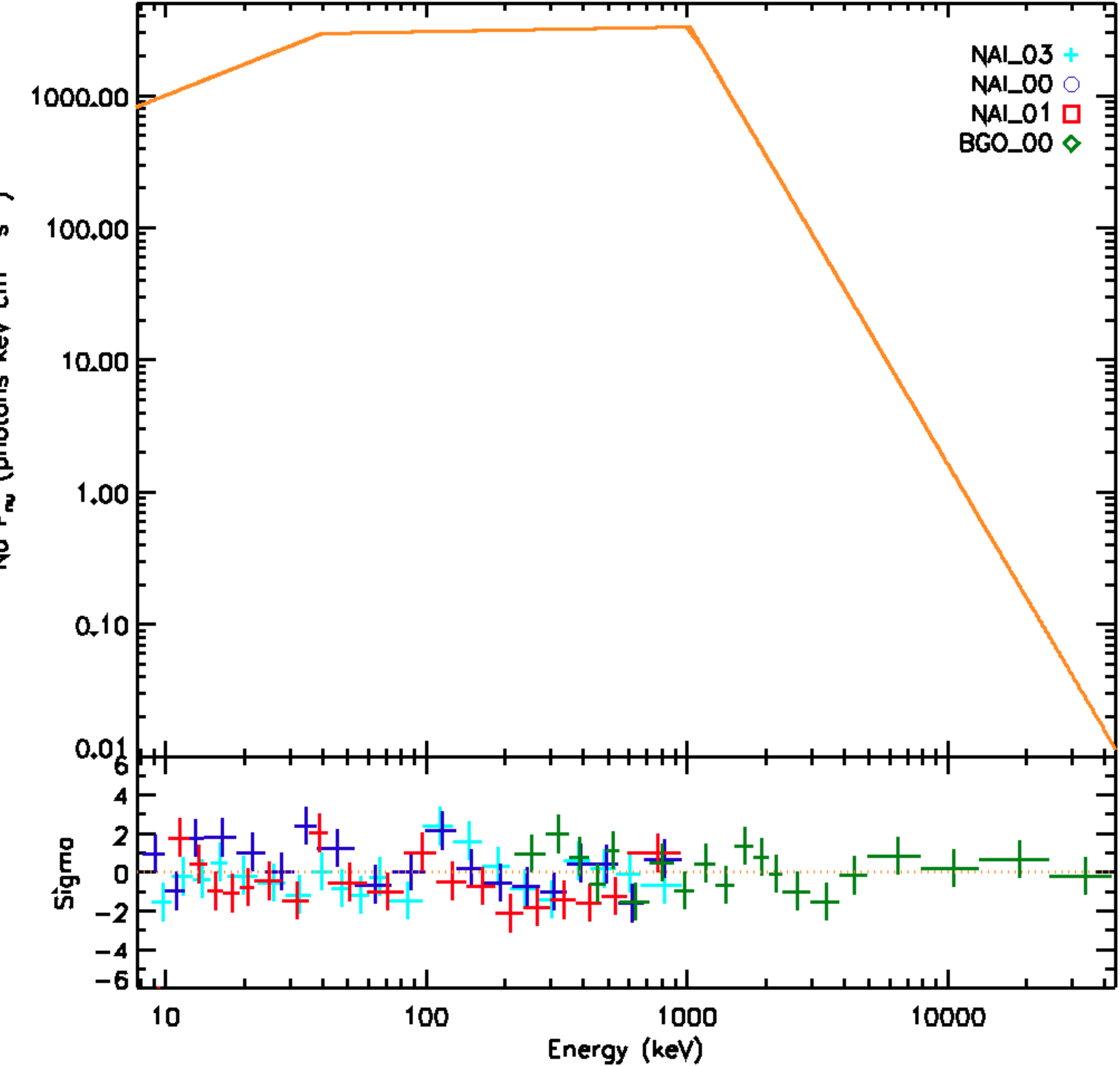}

\includegraphics[totalheight=0.174\textheight,clip]{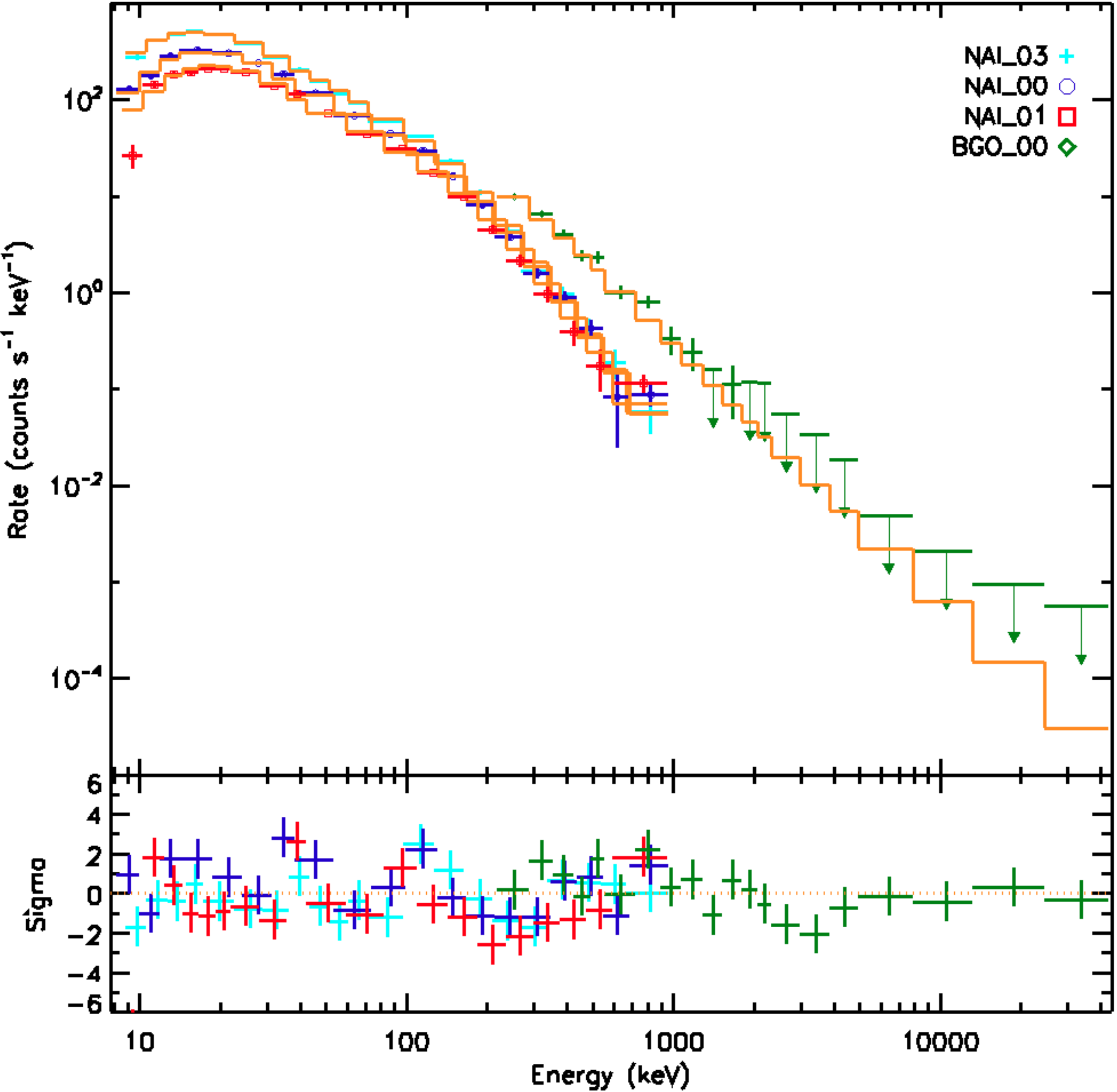}
\includegraphics[totalheight=0.174\textheight, clip]{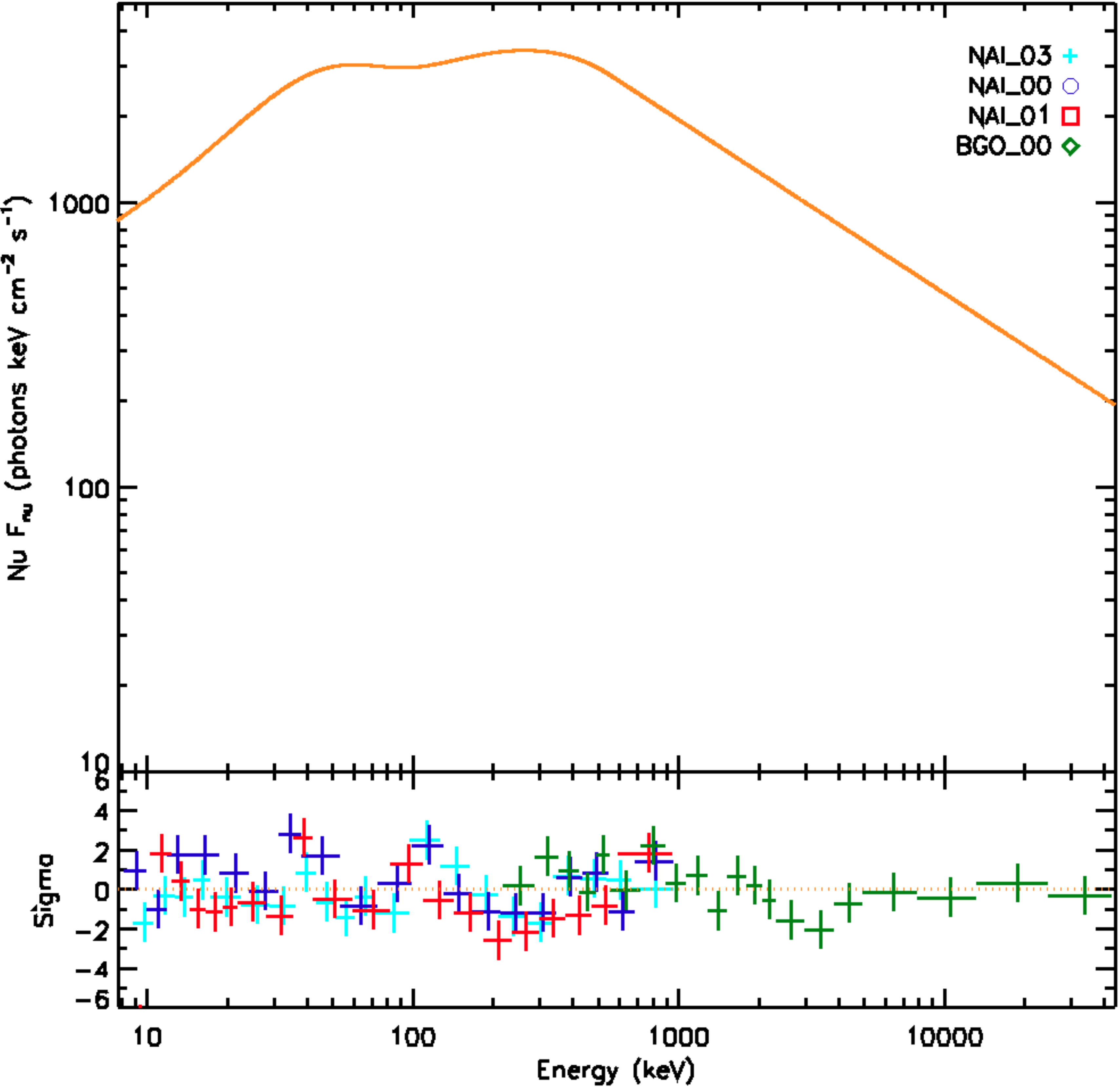}
\caption{\label{fig:GRB120323A_integrated_spectra}Time-integrated spectral analysis from T$_\mathrm{0}$-0.016s to T$_\mathrm{0}$+0.548s. The figures on the left side show count spectra and the figures on the right show the corresponding model in the $\nu$F$_{\nu}$ space. The models used are from top to bottom, Band, B+Cutoff, 2BPL, and B+BB. For each figure, the lower panel correspond to the residuals of the fit. For presentation purposes we rebined the energy channels in the figures, although the fit procedure is performed using the full spectral resolution of the instruments.}
\end{center}
\end{figure}

\begin{table*}
\caption{\label{tab:GRB120323A_integrated_spectra}Time-integrated spectral fit of \grb from T$_\mathrm{0}$-0.016\,s to T$_\mathrm{0}$+0.548\,s using the detectors n0, n1, n3 and b0 (see section~\ref{section:Time-integrated spectral analysis}).}
\begin{center}
{%\tiny
%\small
\begin{tabular}{|l|l|l|l|l|l|l|l|l|l|l|l|l|l}
\hline
\multicolumn{1}{|c|}{Models} &\multicolumn{6}{|c|}{Standard Model} & \multicolumn{3}{|c|}{Additional Model} & \multicolumn{1}{|c|}{Cstat/dof} \\
\hline
     &\multicolumn{6}{|c|}{Band, Compt or 2BPL} & \multicolumn{3}{|c|}{ BB, Compt or Band}  & \multicolumn{1}{|c|}{}\\
\hline
 \multicolumn{1}{|c|}{Parameters} & \multicolumn{1}{|c|}{E$_{\rm peak}$} & \multicolumn{1}{|c|}{$\alpha$} & \multicolumn{1}{|c|}{$\beta$} & \multicolumn{1}{|c|}{E$_{\rm b}$} & \multicolumn{1}{|c|}{E$_{\rm f}$} & \multicolumn{1}{|c|}{index} & \multicolumn{1}{|c|}{$\alpha$} & \multicolumn{1}{|c|}{$\beta$} & \multicolumn{1}{|c|}{kT or E$_{\rm 0}$} & \\
\hline
 \multicolumn{11}{|c|}{ }   \\  
 Band & 71 & -0.92 & -2.06 & -- & -- & -- & -- & -- & -- & 600/470 \\
     & $\pm5$ & $\pm0.07$ & $\pm0.02$ & -- & -- & -- & -- & --  & -- & -- \\
2BPL & 40 & -1.22 & -1.96 & 1024 & -- & -5.35 & -- & --  & -- & 540/468 \\
            &  $\pm2$ & $\pm0.03$ & $\pm0.02$ & $^\mathrm{+282}_\mathrm{-213}$ & -- & $^\mathrm{+1.75}_\mathrm{-19.4}$ & -- & -- & -- & -- \\
 B+Cut & 62 & -0.82 & -1.98 &  764 & 234 & -- & -- & -- & -- & 551/468 \\
              & $\pm4$ & $\pm0.08$ & $\pm0.02$ &  $^\mathrm{+308}_\mathrm{-450}$ & $_\mathrm{-279}^\mathrm{+592}$ & -- & -- & -- & -- & -- \\
 B+BB & 263 & -1.44 & $<-2.36$ & -- & --  & -- & -- & -- & 11.29 & 568/468 \\
     & $^\mathrm{+80}_\mathrm{-44}$ & $^\mathrm{+0.05}_\mathrm{-0.07}$ & -- &  -- & -- & -- & -- & -- & $^\mathrm{+1.21}_\mathrm{-0.68}$ & -- \\
 C+BB & 307 & -1.48 & -- & -- &  -- & -- & -- & -- & 11.72 & 567/469 \\
     & $^\mathrm{+24}_\mathrm{-20}$ & $\pm0.03$ & -- & --  & -- & -- & -- & -- & $\pm0.50$ & -- \\
 B+C2 & 239 & -1.45 & -2.49 & --  & -- & -- & +2.45 & -- & 9.61 & 568/467 \\
     &  $\pm25$ & $\pm0.03$ & $^\mathrm{+0.15}_\mathrm{-0.28}$ & --  & -- & -- & $^\mathrm{+0.57}_\mathrm{-0.51}$  & -- & $^\mathrm{+1.51}_\mathrm{-1.27}$ & -- \\
 B+B2 & 369 & -1.60 & $<-10$ & --  & -- & -- & +1.64 & -2.46 & 12.51 & 549/466 \\
     &  $^\mathrm{+48}_\mathrm{-0.43}$ & $\pm0.05$ & --  & -- & -- & -- & $\pm1.05$  & $^\mathrm{+0.19}_\mathrm{-0.29}$ & $^\mathrm{+6.25}_\mathrm{-3.21}$ & -- \\
 C+C2 & 433 & -1.27 & -- & --  & -- & -- & +0.68 & -- & 38.93 & 559/468 \\
     & $^\mathrm{+0.27}_\mathrm{-0.15}$ & $\pm0.05$ & -- & --  & -- & -- & $^\mathrm{+0.49}_\mathrm{-0.22}$ & -- & $^\mathrm{+10.3}_\mathrm{-11.3}$ & -- \\
 \multicolumn{11}{|c|}{ }   \\       
\hline
\end{tabular}
}
\end{center}
\end{table*}

We first analyzed the spectrum of \grb over the entire duration of the GBM prompt emission, from T$_\mathrm{0}$-0.016s to T$_\mathrm{0}$+0.548s. In Table~\ref{tab:GRB120323A_integrated_spectra} and in Figure~\ref{fig:GRB120323A_integrated_spectra} we report the results for the various acceptable combinations of the models we tested. We discuss our results below, progressing from single to multiple component fits. 

A single component Band function fit gives a value for E$_\mathrm{peak}$ of $\sim$70 keV, which is in the tail (2-3\%) of the E$_\mathrm{peak}$ distribution for GRBs observed with GBM and BATSE~\citep{Goldstein:2012a,Paciesas:1999}. This low value of E$_\mathrm{peak}$ is very unusual for intense short GRBs, whose E$_\mathrm{peak}$ values are typically much higher than those of long ones~\citep{Paciesas:1999, Guiriec:2010}. A high redshift (z) for this GRB could reconcile the observed low E$_\mathrm{peak}$ value with the typical E$_\mathrm{peak}$ distribution. However, while E$_\mathrm{peak}$ evolves as (1+z), the luminosity evolves as 4$\pi$D$_\mathrm{L}^\mathrm{2}$~$\propto$~z$^\mathrm{2}$. Thus, a large distance would also increase dramatically the intrinsic luminosity of this GRB which is already extremely high. The energy flux (8 keV - 1 MeV) of \grb in the observer frame computed from T$_\mathrm{0}$-0.016s to T$_\mathrm{0}$+0.548s is  (1.95$\pm$0.02)$\times$10$^\mathrm{-5}$ erg cm$^\mathrm{-2}$.s$^\mathrm{-1}$, which is above all the values reported in \citet{Goldstein:2012a}. In addition, as we show in Section~\ref{section:Flux-Epeak correlation}, a redshift of $\sim$1 for \grb is compatible with the observations.

The Band function alone is not sufficient to describe adequately the time-integrated spectrum. We find that more complex models significantly improve the fits. A double broken power law (2BPL) and a Band function with a cutoff in the high energy power law (B+Cut) give the largest Cstat improvement over Band alone (60 and 49 units, respectively, for two additional degrees of freedom - dof), which suggests the existence of two spectral breaks, one around few tens of keV and the other around 1 MeV.

We next used a combination of a Band function with a BB (B+BB) as proposed in~\citet{Guiriec:2011a} and found that it significantly improves the Band-only fit by 34 units of Cstat for two additional dof. Interestingly, the BB affects the parameters of the Band function similarly to what was already reported in~\citet{Guiriec:2011a} for GRB~100724B: both $\alpha$ and $\beta$ are shifted towards lower values, and E$_\mathrm{peak}$ changes from $\sim$70 keV to $\sim$200-300 keV, a more typical value for a short GRB. We also notice that the temperature of the BB, $kT\sim 10-13$ keV, corresponds to a $\nu$F$_\nu$ spectrum with a maximum\footnote{A BB spectrum peaks at an energy of  about 3 times its temperature kT} around $30-39$ keV, which matches the first energy break obtained with the 2BPL and B+Cut models. Further, the E$_\mathrm{peak}$ of the Band function is compatible with the second energy break at several hundred keV obtained with 2BPL and B+Cut.

However, with a B+BB model, we can only determine an upper limit for $\beta$, which makes the Band function similar to \fdaigne{a power law with an exponential cutoff~\citep[later CPL -- ][]{Kaneko:2006}}; both fits give similar parameters E$_\mathrm{peak}$ and $\alpha$. This is also evident from the fact that the combination of a \fdaigne{CPL} with a BB (C+BB) which has only one dof difference from Band, leads to the same Cstat improvement as B+BB. The Cstat change ($\Delta$Cstat) per dof between Band and C+BB is the largest of the tested models. The $\Delta$Cstat per dof differences between Band fits only and C+BB, 2BPL, B+Cut and B+BB are 33, 30, 25, and 16 respectively. We discuss below the uniqueness of the various selected components in our fits.

As shown in Figure~\ref{fig:GRB120323A_integrated_spectra} (bottom right panel), the BB component appears like a hump in the low energy power law of the Band function. In order to better explore the intrinsic shape of this hump, we replaced the BB component with less constrained shapes such as another Band (B+B2) or a \fdaigne{CPL} (B+C2) function. With more parameters, the Band (i.e., B2) and \fdaigne{CPL} (i.e., C2) functions have no reason to mimic the shape of a thermal Planck function, which can be approximated with a Band function with $\alpha$ = 1 and a very steep $\beta$, which is also equivalent to a \fdaigne{CPL} function with a power law index value of 1. In addition, due to reprocessing of the thermal emission, a pure black body shape is very unlikely to be obtained. Interestingly, for both B+B2 and B+C2, the $\alpha$ values of B2 or C2 are positive and range between +0.60 and +3.00 and the spectral peaks of B2 and C are compatible with the temperature of the BB. This shows that the hump identified in the low energy power law of the Band function is compatible with a thermal origin and is adequatelly approximated with a BB component.

Next, we tested the effect of the constrained curvature of the Band function by replacing the Band function with a smoothly broken power law (SBPL) with a free break scale, in the B, B+Cut and B+BB models. We obtained a slight improvement of 19, 10 and 10 units of Cstat, respectively, which does not impact drastically the fit results nor the model comparisons. Therefore, in the following we will use the Band function since with one dof less it is easier to derive well constrained fits, especially in shorter time bins and when several components are used.

We note that the global shapes of the three favored models, 2BPL, B+Cut and B+BB (see Figure~\ref{fig:GRB120323A_integrated_spectra}) are very similar and indicate that the time-integrated $\nu$F$_\nu$ spectrum is better fit with two bumps rather than the single one of the Band function. However, none of the tested models lead to a completely satisfactory fit of the time integrated spectrum. Whatever the model, systematic patterns remain in the fit residuals, which are not distributed randomly around zero across the studied energy range (see Figure~\ref{fig:GRB120323A_integrated_spectra}). This indicates that the models tested are either not sufficient enough to describe the data, or that a possible strong spectral evolution during the burst leads to an unsatisfactory description of the spectrum with standard models when integrated over the whole burst duration. It is very likely that our models could present a good description of the data when integrated over time scales encompassing periods with no or less spectral evolution. To investigate such possible issues as well as to follow the evolution of the spectrum during the burst, we performed time-resolved spectral analysis at shorter time scales, presented in Section~\ref{section:Time-resolved spectral analysis}.

We compared the spectral results obtained with GBM to the {\it Konus}-WIND spectra to ensure that there is no major calibration problems. We analyzed GBM data from T$_\mathrm{0}$+0.002\,s to T$_\mathrm{0}$+0.256\,s, which is a similar time interval to the one used in~\citet{Golenetskii:2012B} taking into account the propagation time between the two spacecrafts (V. Pal'shin private communication). Using a power law with exponential cutoff as proposed in~\citet{Golenetskii:2012B}, we obtained an E$_\mathrm{peak}$ value of 251$^\mathrm{+16}_\mathrm{-14}$ keV and a power law index value of -1.65$\pm0.02$. These results are compatible with those reported in~\citet{Golenetskii:2012B} (i.e., E$_\mathrm{peak}$=331$^\mathrm{+64}_\mathrm{-50}$ keV and index=-1.57$\pm0.07$). Such a crosscheck minimizes calibration issues between the two different instruments and reinforces our confidence in the goodness of the data set we used for this analysis.

\section{Time-resolved spectral analysis}
\label{section:Time-resolved spectral analysis}

We perform time-resolved spectroscopy of \grb at two different time scales, and results are presented in Section~\ref{sec:Coarse time resolved}~and~~\ref{sec:Fine time resolved}. Our goals are (i) to investigate the effects of possible spectral evolution during the burst, and (ii) to track the evolution of the various spectral parameters of the fit components, such as power law indices, break energies and temperatures.

\subsection{Coarse time-resolved spectral analysis}
\label{sec:Coarse time resolved}

First, we defined four broad time intervals based on the main structures identified between 20 and 150 keV in the light curves presented in Figure~\ref{fig:GRB120323A_LCs}. The first time interval from T$_\mathrm{0}$-0.018\,s to T$_\mathrm{0}$+0.058\,s includes the first peak of the light curve, the second from T$_\mathrm{0}$+0.058\,s to T$_\mathrm{0}$+0.100\,s covers the most intense part of the second peak, and the time intervals from T$_\mathrm{0}$+0.100\,s to T$_\mathrm{0}$+0.174\,s and from T$_\mathrm{0}$+0.174\,s to T$_\mathrm{0}$+0.600\,s correspond to the decay phase of the light curve. We fitted each time interval with the models used in section~\ref{section:Time-integrated spectral analysis}; the fit results are shown in Table~\ref{tab:GRB120323A_resolved_spectra_large}.

We find that while the 2BPL and B+Cut models showed evidence for a high energy cutoff in the time-integrated spectrum, they do not lead to the same conclusion when fitting time-resolved spectra. The Cstat improvement obtained with these models compared to a Band function only is modest in all four time intervals. This leads to the conclusion that the cutoff measured in the time-integrated spectrum is an artifact due to the strong spectral evolution present in \grbnos. This result also demonstrates that the measured `spectral cutoff' at high energies based on the extrapolation of GBM time-integrated spectral fits to the LAT energy range (as proposed in e.g., ~\citet{Ackermann:2012}) should be interpreted with caution.

B+BB or C+BB lead to the largest Cstat improvement compared to Band-only, with a $\Delta$Cstat between Band and C+BB of 46 and 18 units for 1 dof difference in the second and third time intervals, respectively. We note here that with Band-only fits, we find unusually low values of E$_\mathrm{peak}$ (a few tens of keV) in the first two time intervals with the highest intensity, while the third one corresponding to the global decay phase of the light curve has an E$_\mathrm{peak}$ around 400 keV. The low E$_\mathrm{peak}$ values of the first two intervals are accompanied with high (positive) $\alpha$ values. The second and third time intervals exhibit lower values for $\alpha$ with power law slopes steeper than -1.5. In contrast, with a two component fit, B+BB or C+BB, E$_\mathrm{peak}$ is shifted to higher energies. This is particularly obvious for the second time interval, for which E$_\mathrm{peak}$ is shifted from $\sim$40 keV to $\sim$500 keV. The addition of the BB to the Band function also leads to a lower value of $\alpha$ and $\beta$. When the BB is replaced with a Band or a \fdaigne{CPL} function like in B+C2, C+B2 and C+C2, the parameters of this function are similar to those of the BB function with similar Cstat value. These results confirm the existence of a hump in the low energy part of the spectrum.

\begin{figure*}
\includegraphics[totalheight=0.20\textheight, clip,viewport=0 42 512 395]{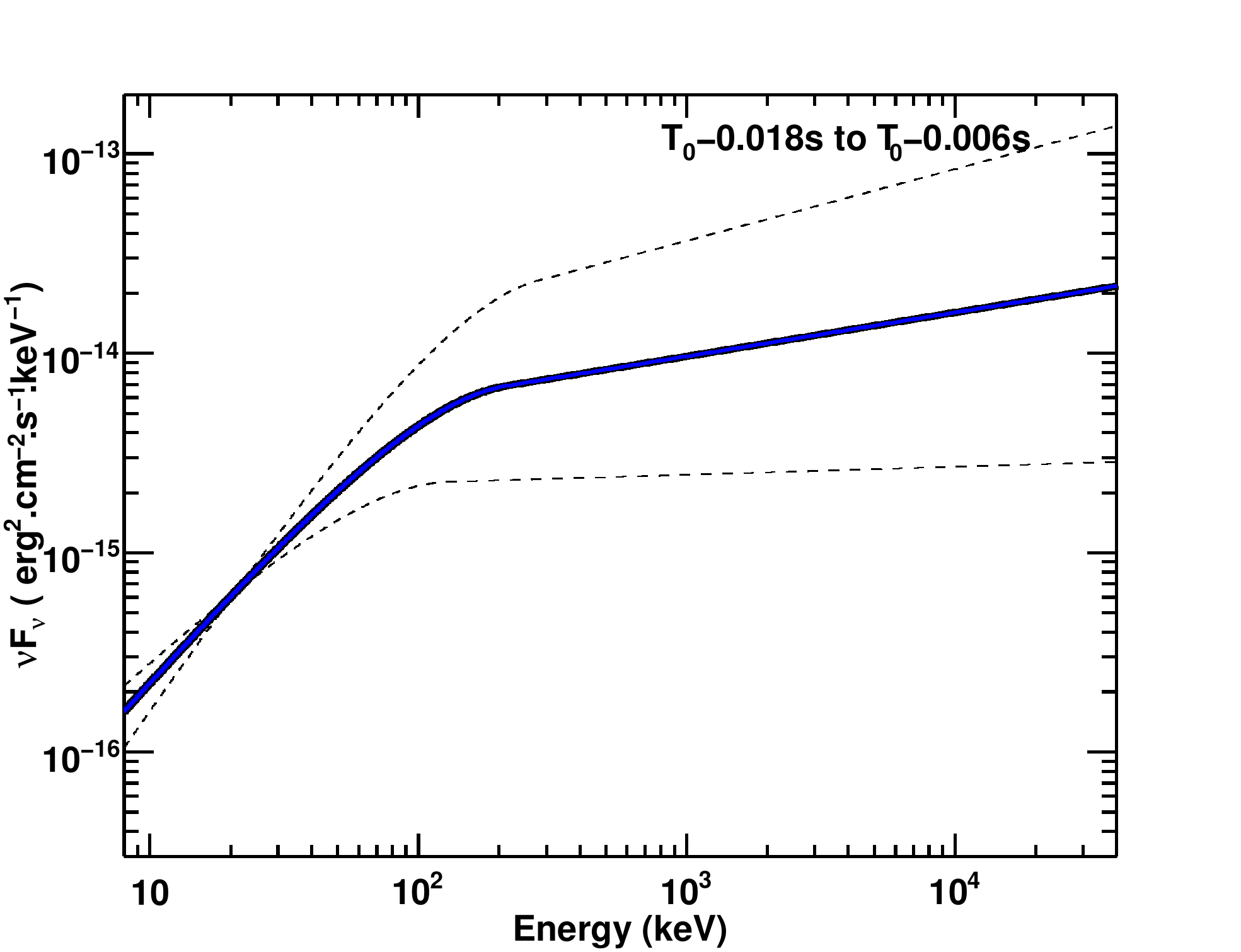}
\includegraphics[totalheight=0.20\textheight, clip,viewport=56 42 512 395]{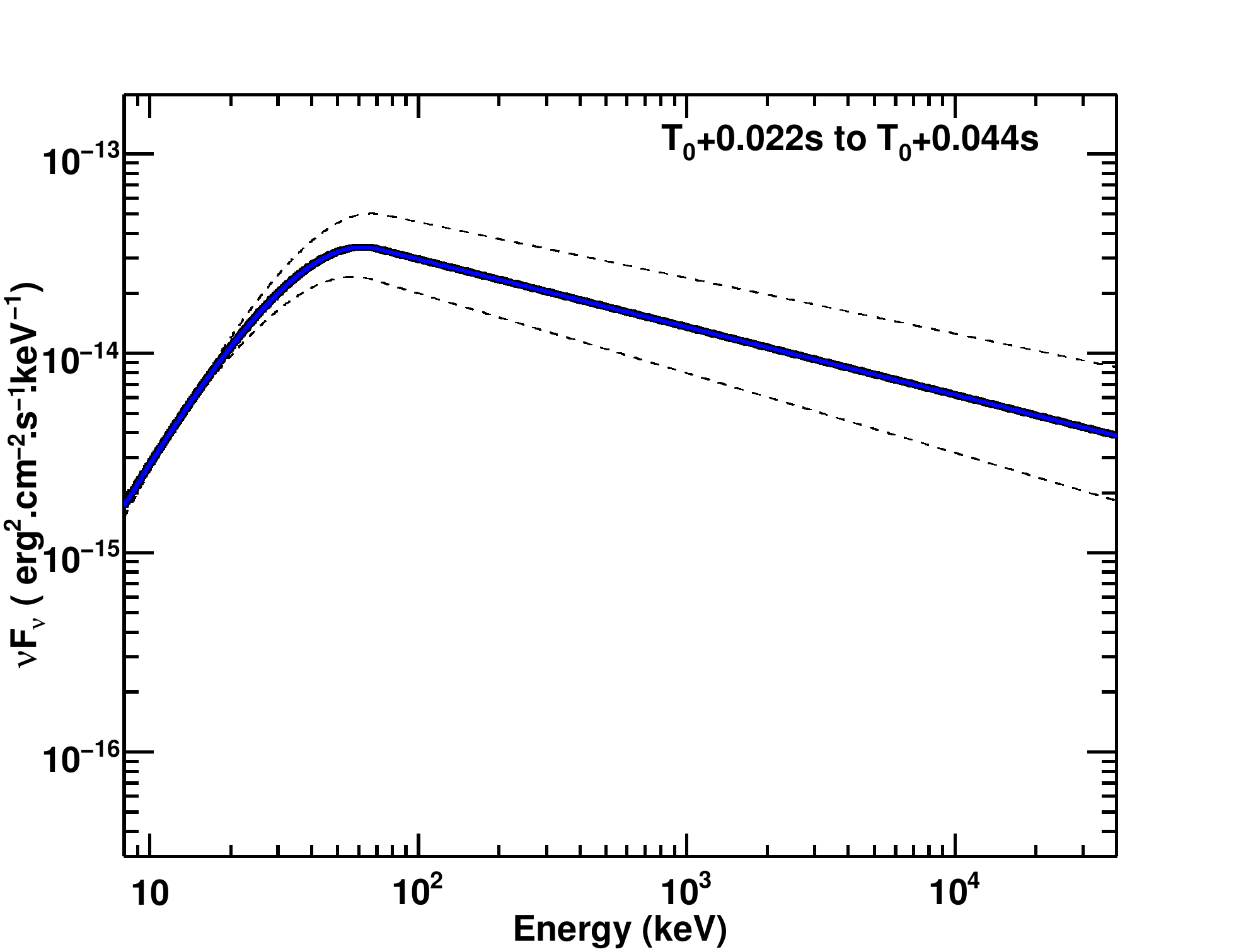}
\includegraphics[totalheight=0.20\textheight, clip,viewport=56 42 512 395]{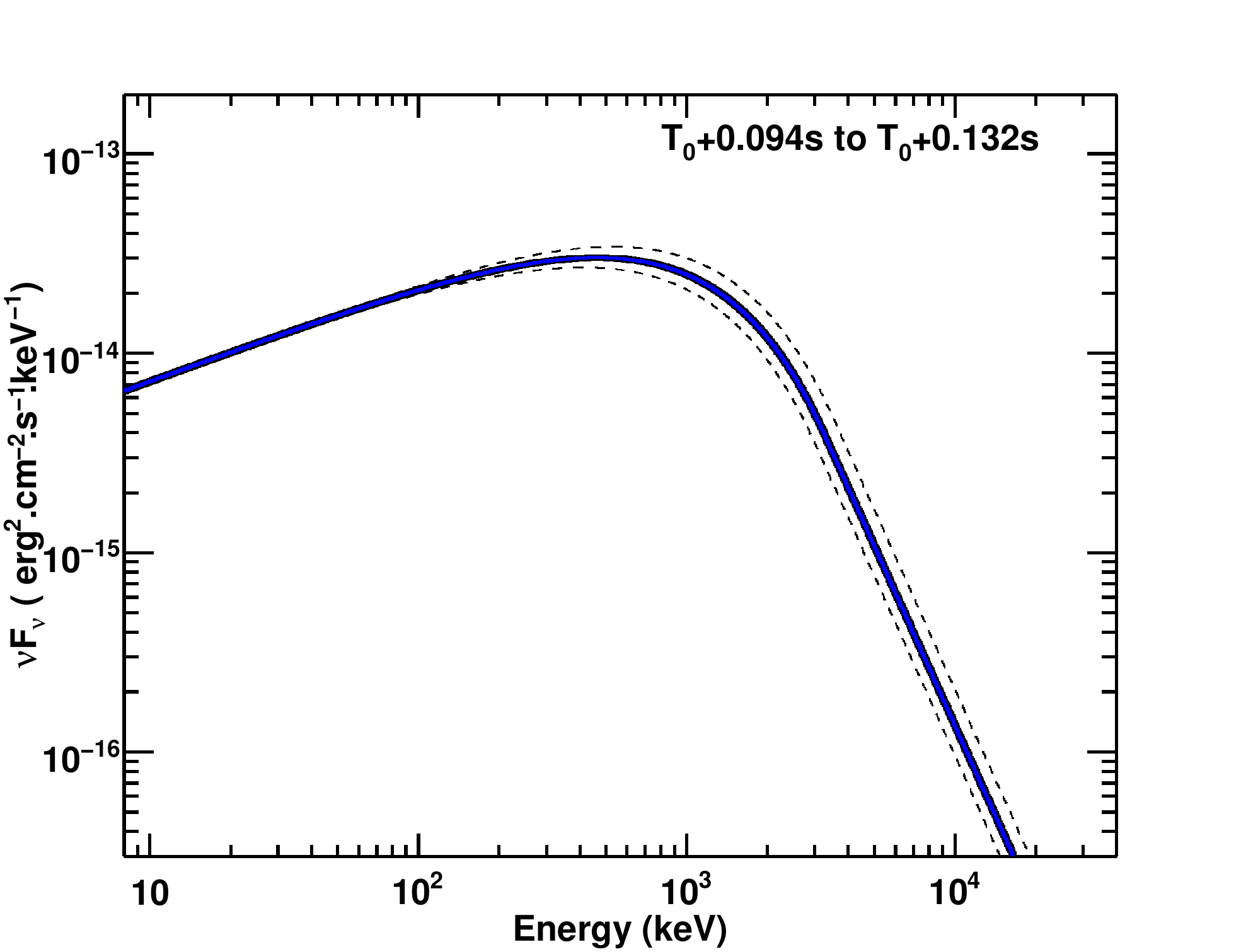}

\includegraphics[totalheight=0.20\textheight, clip,viewport=0 42 512 395]{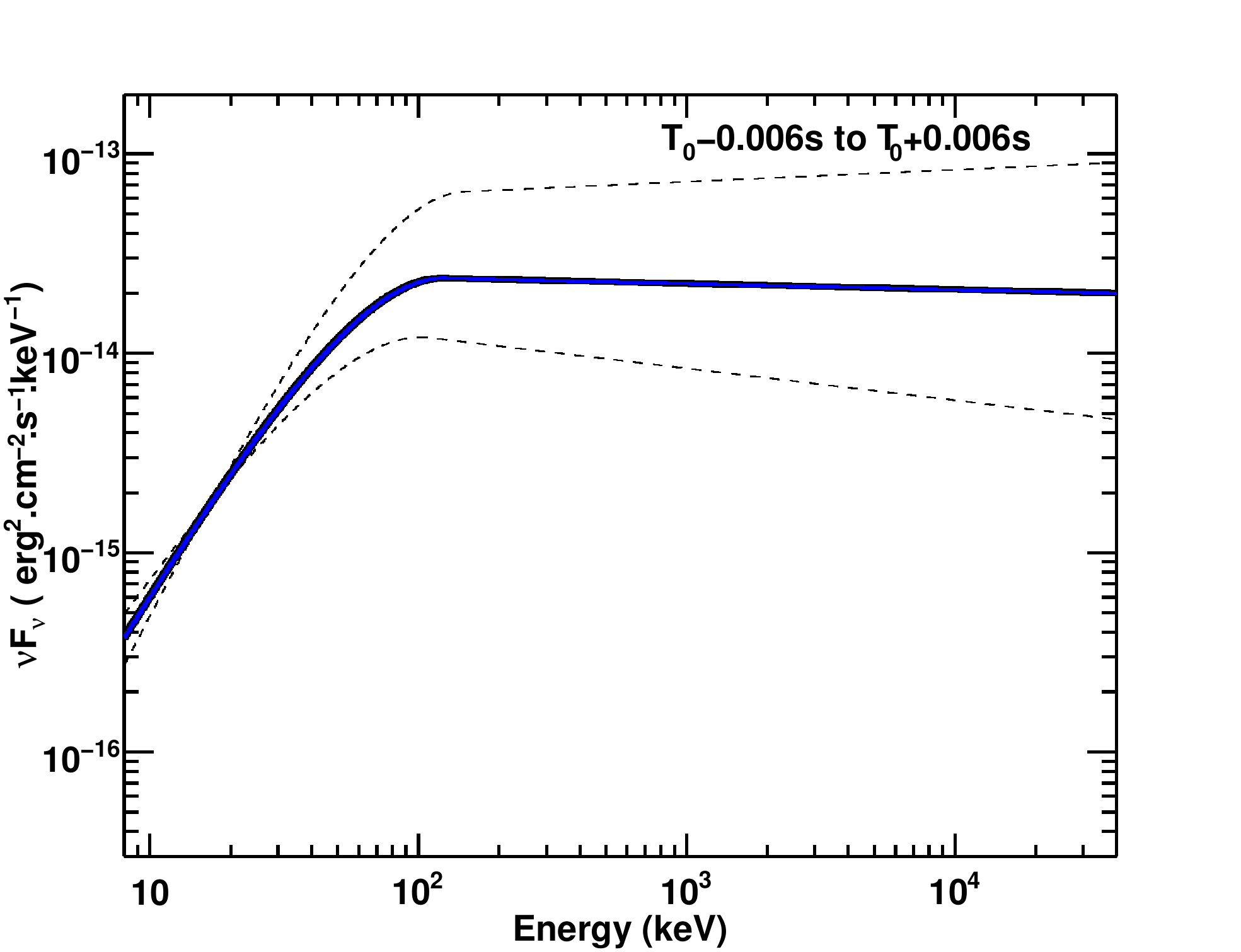}
\includegraphics[totalheight=0.20\textheight, clip,viewport=56 42 512 395]{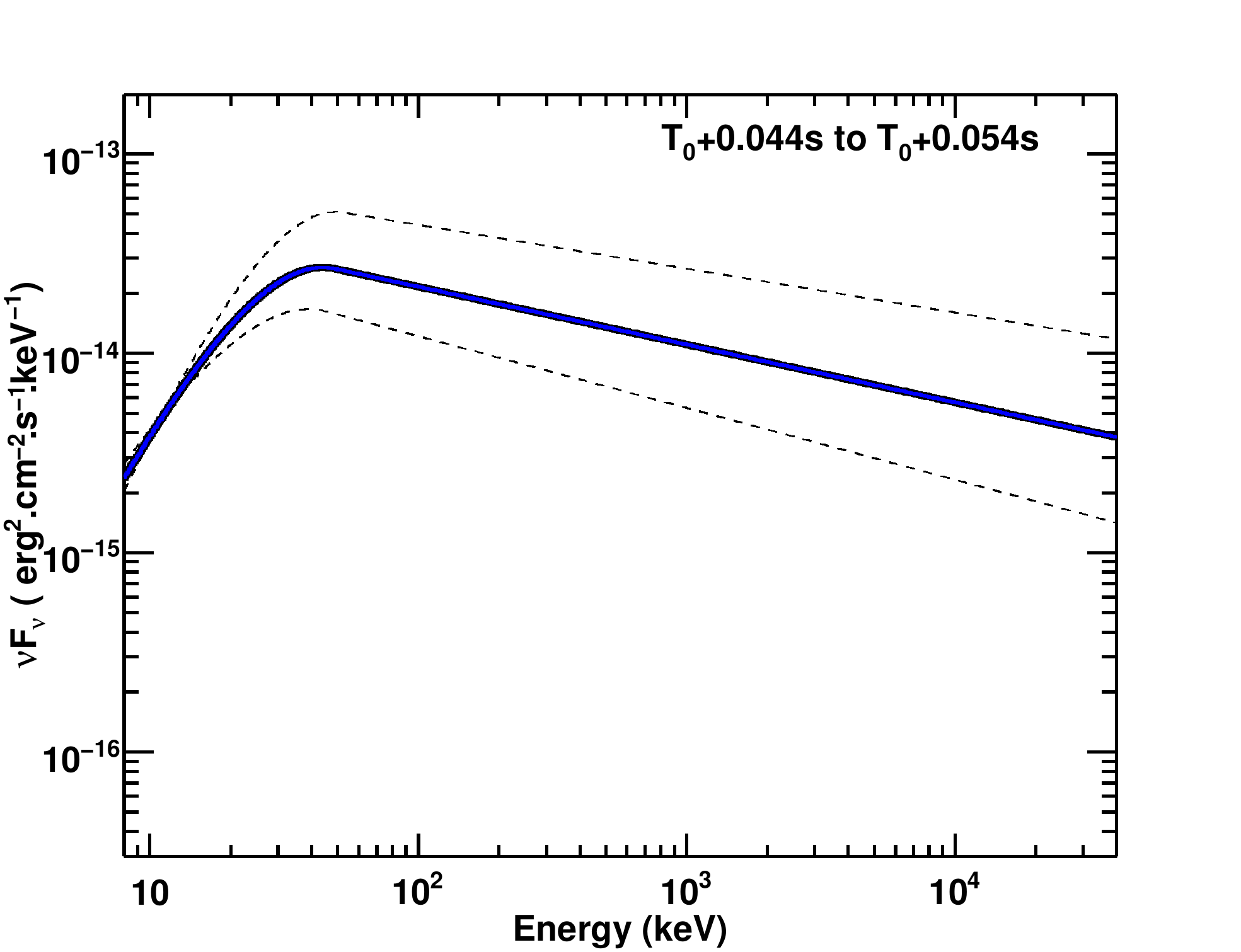}
\includegraphics[totalheight=0.20\textheight, clip,viewport=56 42 512 395]{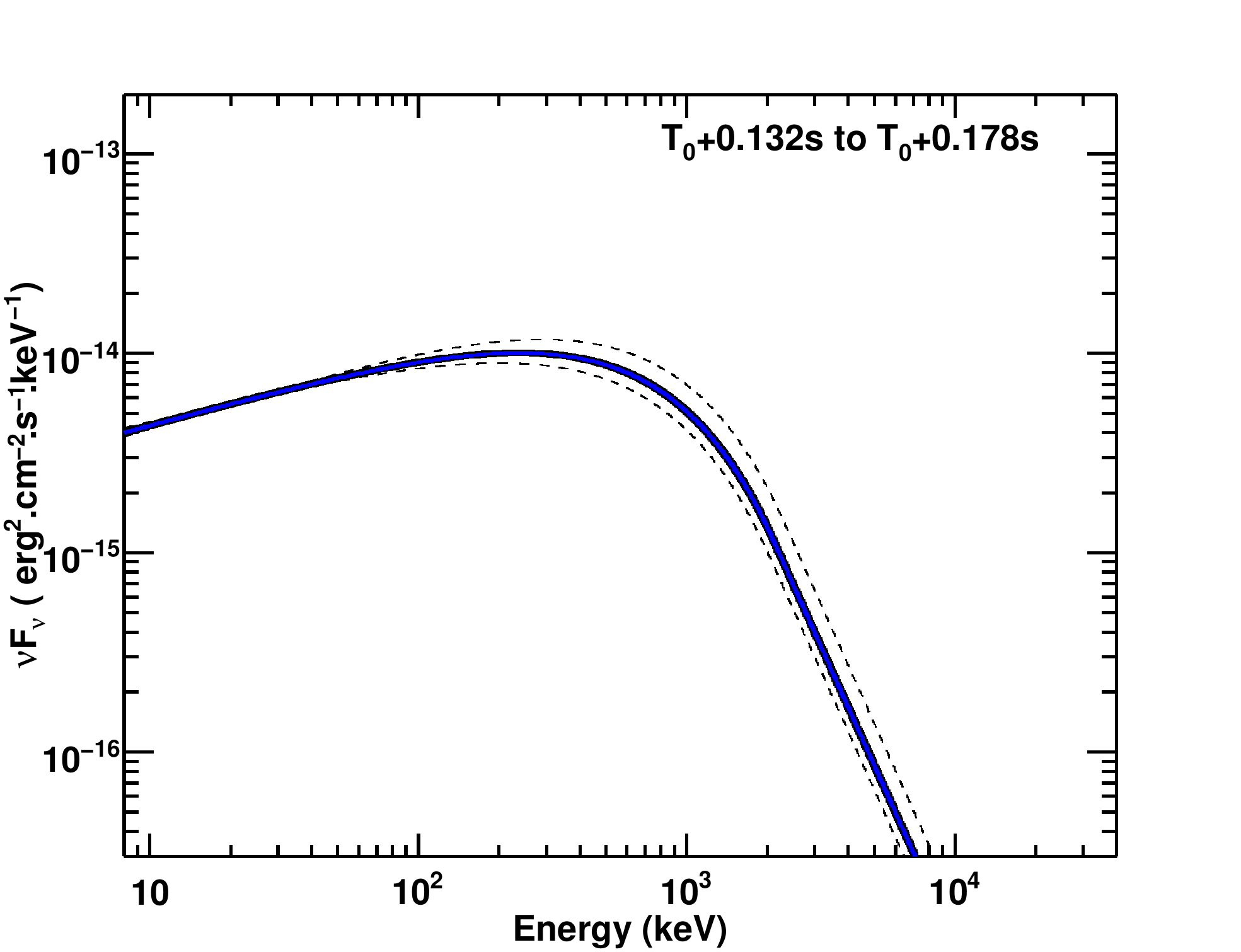}

\includegraphics[totalheight=0.20\textheight, clip,viewport=0 42 512 395]{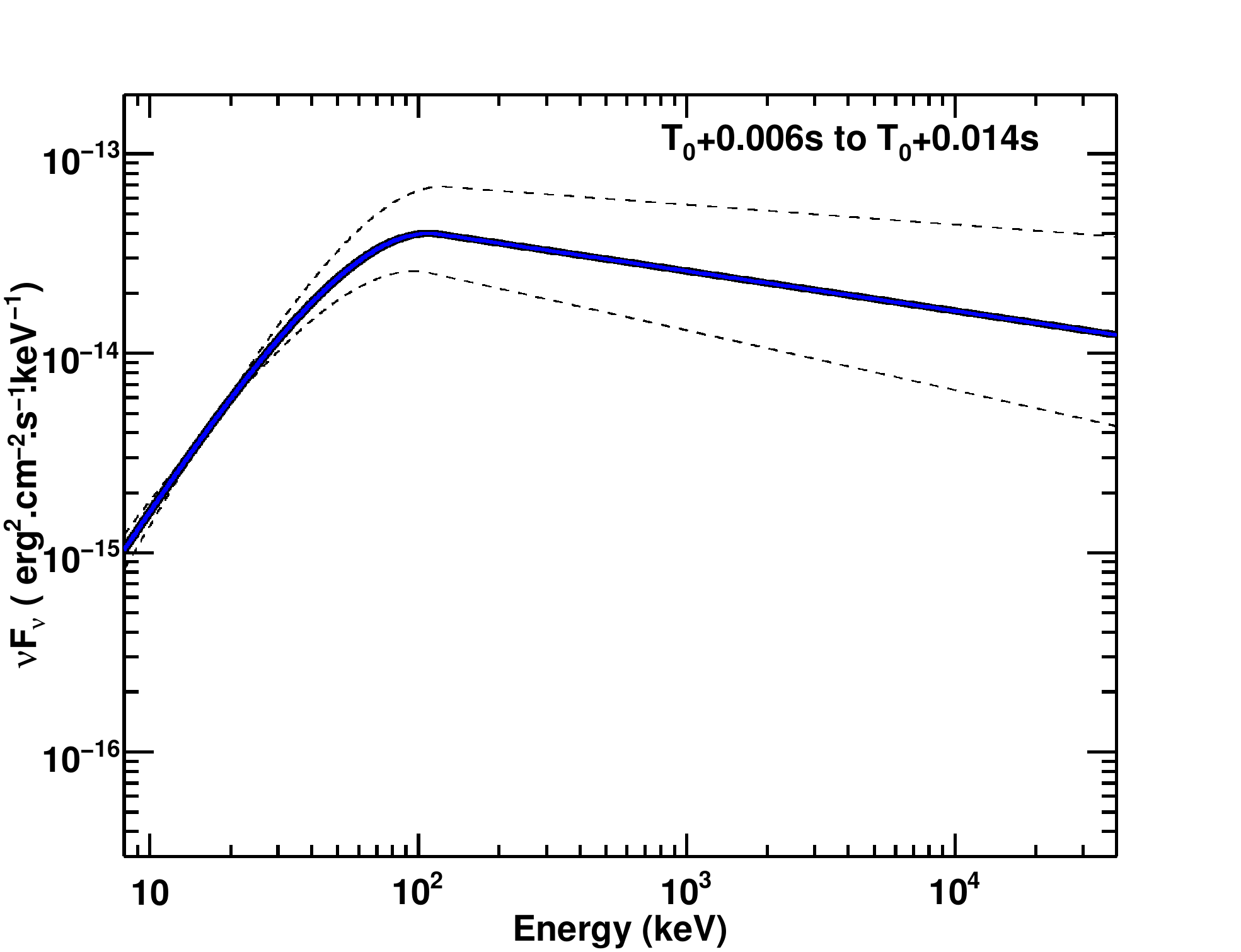}
\includegraphics[totalheight=0.20\textheight, clip,viewport=56 42 512 395]{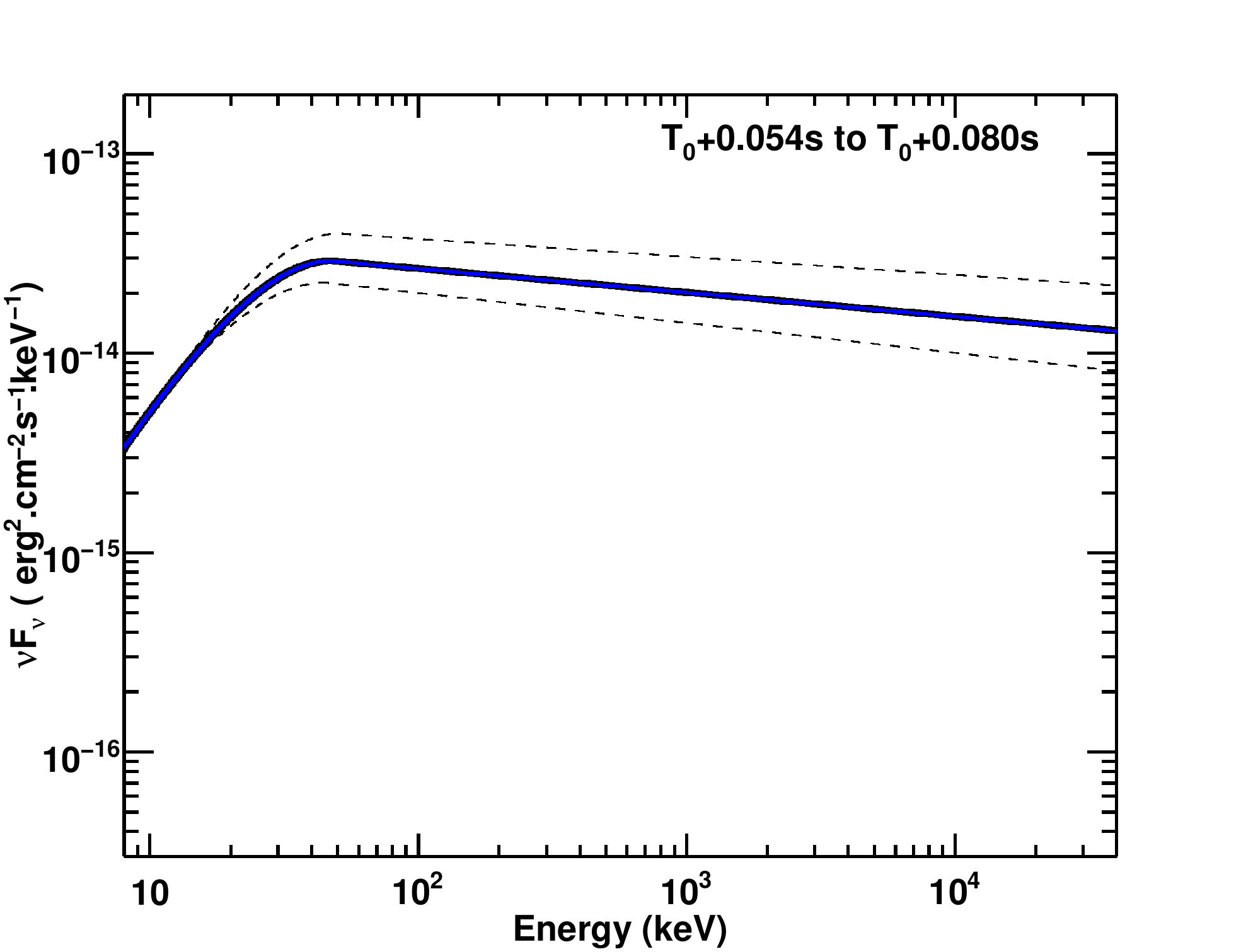}
\includegraphics[totalheight=0.20\textheight, clip,viewport=56 42 512 395]{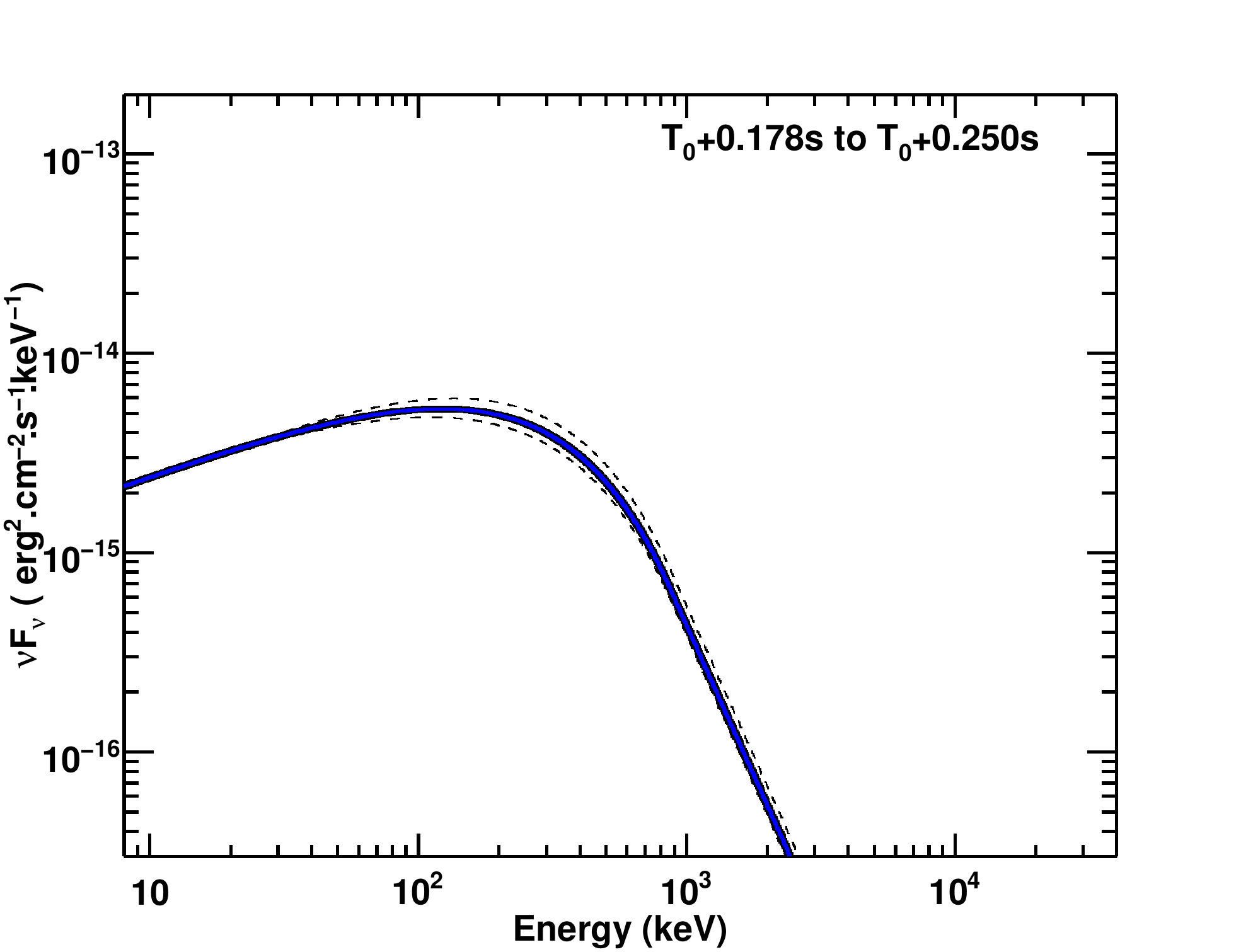}

\includegraphics[totalheight=0.224\textheight, clip,viewport=0 0 512 395]{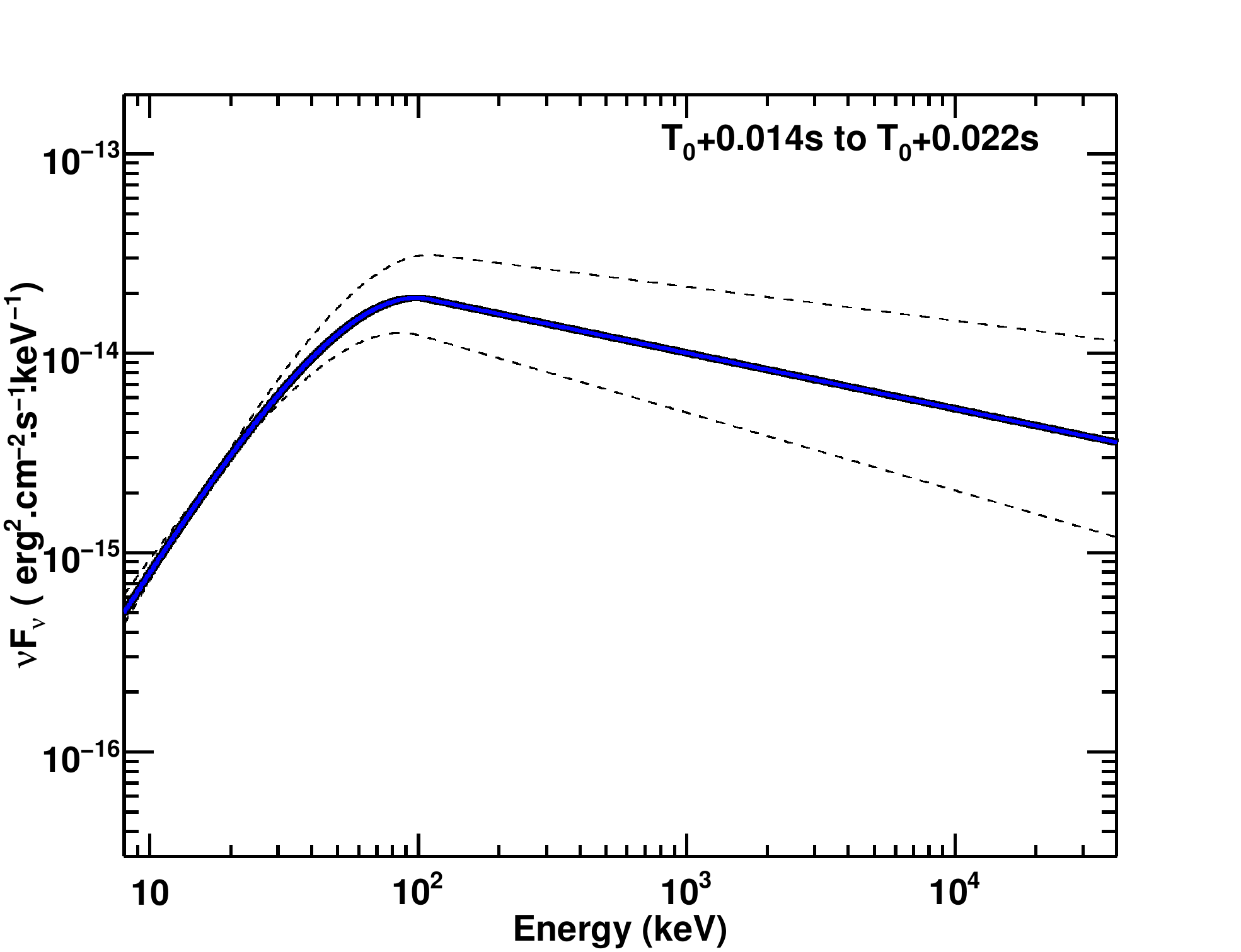}
\includegraphics[totalheight=0.224\textheight, clip,viewport=56 0 512 395]{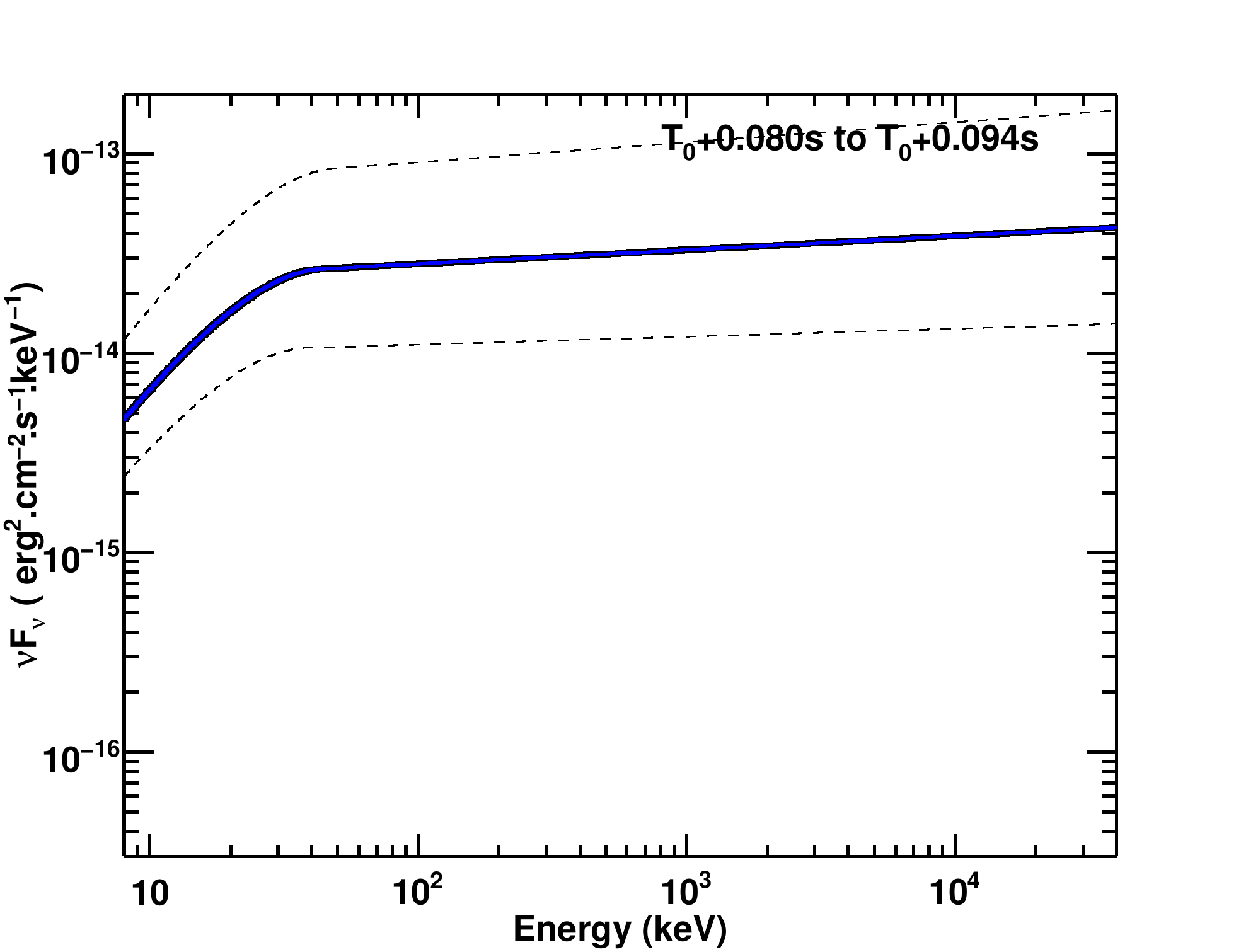}
\includegraphics[totalheight=0.224\textheight, clip,viewport=56 0 512 395]{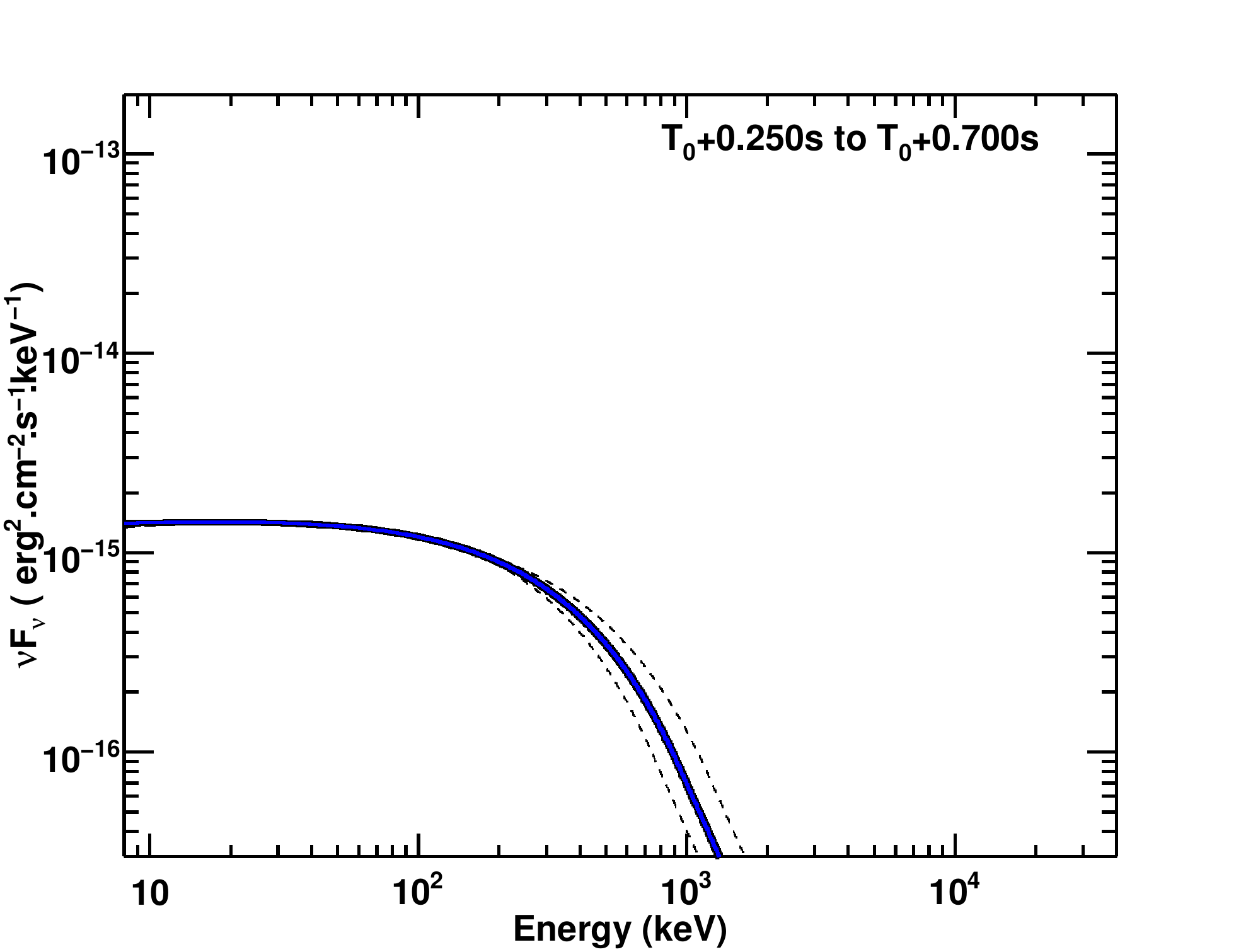}

\caption{\label{fig:GRB120323A_resolved_spectra_B}Fine time-resolved $\nu$F$_\nu$ spectra using Band function only. The solid line corresponds to the model obtained with the best parameters from the fit, and the thin lines correspond to the 1 $\sigma$ uncertainty on the best fit.}
\end{figure*}

\begin{figure*}
%%\hspace{-1.0cm}
\includegraphics[totalheight=0.20\textheight, clip,viewport=0 42 512 395]{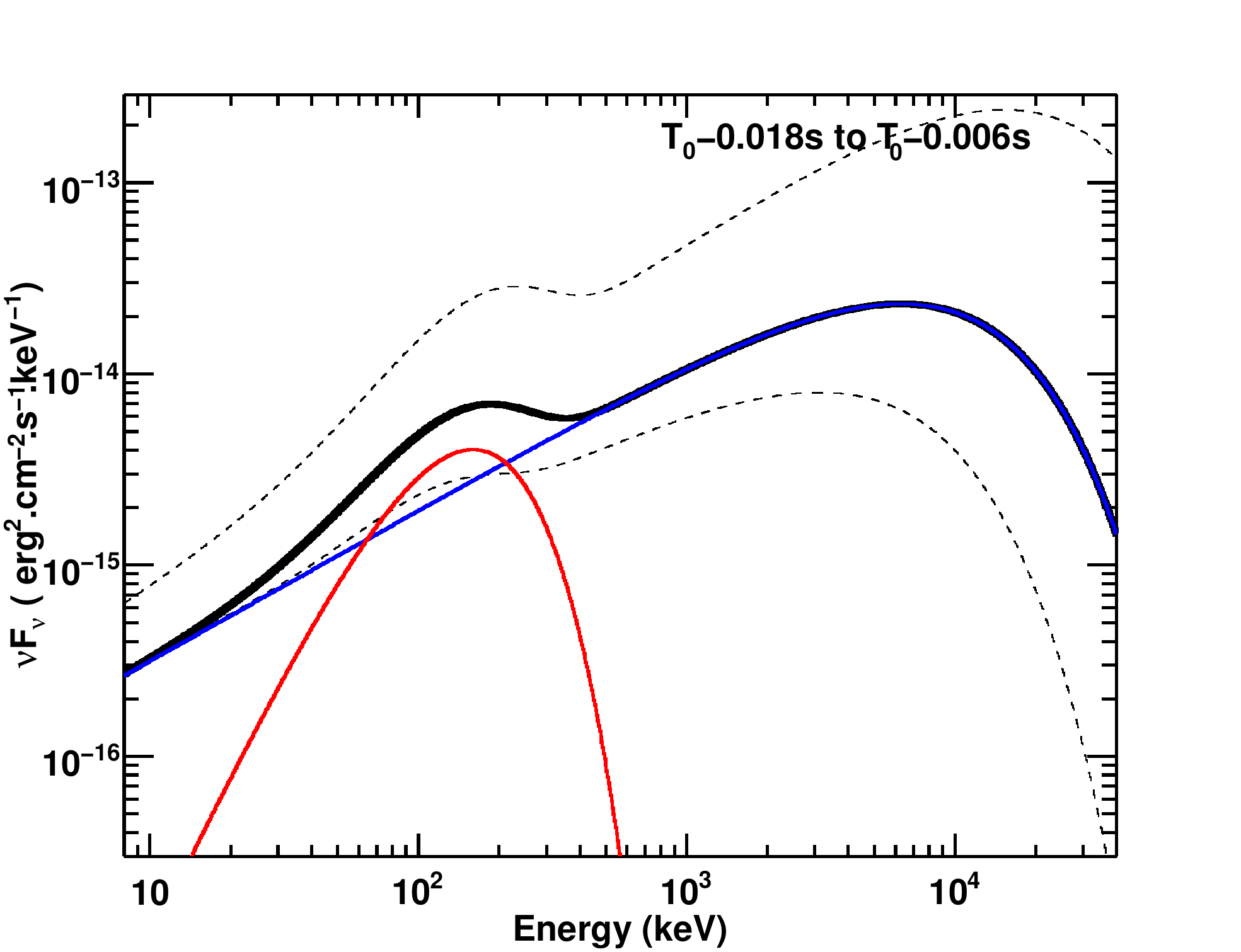}
\includegraphics[totalheight=0.20\textheight, clip,viewport=56 42 512 395]{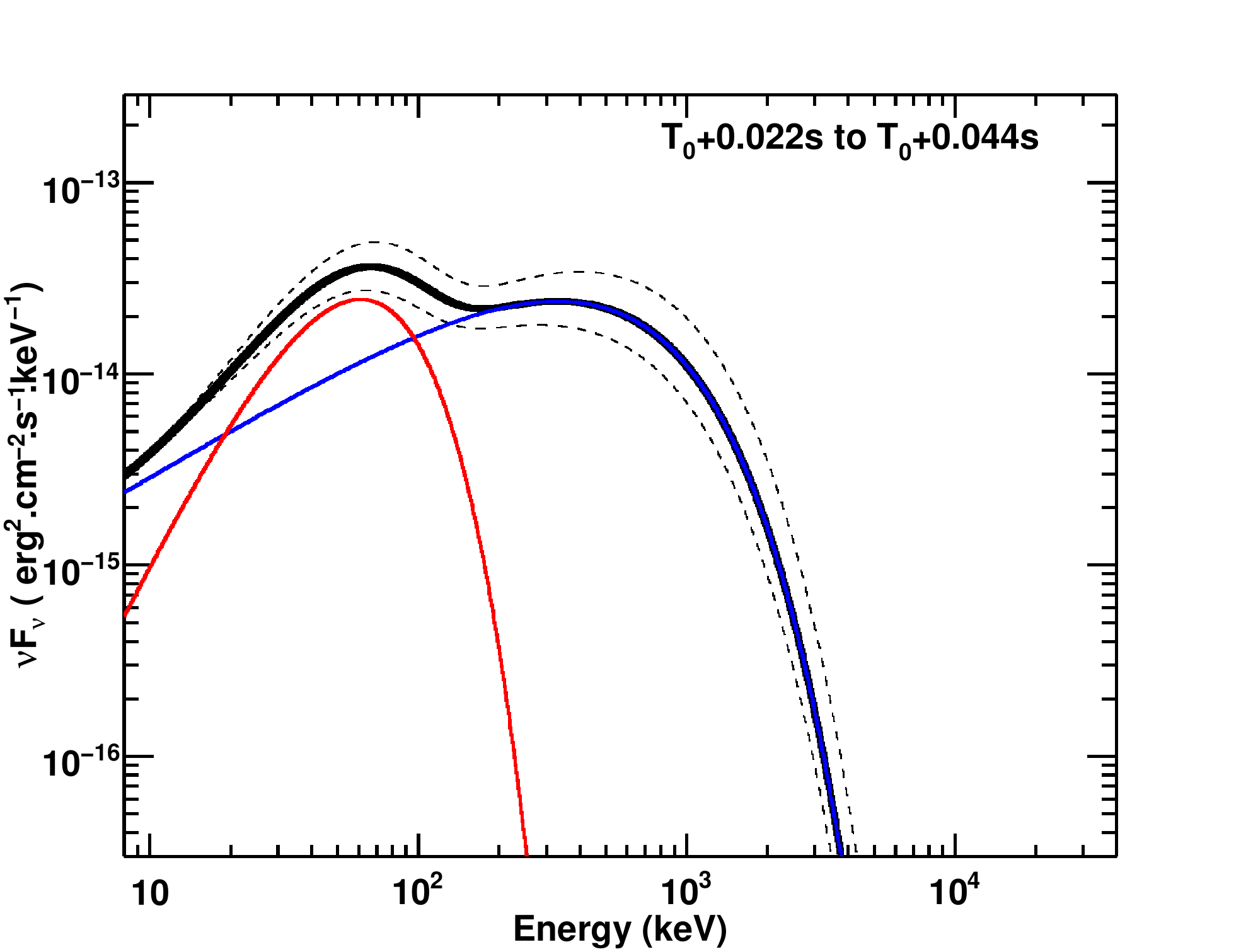}
\includegraphics[totalheight=0.20\textheight, clip,viewport=56 42 512 395]{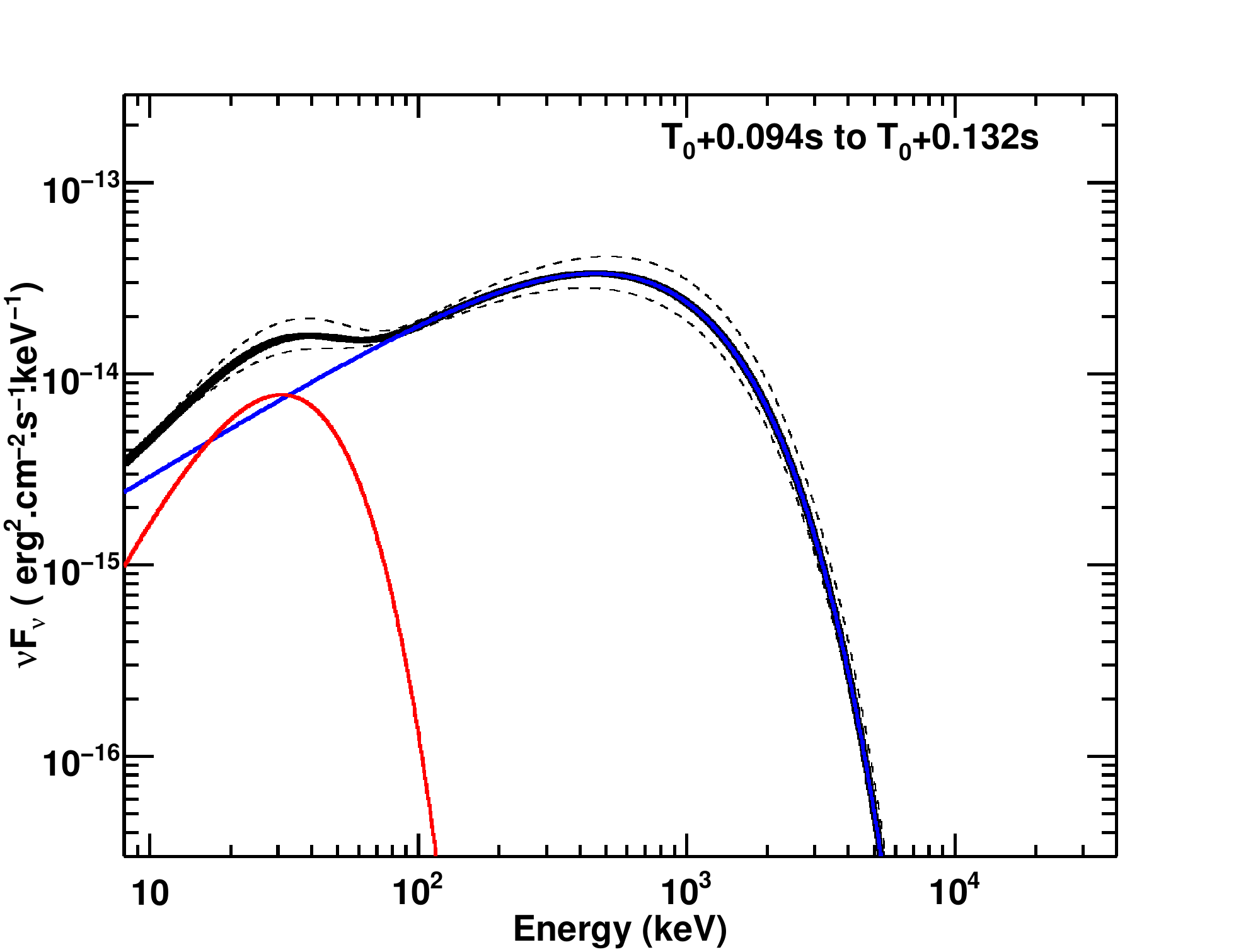}

\includegraphics[totalheight=0.20\textheight, clip,viewport=0 42 512 395]{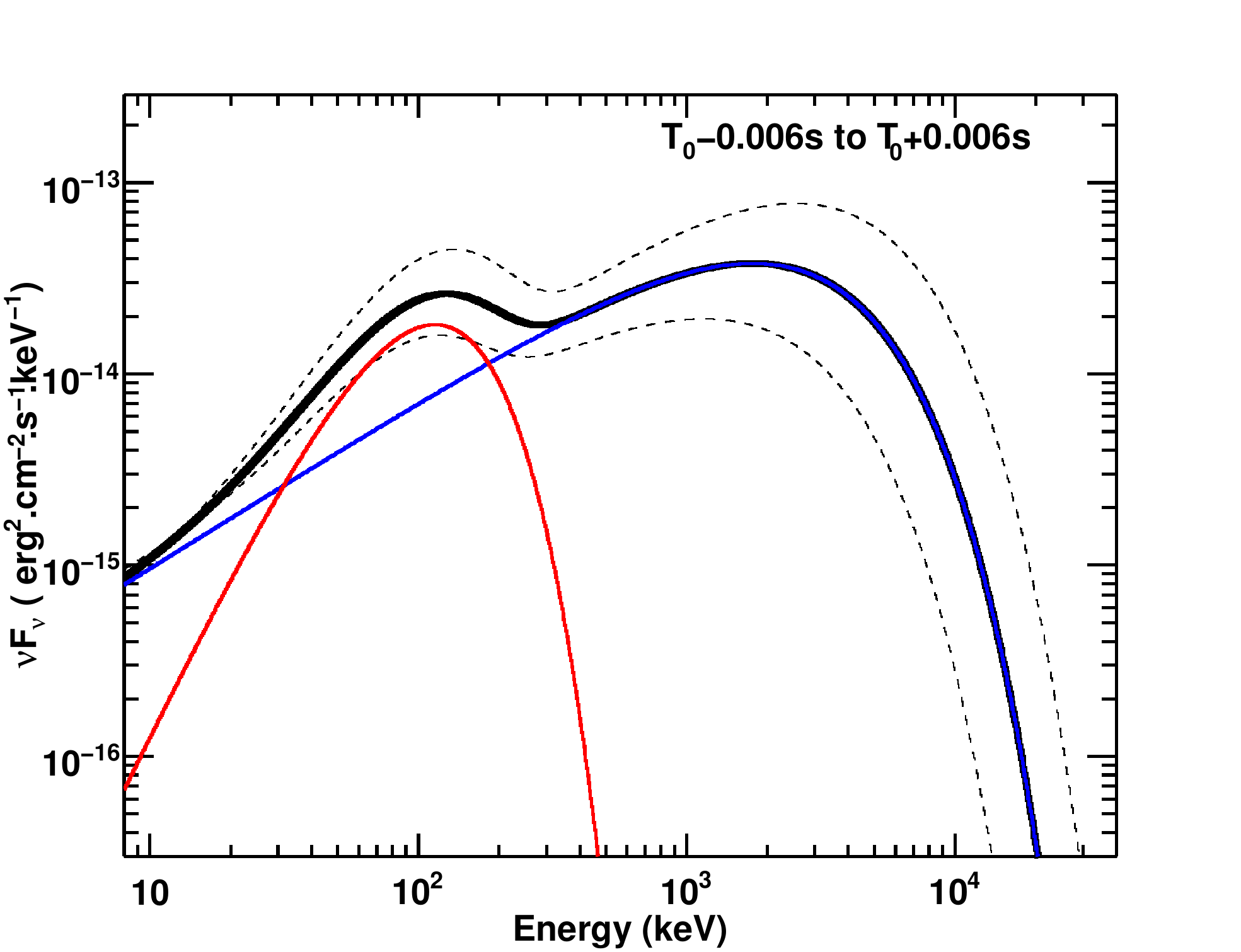}
\includegraphics[totalheight=0.20\textheight, clip,viewport=56 42 512 395]{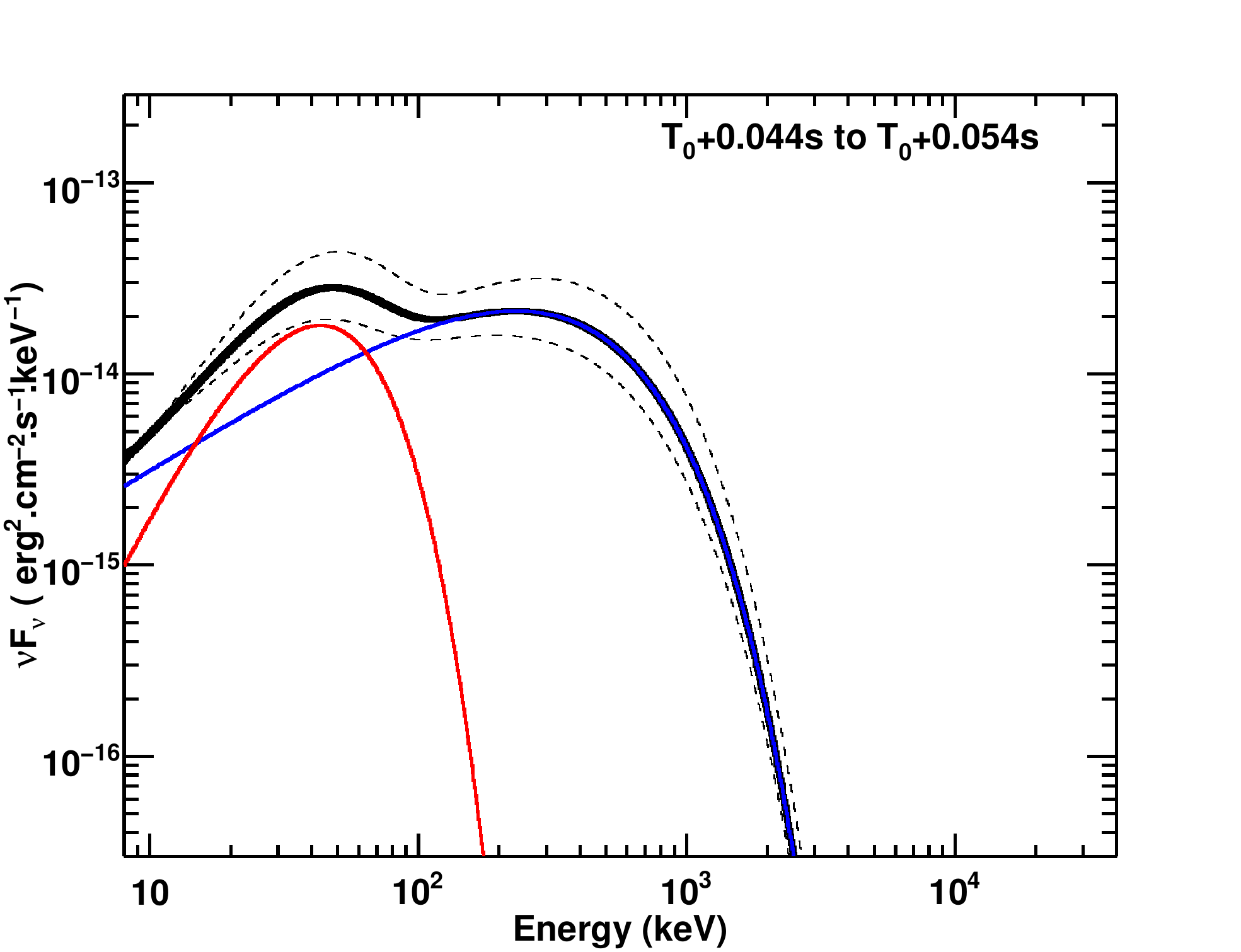}
\includegraphics[totalheight=0.20\textheight, clip,viewport=56 42 512 395]{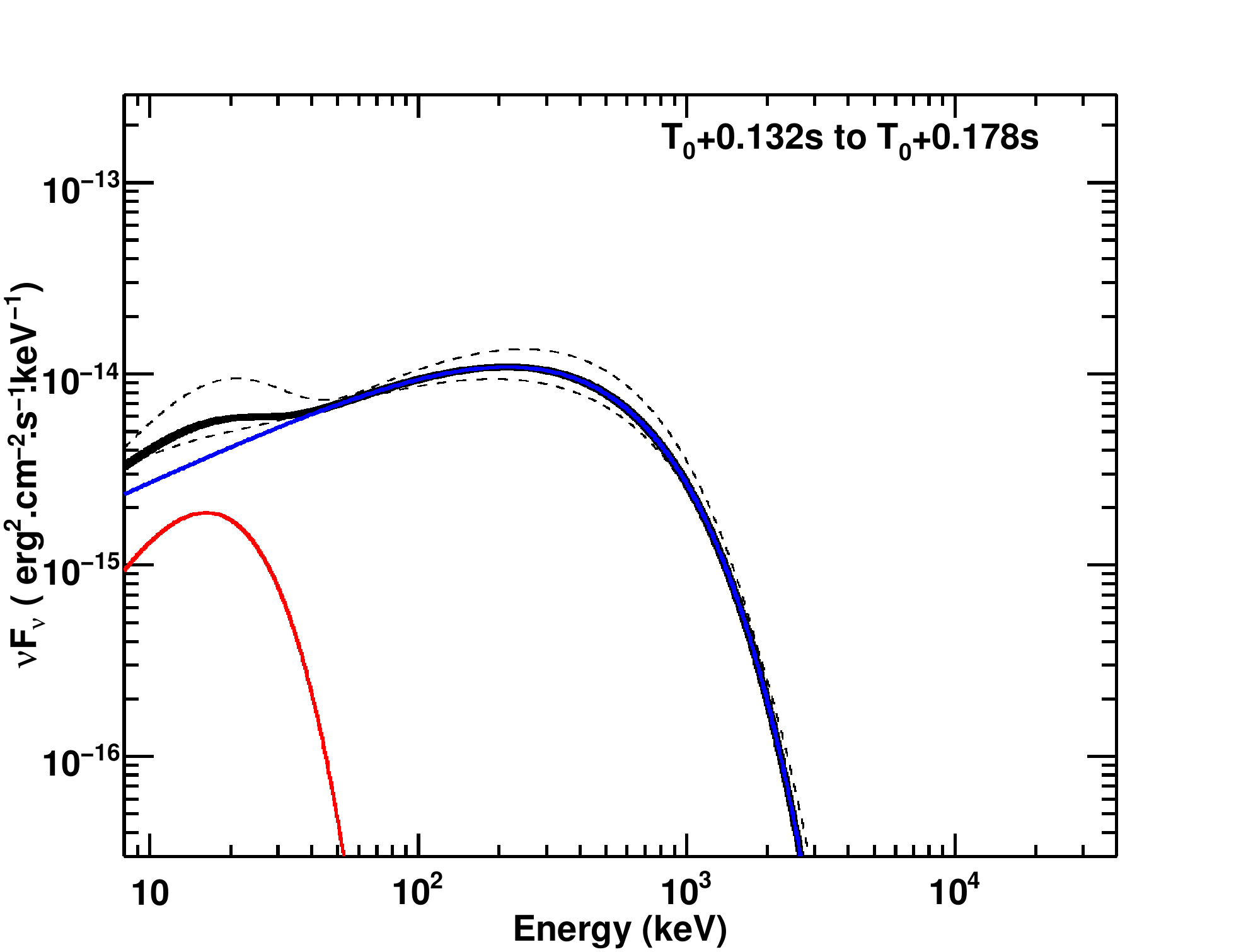}

\includegraphics[totalheight=0.20\textheight, clip,viewport=0 42 512 395]{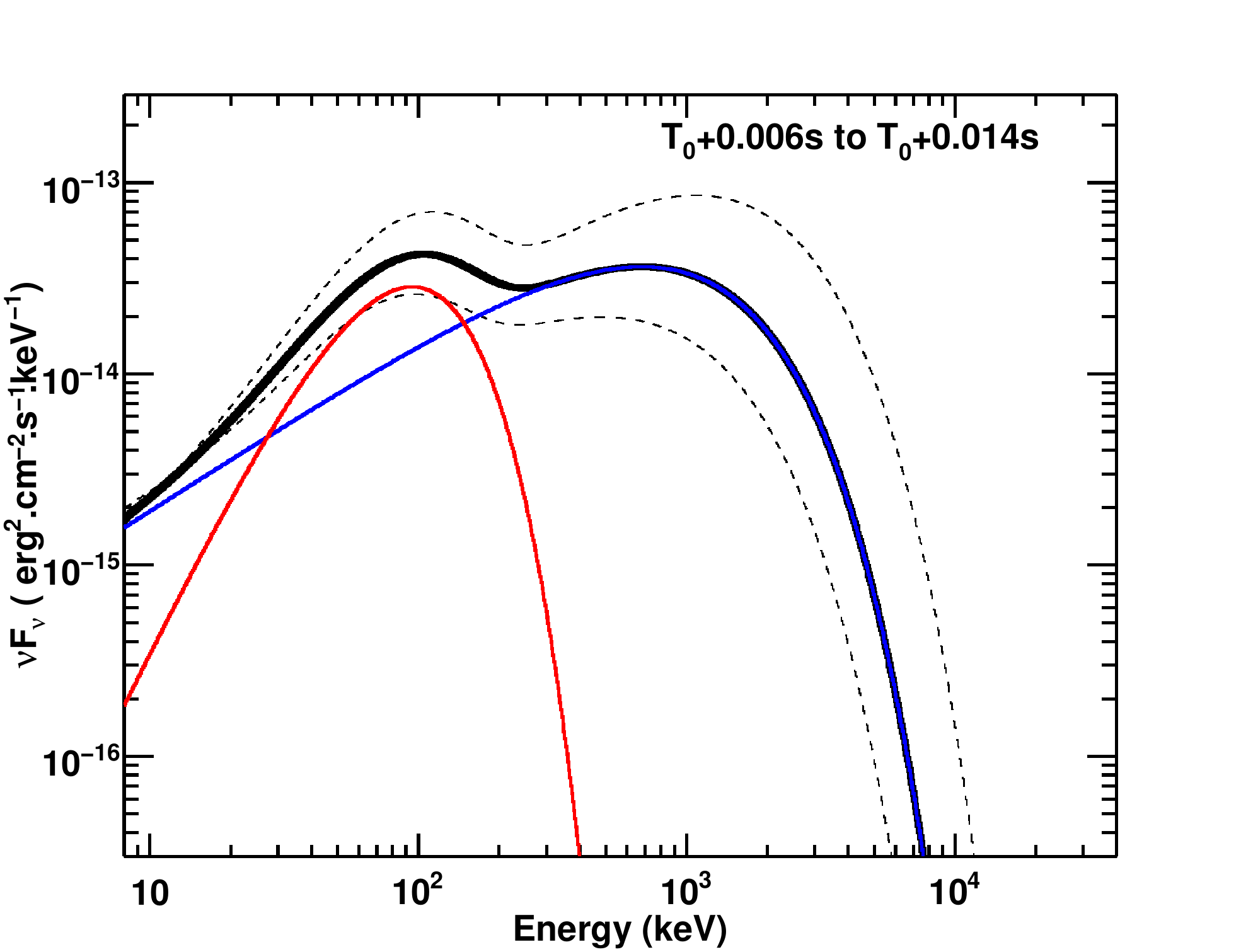}
\includegraphics[totalheight=0.20\textheight, clip,viewport=56 42 512 395]{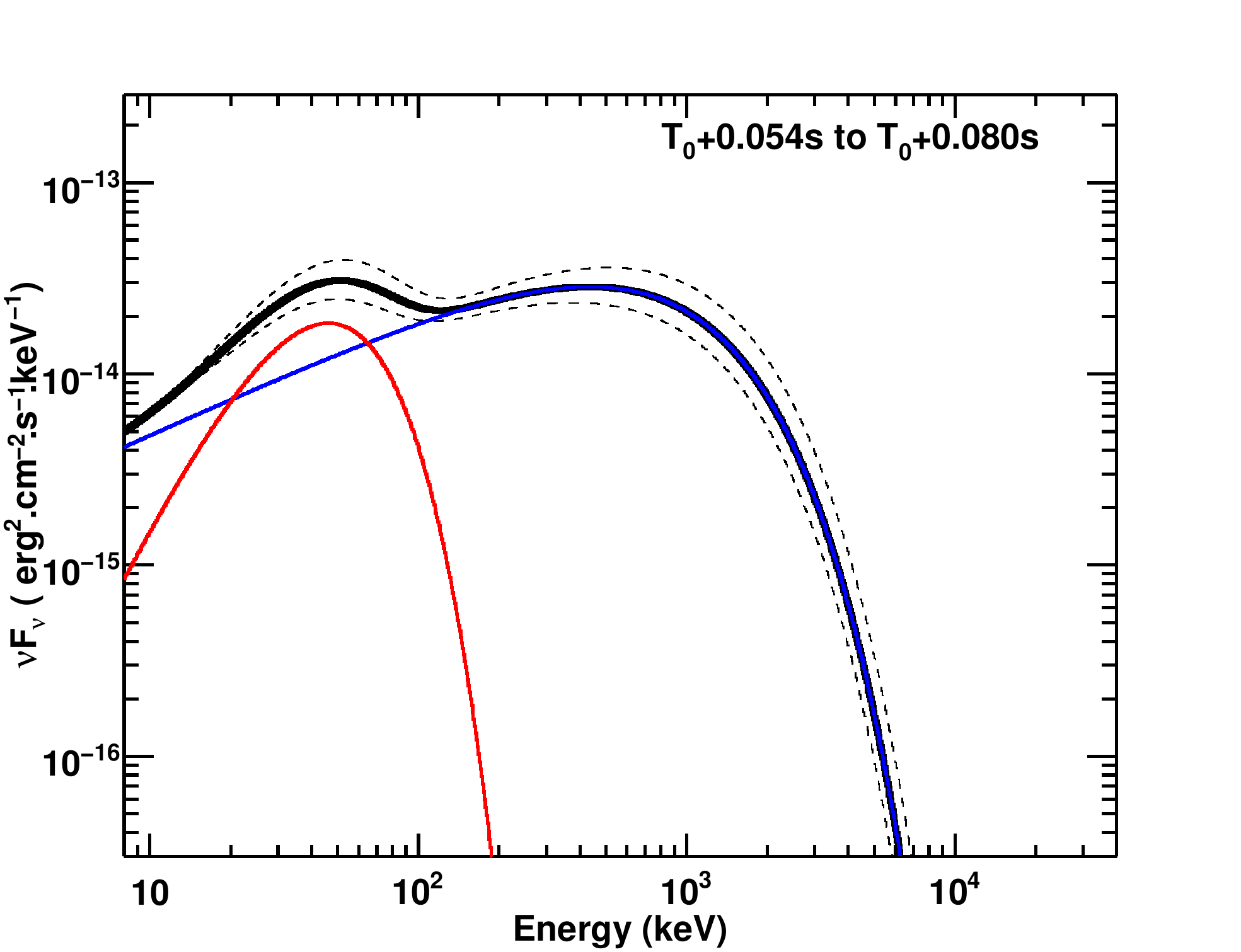}
\includegraphics[totalheight=0.20\textheight, clip,viewport=56 42 512 395]{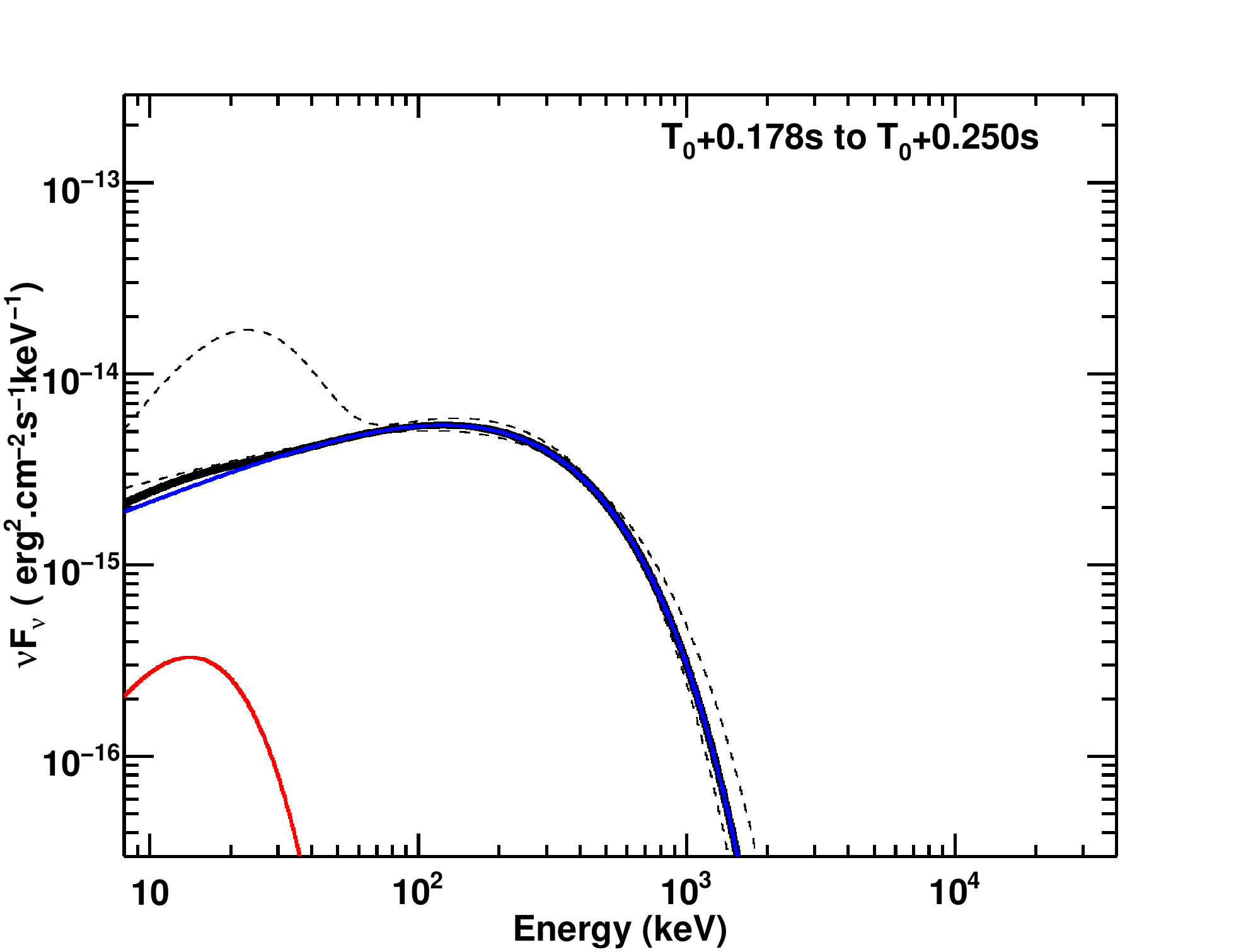}

\includegraphics[totalheight=0.224\textheight, clip,viewport=0 0 512 395]{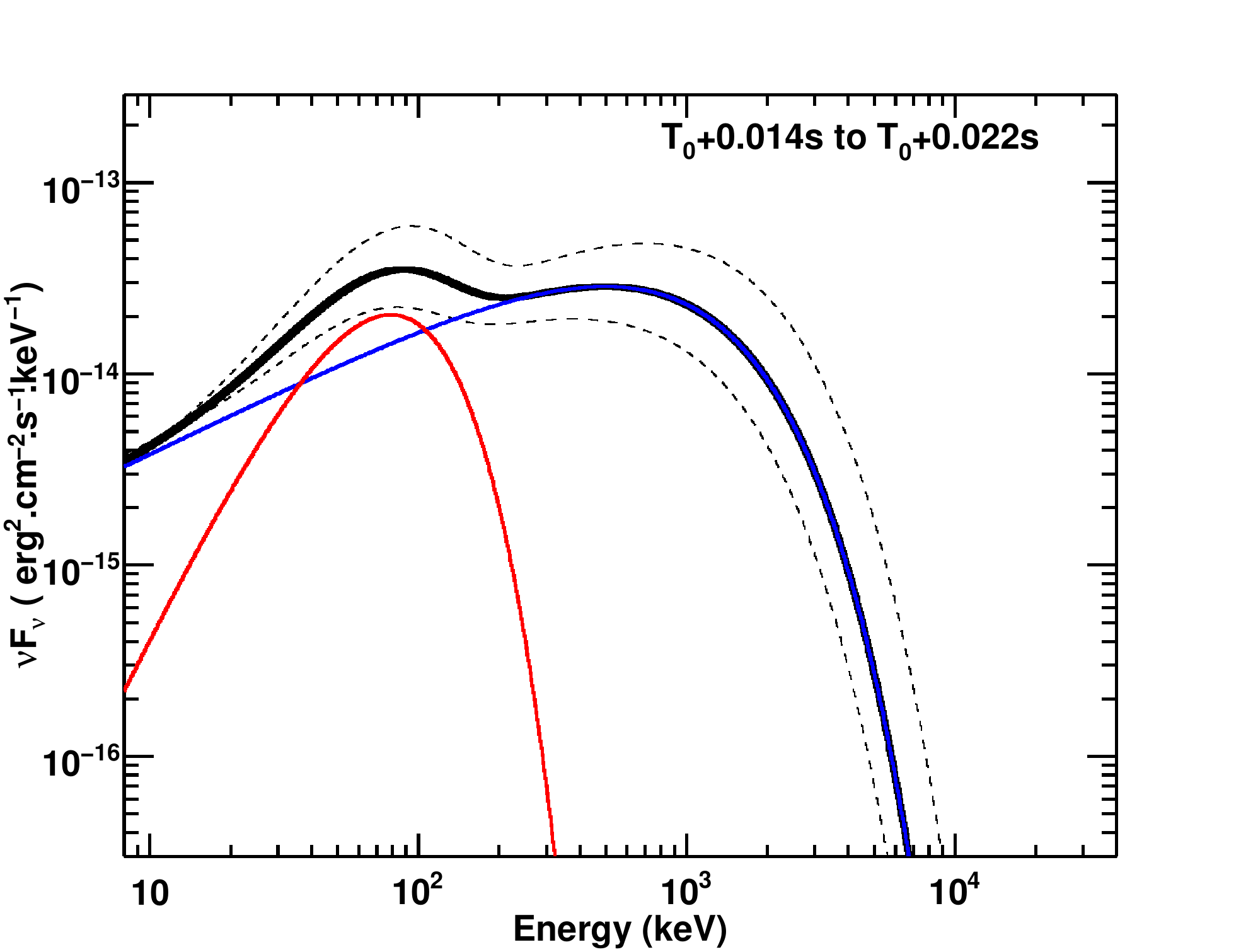}
\includegraphics[totalheight=0.224\textheight, clip,viewport=56 0 512 395]{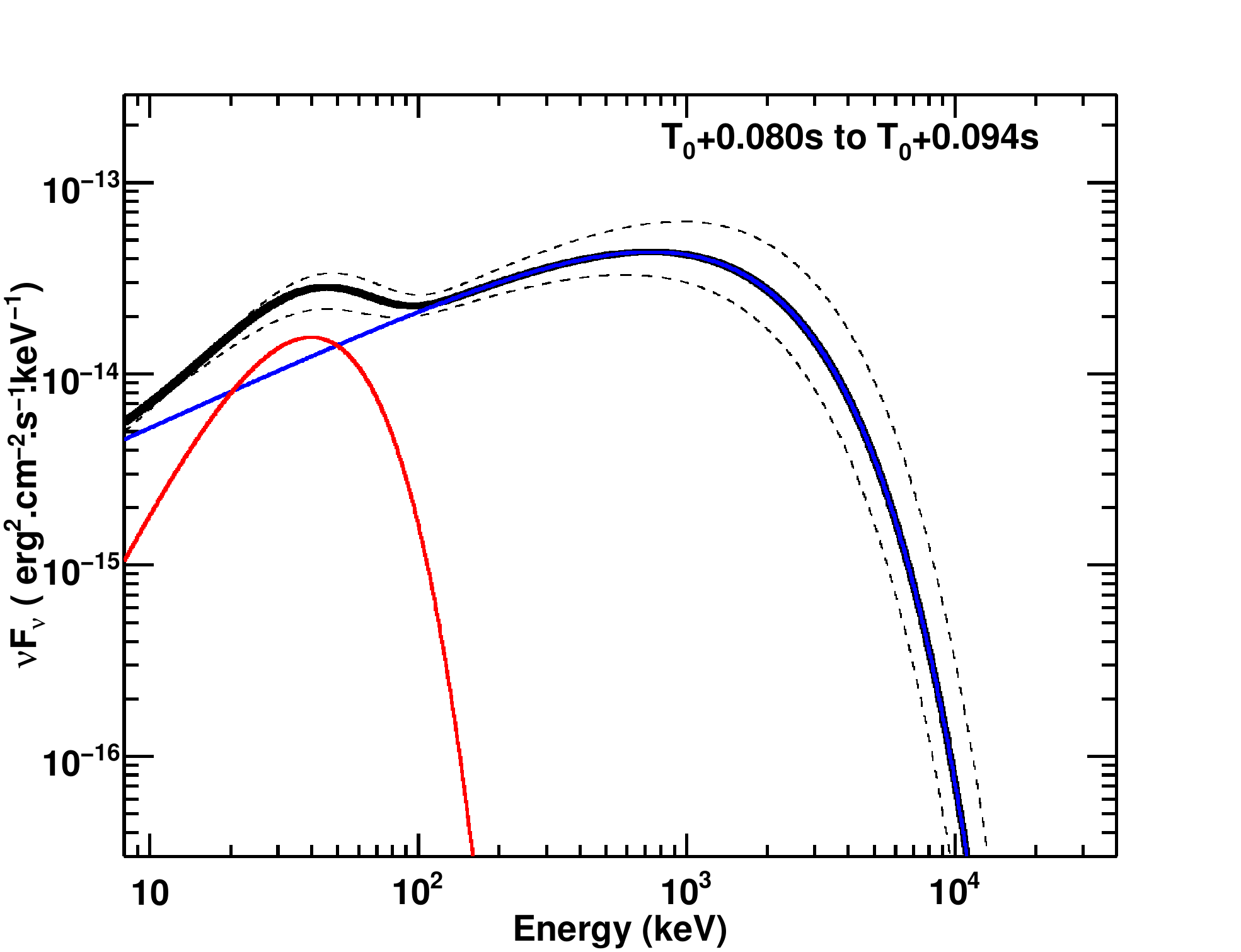}
\includegraphics[totalheight=0.224\textheight, clip,viewport=56 0 512 395]{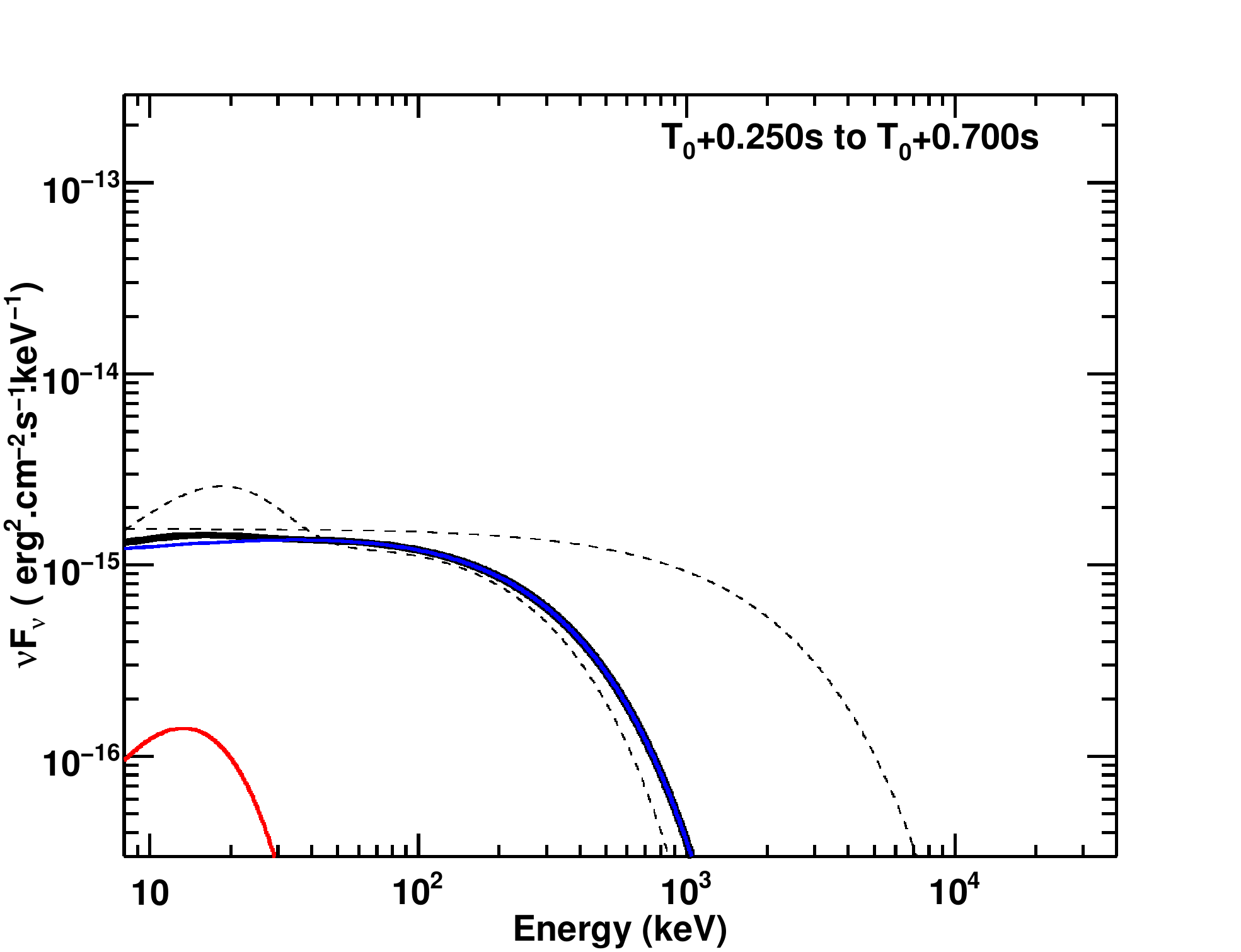}

\caption{\label{fig:GRB120323A_resolved_spectra_B+BB}Fine time-resolved $\nu$F$_\nu$ spectra using C+BB (equivalent to B+BB). The blue lines correspond to the Band function, the red ones to the BB component and the black lines are the sum of the two components. The solid black lines correspond to the model obtained with the best parameters from the fits, and the thin black lines correspond to the 1 $\sigma$ uncertainty on the best fits.}
\end{figure*}

\subsection{Spectral analysis of hardness-ratio selected time intervals}
\label{sec:Fine time resolved}

We showed in Section~\ref{section:Time-integrated spectral analysis} and~\ref{sec:Coarse time resolved} that spectral evolution within a burst can affect dramatically spectral fit results, potentially leading to a misinterpretation of the physics of the observed emission.

In with section, we defined time intervals as short as possible to reduce effects due to spectral evolution but still large enough to be able to adequately fit at least a Band or a \fdaigne{CPL} function to the data. We propose a novel method to determine the time intervals for time-resolved spectroscopy. Similar to the idea by~\citet{Scargle:1998} of characterizing flux variation with Bayesian statistics, which is referred as Bayesian Blocks method (BBM), we apply BBM to the evolution of the GRB light curve hardness ratio (HR), which is a good proxy to its spectral evolution. First, HRs were calculated between 8 - 100 keV and 105 keV - 10 MeV for combined NaI and BGO data in a base bin, which is the finest possible bin with at least 25 counts in each energy band to ensure the Gaussian statistics of HR. The errors in the HR were propagated from the count errors. Then, BBM was applied to the HR profile to find its change points. The prior for BBM is chosen in order to have a sufficient number of counts to perform spectral fitting and to avoid individual time intervals including too much spectral evolution. This analysis resulted in 12 time intervals which are used to generate the light curve presented in the right panel of Figure~\ref{fig:GRB120323A_LCs}. We verified our ability to reconstruct properly a Band function and a B+BB model in all these short time intervals with simulations following the procedure described in Appendix~\ref{section:Simulations}.

\begin{figure*}
\begin{center}
\includegraphics[totalheight=0.25\textheight, clip]{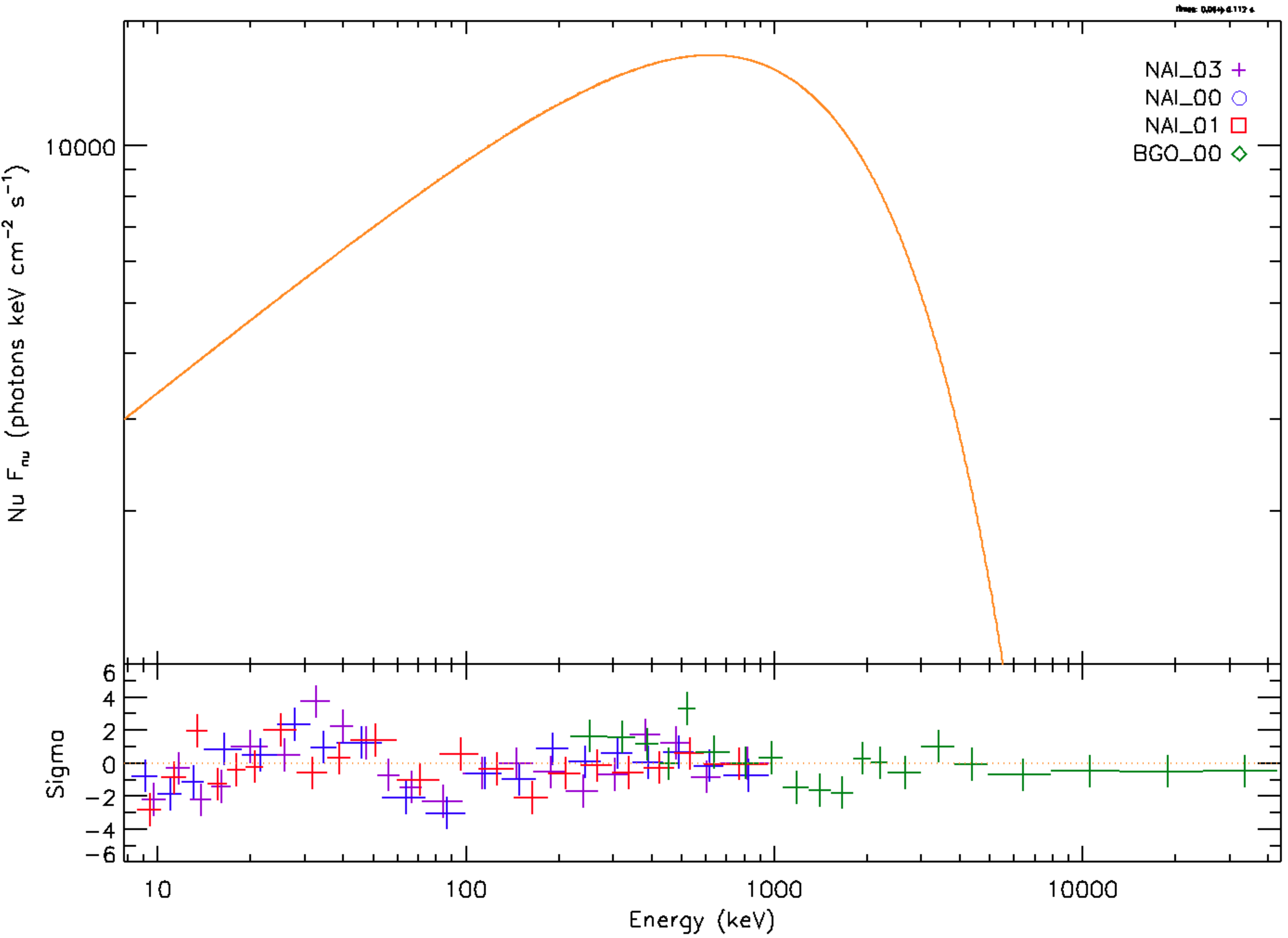}
\includegraphics[totalheight=0.25\textheight, clip]{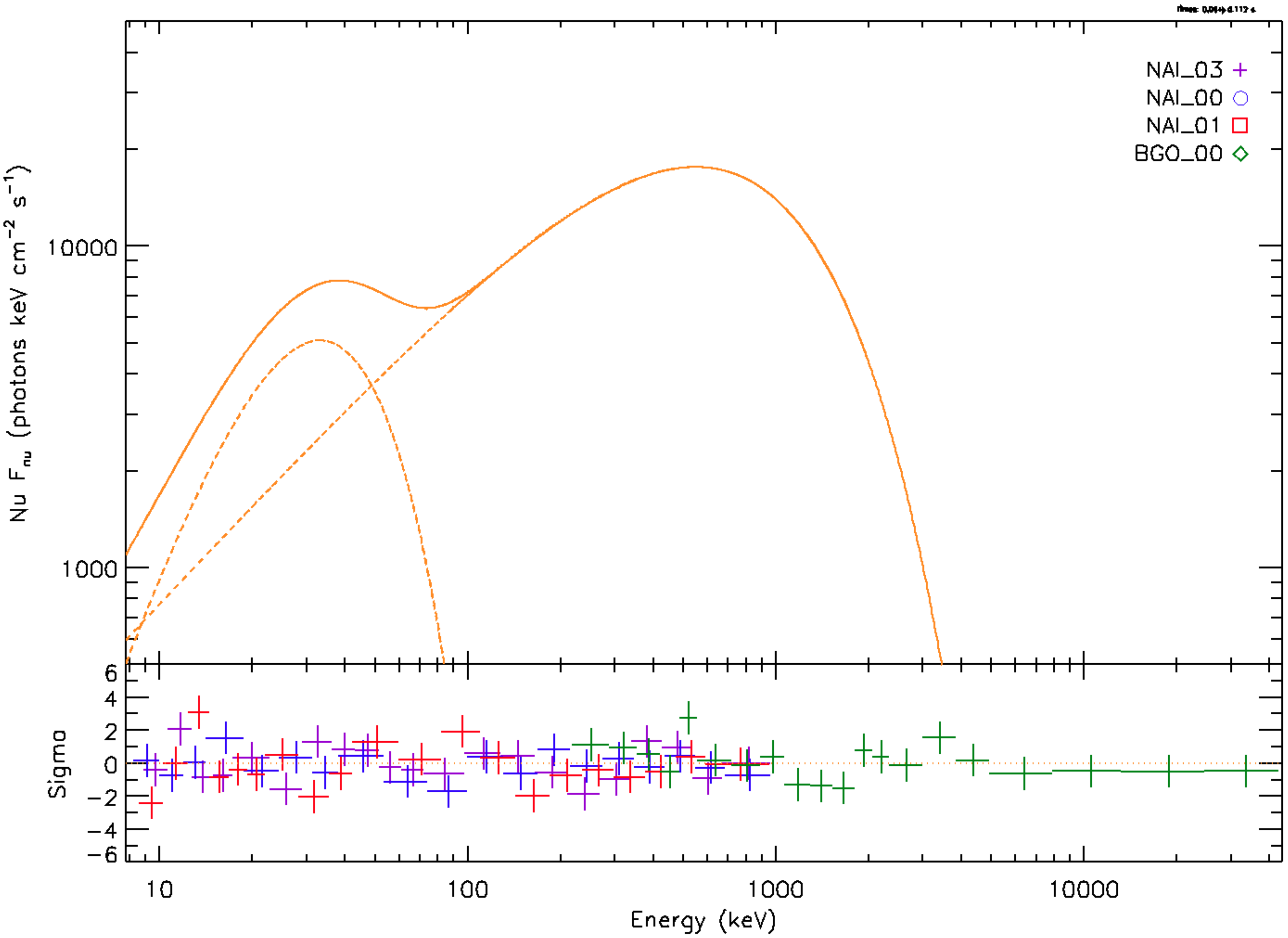}
\caption{\label{fig:GRB120323A_interval8}The $\nu$F$_\nu$ spectrum in the time interval from T$_\mathrm{0}+0.094$\,s to T$_\mathrm{0}+0.132$\,s using a Band function (left) and B+BB (right). The lower panel for each figure shows to the residuals of the fits.}
\end{center}
\end{figure*}

We fitted each of these intervals with the models presented in section~\ref{sec:Coarse time resolved} (i.e., B, C, B+BB and C+BB) even if the Cstat improvement were not statistically significant based on the number of dof differences for the various models. These results are presented in Table~\ref{tab:GRB120323A_resolved_spectra_fine} as well as in Figures~\ref{fig:GRB120323A_resolved_spectra_B} and~\ref{fig:GRB120323A_resolved_spectra_B+BB}. When one model was clearly not adequate to fit the spectrum in a time interval based on the Cstat value, it was not included in Table~\ref{tab:GRB120323A_resolved_spectra_fine}. When the high energy power law index of the Band function could not be constrained, we estimated its 1$\sigma$ upper limit.

The Band model gives a satisfactory fit to the data in all the short time intervals. However, we can constrain the parameters of the B+BB (or C+BB) model in all these time intervals, and C+BB is statistically significantly\footnote{Based on the procedure described in Section~\ref{section:Simulations}.} better than Band alone in the three time intervals T$_\mathrm{0}$+0.054\,s to T$_\mathrm{0}$+0.080\,s, T$_\mathrm{0}$+0.080\,s to T$_\mathrm{0}$+0.094\,s and T$_\mathrm{0}$+0.094\,s to T$_\mathrm{0}$+0.132\,s with a Cstat improvement of 9, 25, and 51 units for 1 dof difference between models, respectively. In principle this could be over-fitting our data, in which case we would expect the BB component to pick up random statistical fluctuations in the spectrum, with erratic changes of temperature and normalization from one time interval to the other. Instead, the temperature follows a constant cooling trend identical to the evolution reported from the coarser time interval analysis. This observational result can be hardly explained with random statistical fluctuations of the number of counts in the various energy channels of the detectors.

The three time intervals for which the addition of the BB leads to the greatest Cstat improvement correspond to the second peak of the light curve where the BB was also a statistically significant improvement in the coarse time-resolved spectral analysis (see Section~\ref{sec:Coarse time resolved}). Figure~\ref{fig:GRB120323A_interval8} shows the ${\nu}F_\nu$ spectra resulting from the fit to the data during the time interval T$_\mathrm{0}$+0.094\,s to T$_\mathrm{0}$+0.132\,s using a Band function (left) and B+BB (right). The systematic pattern observed in the residuals of the Band fit is clearly flattened when adding the BB.

\section{One or multiple components ?}
\label{Model description}

In this section, we describe the spectral evolution within the burst resulting from the fine time resolved spectral analysis presented in Section~\ref{sec:Fine time resolved}, and discuss the best two spectral fit models, a single component (Band or \fdaigne{CPL}) and a two component scenario (B+BB or C+BB).
\begin{figure*}
\begin{center}
\includegraphics[totalheight=0.286\textheight, clip,viewport=0 0 520 403]{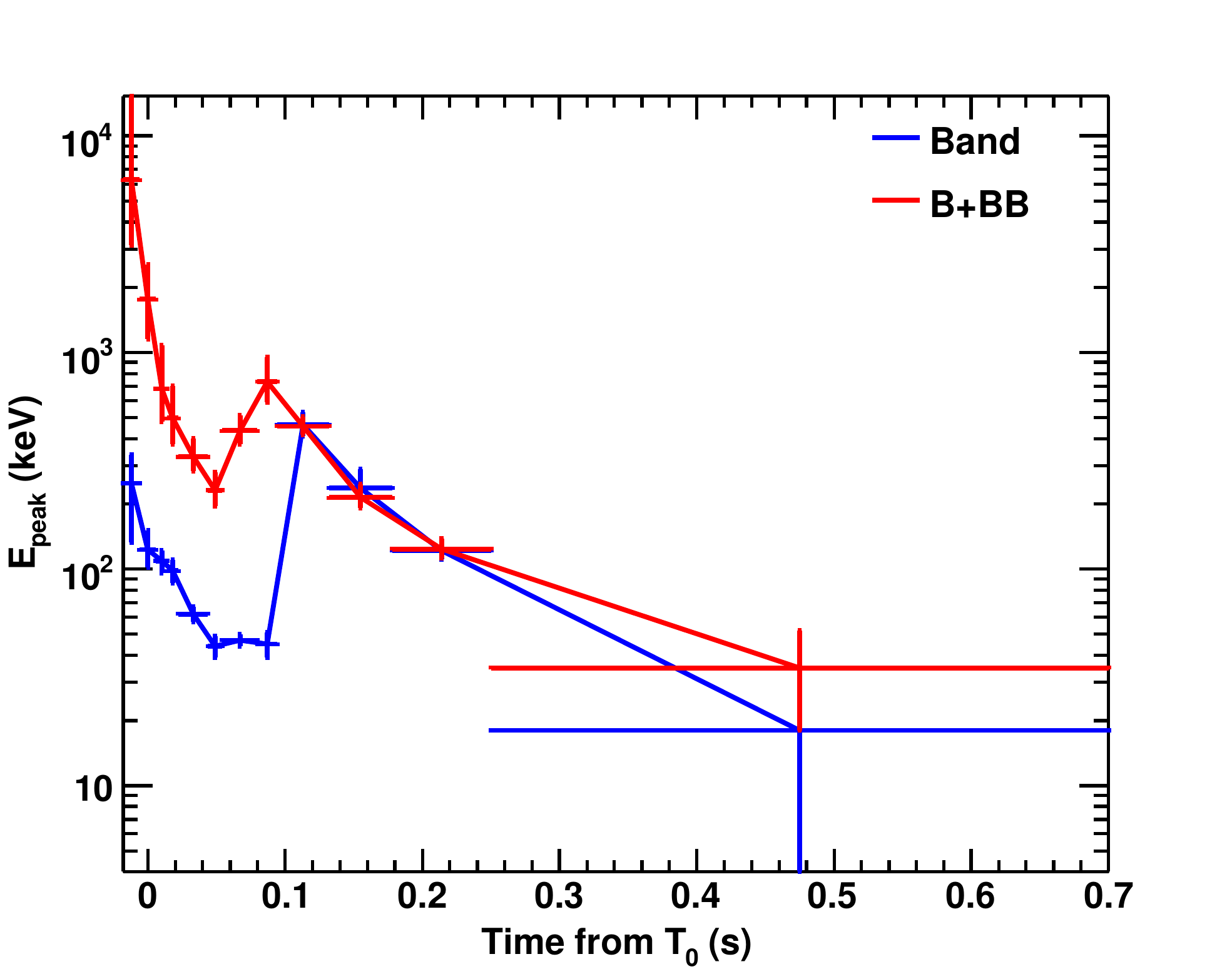}
\includegraphics[totalheight=0.286\textheight, clip,viewport=0 0 520 395]{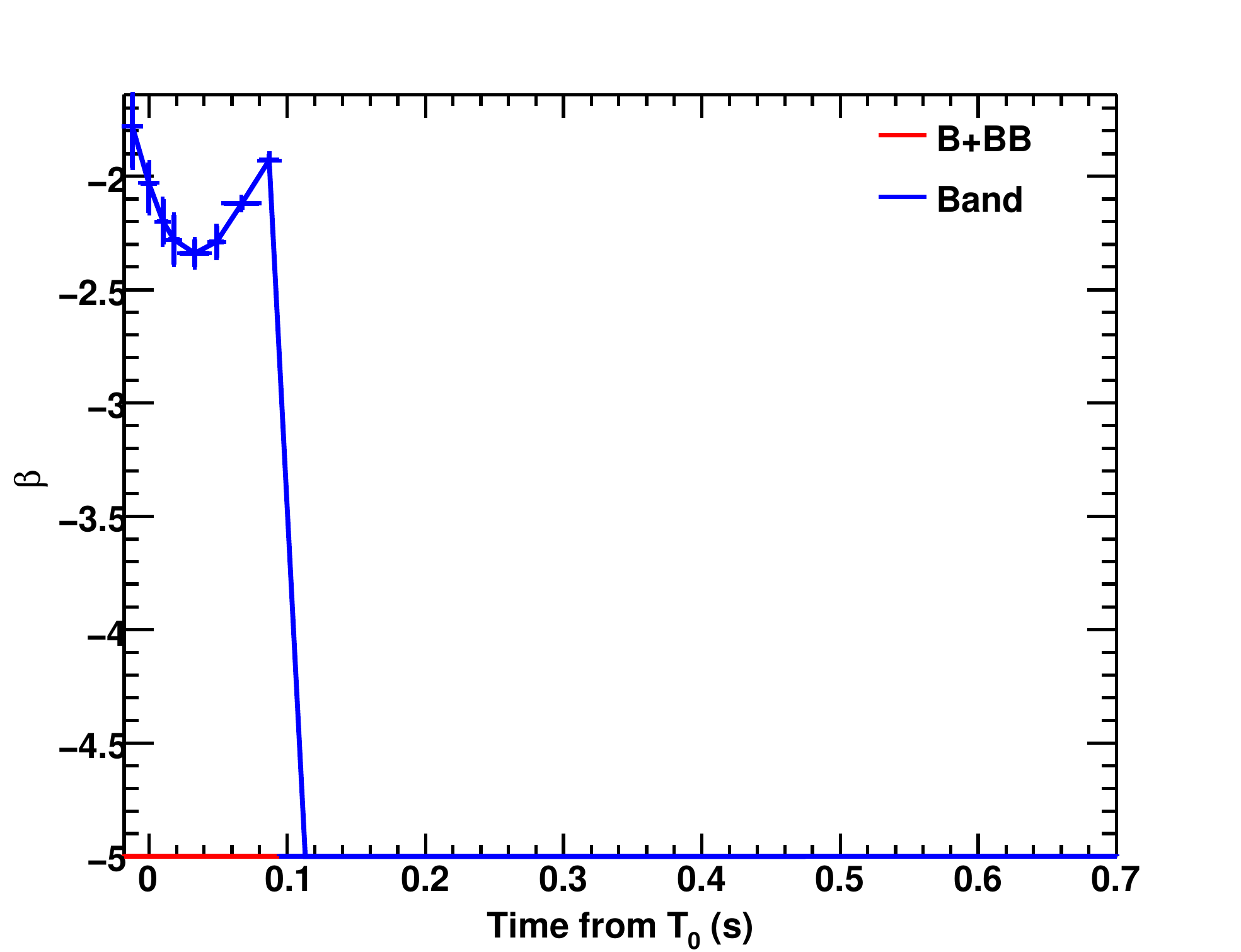}

\includegraphics[totalheight=0.286\textheight, clip,viewport=0 0 520 395]{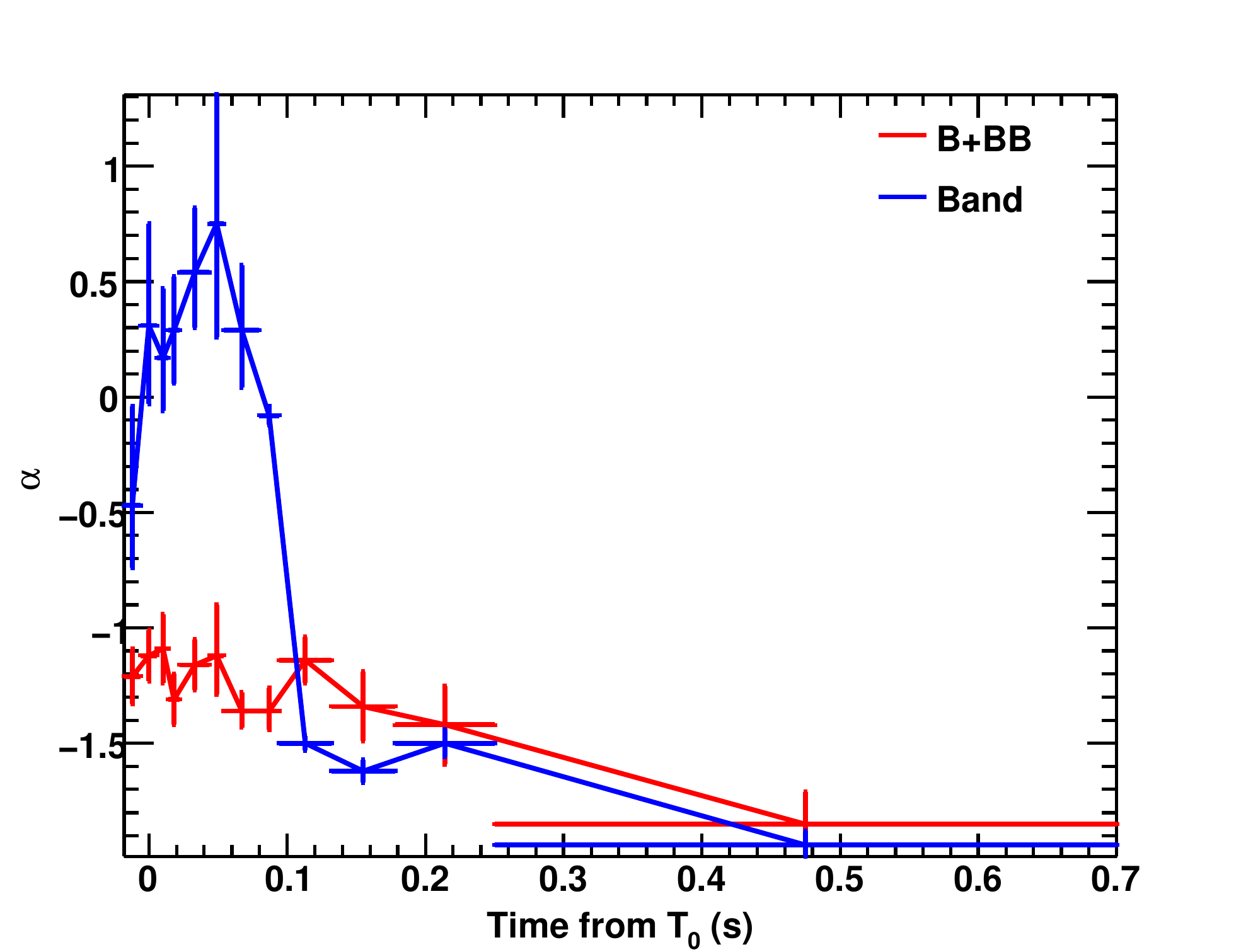}
\includegraphics[totalheight=0.286\textheight, clip,viewport=0 0 520 395]{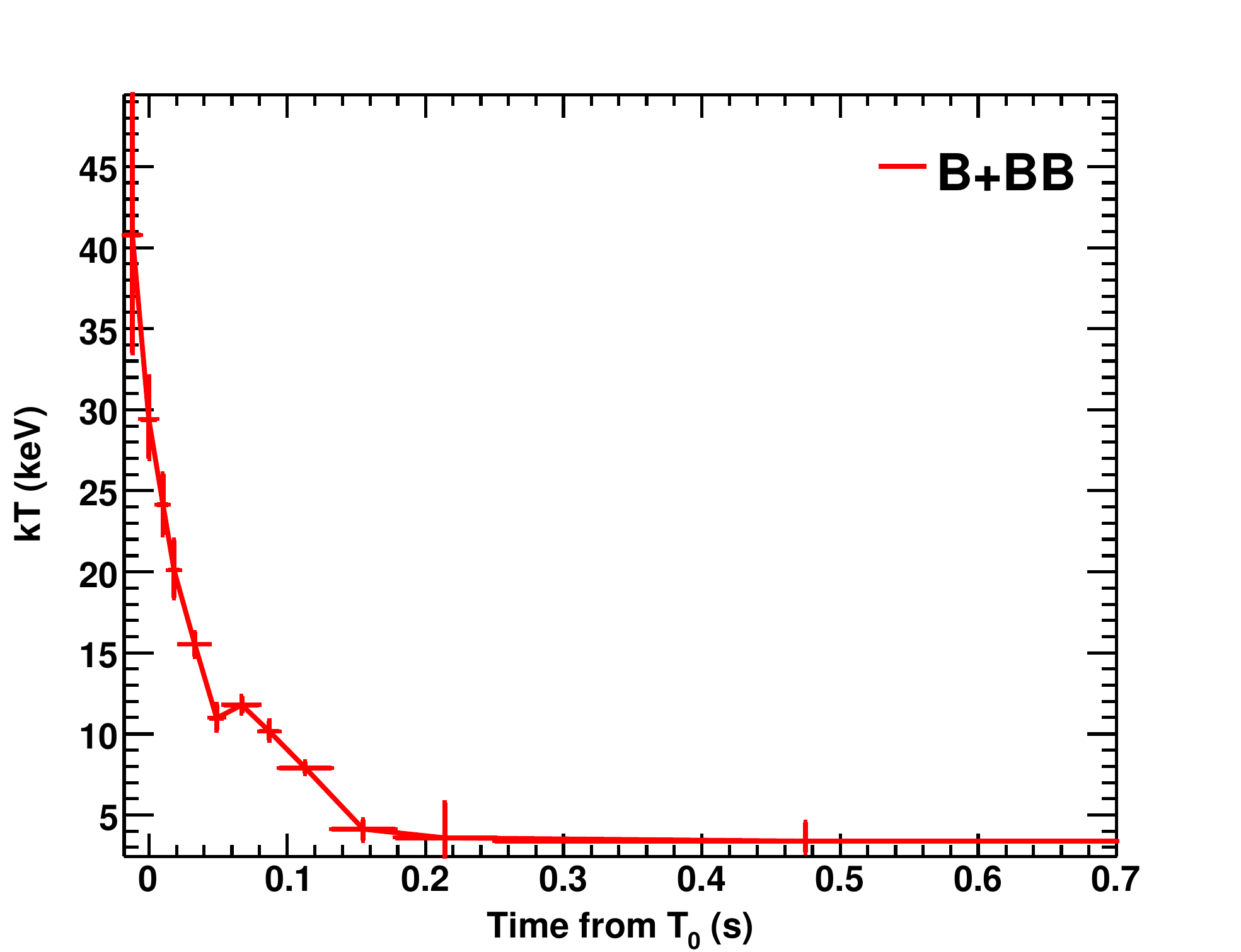}
\caption{\label{fig:GRB120323A_Spectral_parameter_evolution}Evolution with time of the parameters of the spectral fits using the shortest time intervals. The blue lines correspond to the Band-only fit to the data and the red lines correspond to the B+BB fits.}
\end{center}
\end{figure*}
Figure~\ref{fig:GRB120323A_Spectral_parameter_evolution} shows the evolution of the parameters of the various spectral components with time. The blue curves correspond to the parameters obtained when fitting Band-only to the data, while the red ones are obtained when fitting B+BB.

\subsection{Single component: Band function}
\label{sec:SinglecomponentBandfunction}

When fit with a Band function only, the E$_\mathrm{peak}$ of \grbnos~ tracks the burst flux, especially the two peaks identified in the light curves above 20 keV (see Figure~\ref{fig:GRB120323A_LCs}) as also seen in previous burst spectra~\citep{Ford:1995}. However, while E$_\mathrm{peak}$ is usually harder in the earliest high intensity peaks, especially in cases of simply structured light curves like \grbnos, here the first peak is rather soft with values ranging between $\sim$40 and 200 keV, while the second peak is much harder, with E$_\mathrm{peak}$ values reaching $\sim$600 keV. Each pulse exhibits an intrinsic hard to soft evolution. The evolution of $\alpha$ shows a striking discontinuity at $\sim$T$_\mathrm{0}$+0.094\,s. During the first intensity peak of the light curve, $\alpha$ is mostly positive with values above $+0.2$ in some cases, while during the second one, the $\alpha$ values drop below $-1.5$. Similarly, the values of $\beta$ are well constrained between $-1.6$ and $-2.4$ until 0.094\,s after trigger time, while only upper limits below $-2.7$ can be measured thereafter. We note that the discontinuity in the evolution of the parameter values appears simultaneously for all the parameters of the Band function. Figure~\ref{fig:GRB120323A_resolved_spectra_B} shows the evolution of the Band function with time.

{Interestingly, Rmfit converges towards two different minima when fitting a Band function to the data in the time interval T$_\mathrm{0}$+0.080\,s to T$_\mathrm{0}$+0.094\,s (see Table~\ref{tab:GRB120323A_resolved_spectra_fine}). The lowest Cstat is obtained for a low E$_\mathrm{peak}$ value and a high $\alpha$ one. The other local minimum is obtained for a much higher E$_\mathrm{peak}$ value and a very steep $\alpha$. We will describe in section~\ref{sec:comparison} the impact of this result when comparing the single component to the two components scenario.

\subsection{Two components: B+BB Model}
\label{sec:twocomps}

\begin{figure*}
\begin{center}
\includegraphics[totalheight=0.61\textheight, clip, viewport=0 0 286 439]{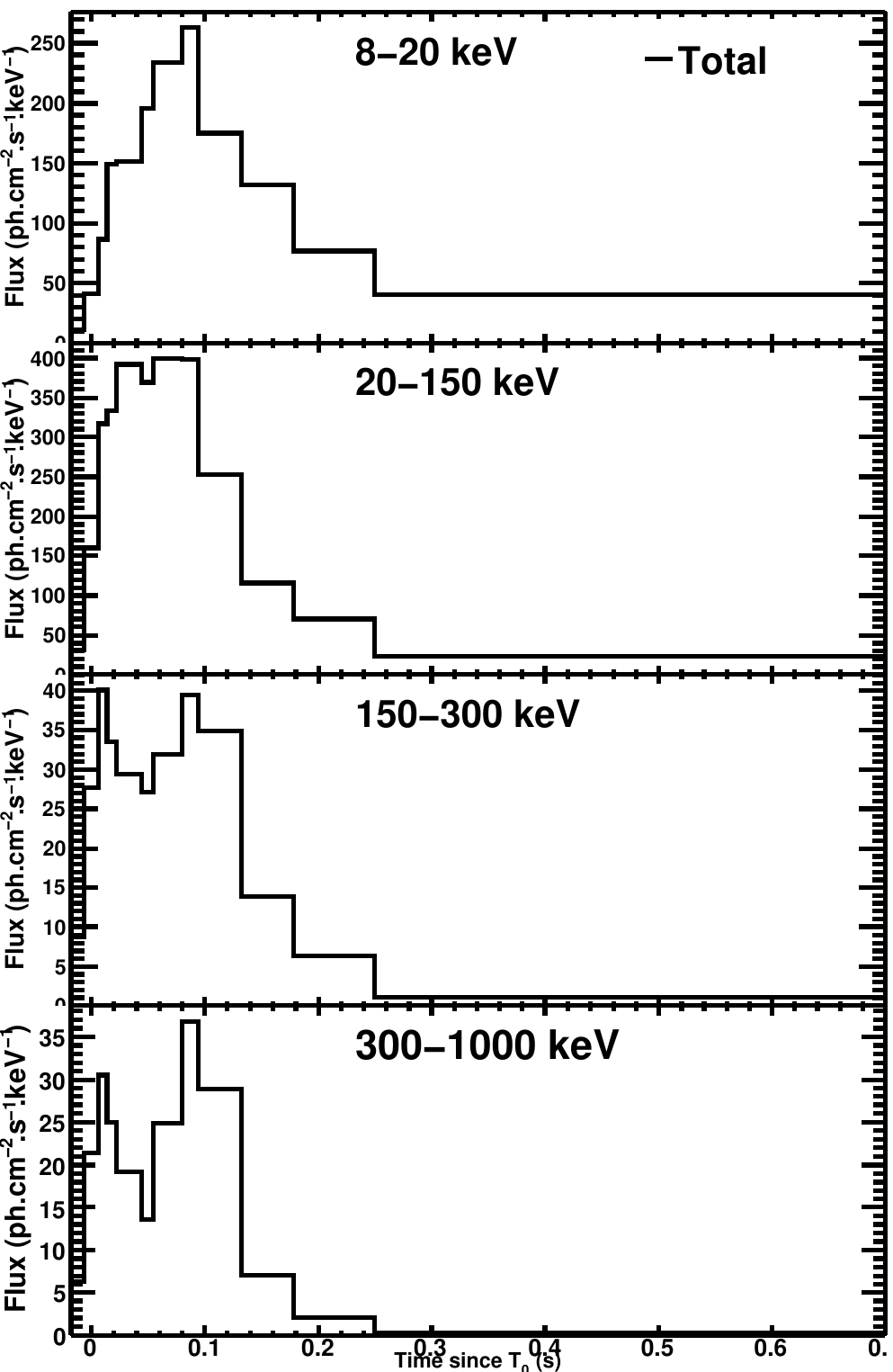}
\includegraphics[totalheight=0.61\textheight, clip, viewport=292 0 542 439]{./figure8.pdf}

\caption{\label{fig:GRB120323A_reconstructed_LC_B+BB}Photon light curves reconstructed from the B+BB fits. The light curves are drawn in the same energy range as the counts light curves presented in Figure~\ref{fig:GRB120323A_LCs}. The right figures shows the photon light curves corresponding to the Band function in blue and the BB in green. The figures on the left side are the sum of the two components.}
\end{center}
\end{figure*}

\begin{figure*}
\begin{center}
\includegraphics[totalheight=0.75\textheight, clip, viewport=140 260 1650 1760]{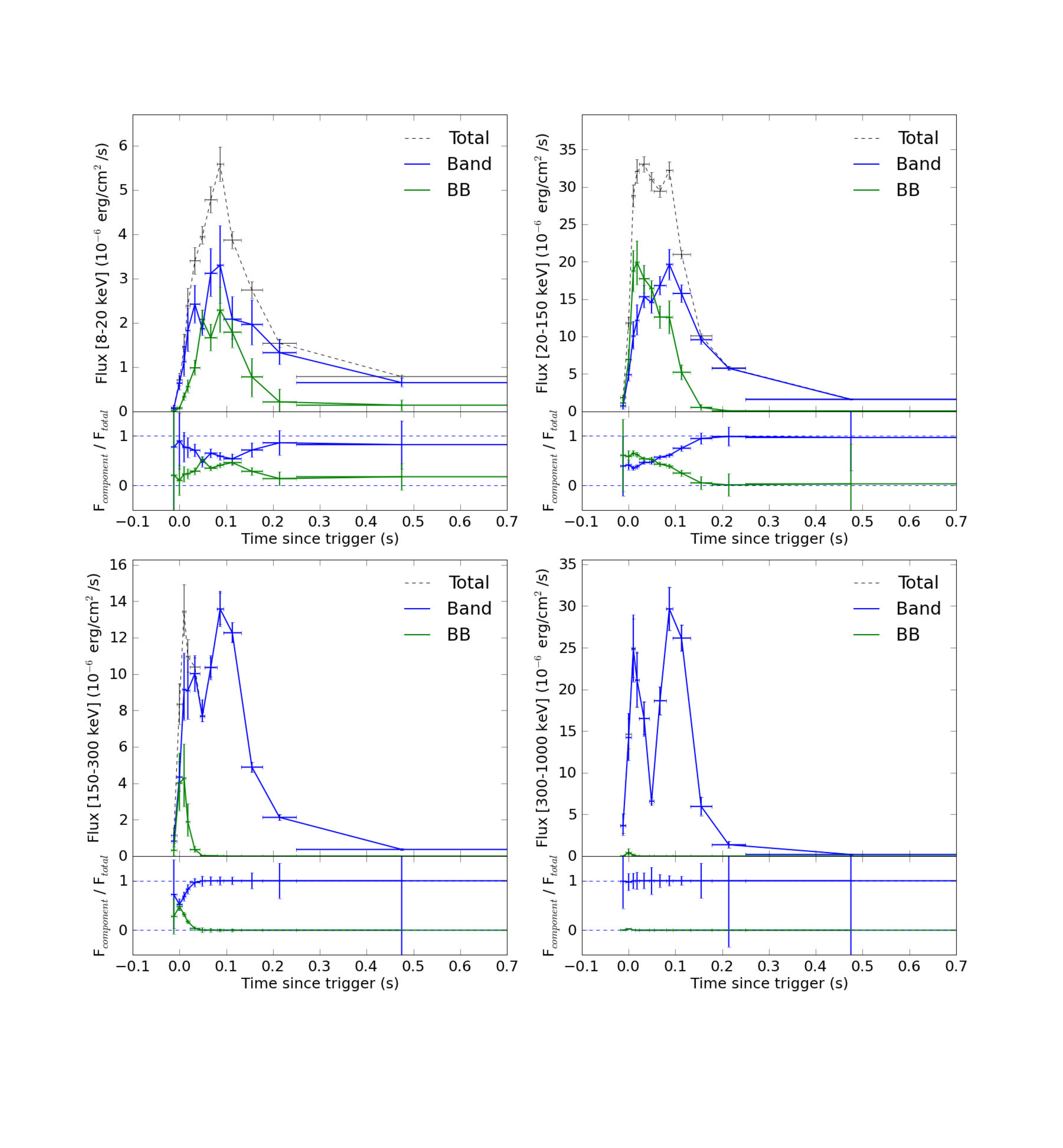}
\caption{\label{fig:GRB120323A_energy_LC}Energy flux evolution when using B+BB model in the same energy bands as the count light curves presented in Figure~\ref{fig:GRB120323A_LCs}. For each figure, the top panel corresponds to the total energy flux (i.e., B+BB in black) as well as the energy flux per component (i.e., Band in blue and BB in green) and the bottom panel corresponds to the contribution of each component to the total energy flux in the corresponding energy band (i.e., Band in blue and BB in green).}
\end{center}
\end{figure*}

When a B+BB model is fitted to the data, the E$_\mathrm{peak}$ of the Band function undertakes a global hard to soft evolution all across the burst with values evolving from $\geq$3 MeV during the beginning of the burst to $\sim$30 keV during the late intensity decay phase. However, E$_\mathrm{peak}$ tracks strongly the light curve flux with an increase of the values from $\sim$200 keV to $\sim$800 keV corresponding to the flux increase phase of the second peak of the light curve. The values of $\alpha$ remain mostly constant around $-1.3$; only upper limits (always below $-2.4$) can be determined for $\beta$.

The temperature $kT$ of the BB component decreases linearly with time in log-log space from $\sim$40 keV to few keV with a possible plateau or small reheating at the time of the second peak of the light curve. However, it is difficult to confirm this small feature since it could be simply due to a correlation between the parameters of the two components due to the fit process.

Globally, the Band function and the BB component evolve independently. Figure~\ref{fig:GRB120323A_resolved_spectra_B+BB} shows the evolution of the two spectral components with time. The reconstructed photon and energy light curves in the same energy bands as those used for the count light curves in Figure~\ref{fig:GRB120323A_LCs}, are presented in Figures~\ref{fig:GRB120323A_reconstructed_LC_B+BB} and~\ref{fig:GRB120323A_energy_LC}, respectively. The BB component contributes more than half the emitted energy between 20 and 150 keV during the first peak of the burst until $\sim$T$_\mathrm{0}$+0.080\,s. This contribution decreases very quickly to a few percent during the second peak of the burst. However, the BB remains an energetically subdominant component when computed from the time integrated spectrum over the entire GBM energy range (8 keV to 40 MeV), where it only contributes about 10\% of the total radiated energy.

Finally, we replaced the BB with a \fdaigne{CPL} function (C2) in the time interval with the highest statistical significance for the existence of the BB (i.e., from T$_\mathrm{0}$+0.94\,s to T$_\mathrm{0}$+0.132\,s), to investigate the shape of this low energy excess. C+C2 leads to the same Cstat values. The parameter $kT=15.6^\mathrm{+4.6}_\mathrm{-3.5}$ keV of C2 is similar to the temperature $kT=8.4\pm0.4$ keV of the BB in this time interval and the index of the C2 function has a value of $+\mathrm{0.1}^\mathrm{+0.5}_\mathrm{-0.4}$ which is similar with what is expected from a perfect Planck function (i.e., $+1$). As discussed in Section~\ref{section:interpretation1}, a thermal emission component with a low energy slope index around +0.4 is expected due to reprocessing of the photospheric emission, which is compatible with our data set.

\newpage
\subsection{Comparison}
\label{sec:comparison}

The most striking result when comparing Band-only fits with B+BB ones is the dramatic difference in the parameters of the Band function. The strong discontinuity observed in the evolution of the spectral parameters of the Band function around T$_\mathrm{0}$+0.094\,s with Band-only fits does not exist when fitting B+BB. In the B+BB model, E$_\mathrm{peak}$ is systematically shifted towards higher energies, and both $\alpha$ and $\beta$ are shifted towards lower values. In the Band-only scenario, $\alpha$ values exhibit large variations between the first and second peak of the light curve. With B+BB model, $\alpha$ remains constant throughout the burst. Similarly, the $\beta$ values vary during the burst with Band fits only, while the Band function high-energy power-law is always compatible with an exponential slope in the B+BB scenario. Therefore, in the B+BB scenario, the Band function can be replaced with a \fdaigne{CPL} function with no impact on the Cstat value of the fit; this replacement is only possible after $\sim$T$_\mathrm{0}$+0.094\,s in the Band-only fits. This explains the discontinuity in the evolution of the high energy power law indices, $\beta$, of the Band function around T$_\mathrm{0}$+0.094\,s when fitting Band-only to the data as presented in Figure~\ref{fig:GRB120323A_Spectral_parameter_evolution}.

Fundamentally, a comparison between a Band-only to B+BB fits defaults to comparing different global spectral shapes. The former corresponds to a single peak spectrum in the $\nu$F$_\nu$ space while the latter results in a two-peak $\nu$F$_\nu$ spectrum with each peak evolving independently. In the B+BB scenario, the additional BB is a significant component in terms of flux, especially between 20 and 150 keV, where it contributes to more than half of the total emission. The statistical significance of the additional BB component depends, for instance, on the energy separation of the Band and BB $\nu$F$_\nu$ peaks as well as on the relative intensity of the two components.

\fdaigne{In the next two paragraphs, we suggest that the low energy hump, which is well described with the BB component in the B+BB scenario, is responsible for the Band function shape when fitting a Band function alone to the data in the first peak of the light curve (i.e., $\leq$T$_\mathrm{0}$+0.054\,s).}
%In the next two paragraphs, we report the possibility for the low energy hump, which is well described with the BB component in the B+BB scenario, to be driving the Band function shape when fitting a Band function alone to the data in the first peak of the light curve (i.e., $\leq$T$_\mathrm{0}$+0.054\,s).
We then point out the strong similarities between the Band function shape from the Band-only fit with the BB and the Band function shape from the B+BB fit in the first and second peak of the light curve, respectively.

During the first peak of the light curve (i.e., $\leq$T$_\mathrm{0}$+0.054\,s), the E$_\mathrm{peak}$ obtained from Band-only fits decreases constantly from $\sim$200 to $\sim$40 keV. Similarly, the BB in the B+BB fits cools constantly from $\sim$35 to $\sim$10 keV during the same period of time. Since the maximum of the $\nu$F$_\nu$ spectrum of a BB with a temperature kT is $\sim$3$\times$kT, then the maximum of the $\nu$F$_\nu$ spectra of the BB resulting from the B+BB fits decrease from $\sim$105 to $\sim$30 keV. The evolution of  the peak of the BB spectrum from the B+BB fits is then very similar to the evolution of the E$_\mathrm{peak}$ of the Band-only fit. The positive values of $\alpha$ resulting from Band-only fits during the first peak of the light curve are also similar to the positive low energy slope of a Planck function. With both E$_\mathrm{peak}$ and $\alpha$, the Band function from Band-only fits and the BB from B+BB fits have very similar shapes for the low energy part. The difference between the Band function and the BB appears in the values of $\beta$. While a Planck function has very steep high energy spectral slope, the $\beta$ values of the Band-only fits are high during the first light curve peak ($\geq-2.4$). %During the first peak of the light curve where the BB would be the most intense, this component could be driving the spectral shape when fitting Band alone to the data.
\fdaigne{During the first peak of the light curve where the BB is most intense, it could strongly affect the spectral shape when fitting a Band function alone to the data.} This is illustrated in Figure~\ref{fig:GRB120323A_comparison} (top panel) where Band and B+BB fits are overplotted in a time-interval included in the first peak of the light curve.

During the second half of the burst (i.e., $\geq$T$_\mathrm{0}$+0.094\,s), the Band parameters, E$_\mathrm{peak}$, $\alpha$ and $\beta$ obtained either with Band-only or B+BB are very similar. The E$_\mathrm{peak}$ of Band-only fits decreases from $\sim$500 to $\sim$20 keV during the second peak of the light curve, and the temperature of the BB from B+BB cools from $\sim$10 keV down to $\sim$4~keV, which corresponds to a $\nu$F$\nu$ peak decreasing from $\sim$30 to $\sim$12 keV. Thus, conversely to what is observed during the first peak of the light curve, during the second peak, the Band function shape resulting from Band-only fits is very different from the shape of the BB resulting from B+BB model but consistent with the Band component evolution obtained with B+BB fits. This is again illustrated in Figure~\ref{fig:GRB120323A_comparison} (bottom panel) where Band and B+BB fits are this time overplotted in a time-interval included in the second peak of the light curve.

\begin{figure}
\begin{center}
\includegraphics[totalheight=0.286\textheight, clip,viewport=0 0 520 395]{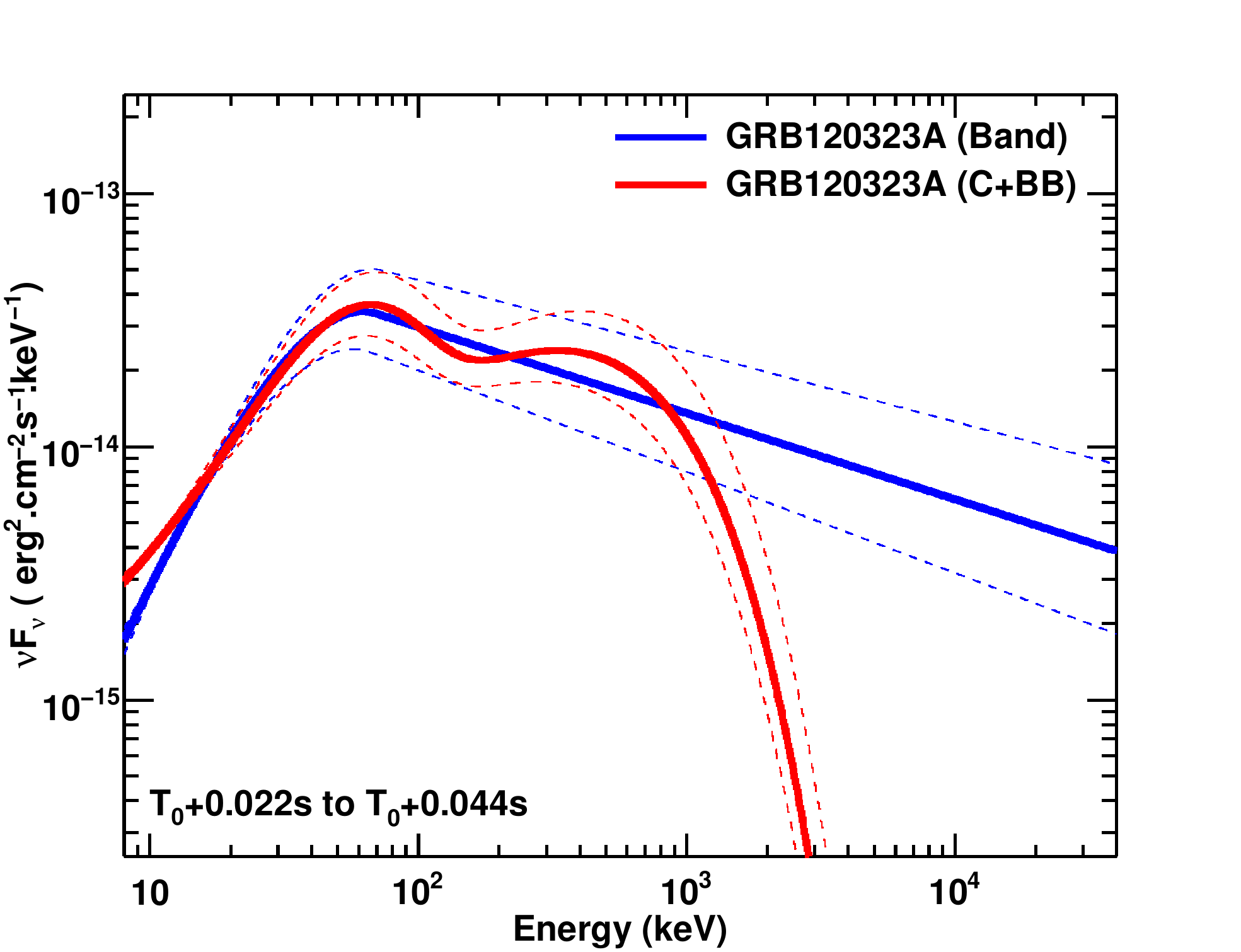}
\includegraphics[totalheight=0.286\textheight, clip,viewport=0 0 520 395]{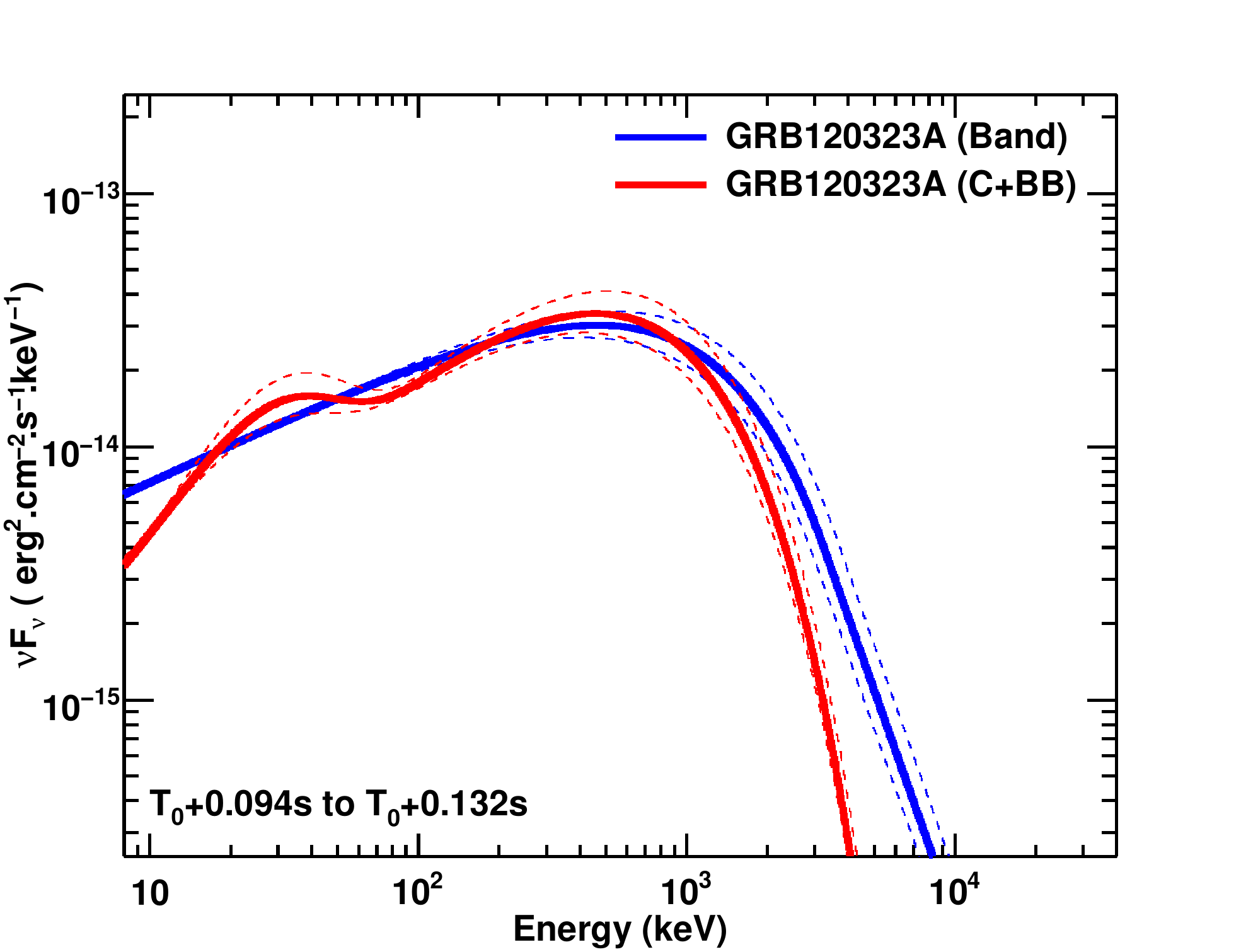}

\caption{\label{fig:GRB120323A_comparison}$\nu$F$_\nu$ spectra of \grb in two time intervals, T$_{0}$+0.022s to T$_{0}$+0.044s (top) and T$_{0}$+0.022s to T$_{0}$+0.044s (bottom), included in the first and the second peak of the light curve, respectively. The solid blue and red lines correspond to the best fits using the Band function only and the two humps model (i.e., C+BB), respectively. The dashed lines correspond to the 1-$\sigma$ uncertainties on the best fit. In the C+BB scenario, the low energy hump correspond to the BB component, while the high energy one correspond to the C function. In the time interval from T$_{0}$+0.022s to T$_{0}$+0.044s, the Band function of the Band-only fit mimic the shape of the BB component from the C+BB model (i.e., low energy hump), while in the time interval from T$_{0}$+0.022s to T$_{0}$+0.044s, the Band function of the Band-only fit mimic the shape of the C component from the C+BB model (i.e., high energy hump). This explains the low values for E$_\mathrm{peak}$ as well as the discontinuities in the evolution of the spectral parameters of the Band-only fits observed in Figure~\ref{fig:GRB120323A_Spectral_parameter_evolution}.}
\end{center}
\end{figure}

In section~\ref{sec:SinglecomponentBandfunction}, we pointed out that Rmfit converges towards two different minima when fitting a Band function to the data in the time interval T$_\mathrm{0}$+0.080\,s to T$_\mathrm{0}$+0.094\,s (see Table~\ref{tab:GRB120323A_resolved_spectra_fine}). The best fit is obtained with a low E$_\mathrm{peak}$ value and a high $\alpha$ one (i.e., B1) ; the other local minimum has a much higher E$_\mathrm{peak}$ value and a very steep $\alpha$ (i.e., B2). It is interesting to notice that with their measured spectral parameters, B1 and B2 mimic the Band and the BB components of the B+BB fit, respectively. Before T$_\mathrm{0}$+0.080\,s, the fit of a single Band function to the data would mimic the BB component of the B+BB fit, while after T$_\mathrm{0}$+0.094\,s, the Band-only fit would mimic the Band component of the B+BB fit. %This supports the hypothesis that the BB component from the B+BB model is very intense during the first part of the burst and would be driving the Band-only fit, while when the intensity of this BB component decays during the second part of the burst, the Band-only fit would be driven by the Band function of the B+BB scenario (see Figure~\ref{fig:GRB120323A_comparison}).
\fdaigne{This supports the hypothesis that the BB component from the B+BB model is very intense during the first part of the burst and would be strongly affecting the Band parameters of the single component (Band function) fit, while when the intensity of this BB component decays during the second part of the burst, these parameters are well determined by the Band function of the B+BB scenario (see Figure~\ref{fig:GRB120323A_comparison})}.

\subsection{Light curve peak overlapping scenario}

We cannot completely exclude that the two humps detected in the $\nu$F$\nu$ spectra when using the coarse time intervals (see Section~\ref{sec:Coarse time resolved}) are an artifact due to spectral evolution when the two peaks of the light curve overlap. Lets consider the time interval from T$_\mathrm{0}$+0.058s to T$_\mathrm{0}$+0.100s from Table~\ref{tab:GRB120323A_resolved_spectra_large} where the BB is statistically the most significant. This time interval includes the decay phase of the first peak of the light curve and the intense part of the second one. When fitting C2+C or C+B2 to the data (see Table~\ref{tab:GRB120323A_resolved_spectra_large}), the component with the lowest E$_\mathrm{peak}$ has spectral parameters (i.e., both E$_\mathrm{peak}$ and $\alpha$) compatible with the trend reported in Section~\ref{sec:Fine time resolved} when fitting Band-only or Compt to the data up to T$_\mathrm{0}$+0.094s (see Table~\ref{tab:GRB120323A_resolved_spectra_fine}). Similarly, the spectral parameters (i.e., both E$_\mathrm{peak}$ and $\alpha$) of the component with the highest E$_\mathrm{peak}$ are compatible with the trend reported in section~\ref{sec:Fine time resolved} when fitting Band-only to the data after T$_\mathrm{0}$+0.094s (see Table~\ref{tab:GRB120323A_resolved_spectra_fine}). In addition, the flux of the component with the lowest E$_\mathrm{peak}$ is compatible with the decaying flux of the first peak of the light curve when fitting Band-only or Compt, and the flux of the component with the highest E$_\mathrm{peak}$ is compatible with the peak intensity of the second peak of the light curve when fitting the simplest models. Thus, the spectrum in this interval could be described as the superposition of the end of the first structure of the light curve and the beginning of the second one.

However, this scenario cannot explain all the observations reported in this article. For instance, the superposition of the two peaks of the light curve can hardly explain the possibility to fit two components with similar intensity at the very beginning of the burst where the contribution of the second peak of the light curve should be very weak. Further we note that a two-component fit at the beginning of the burst cannot be due to random fluctuations because of the monotonic trend of the BB temperature (see Table~\ref{tab:GRB120323A_resolved_spectra_fine} and bottom right panel of Figure~\ref{fig:GRB120323A_Spectral_parameter_evolution}). It is also difficult to explain the evolution of the spectral parameters obtained when using the simplest models such as the sharp discontinuity observed simultaneously for all parameters or the resulting Luminosity-E$_\mathrm{peak}$ relation described in section~\ref{section:Flux-Epeak correlation}.

\section{E$_\mathrm{peak,i}^\mathrm{rest}$-L$_\mathrm{i}^\mathrm{Band}$ relation}
\label{section:Flux-Epeak correlation}

\citet{Golenetskii:1983} reported for the first time a hardness-intensity correlation during the prompt emission of GRBs observed in the Konus experiment on the Venera 13 and 14 spacecraft. \citet{Ryde:2001} \fdaigne{as well as~\citet{Liang:2004}} confirmed this correlation in a sample of BATSE GRBs extending it to the evolution of the Band E$_\mathrm{peak}$ during the burst and its corresponding Luminosity. \citet{Guiriec:2010} showed that the correlation between the Band-function E$_\mathrm{peak}$ evolution and the count light curve increased with the energy range used to define the light curve for three short and bright GRBs observed with GBM. \citet{Ghirlanda:2011a} extended this result to a sample of 13 short GRBs detected with GBM showing the correlation between the Band function luminosities and their instantaneous E$_\mathrm{peak}$ values, and then also to a sample long GBM GRBs in~\citet{Ghirlanda:2010,Ghirlanda:2011b}. More recently, \citet{Lu:2012} reported a similar analysis on a large sample of 62 bright GBM GRBs.

\begin{figure*}
\begin{center}
\includegraphics[totalheight=0.286\textheight, clip,viewport=0 0 520 403]{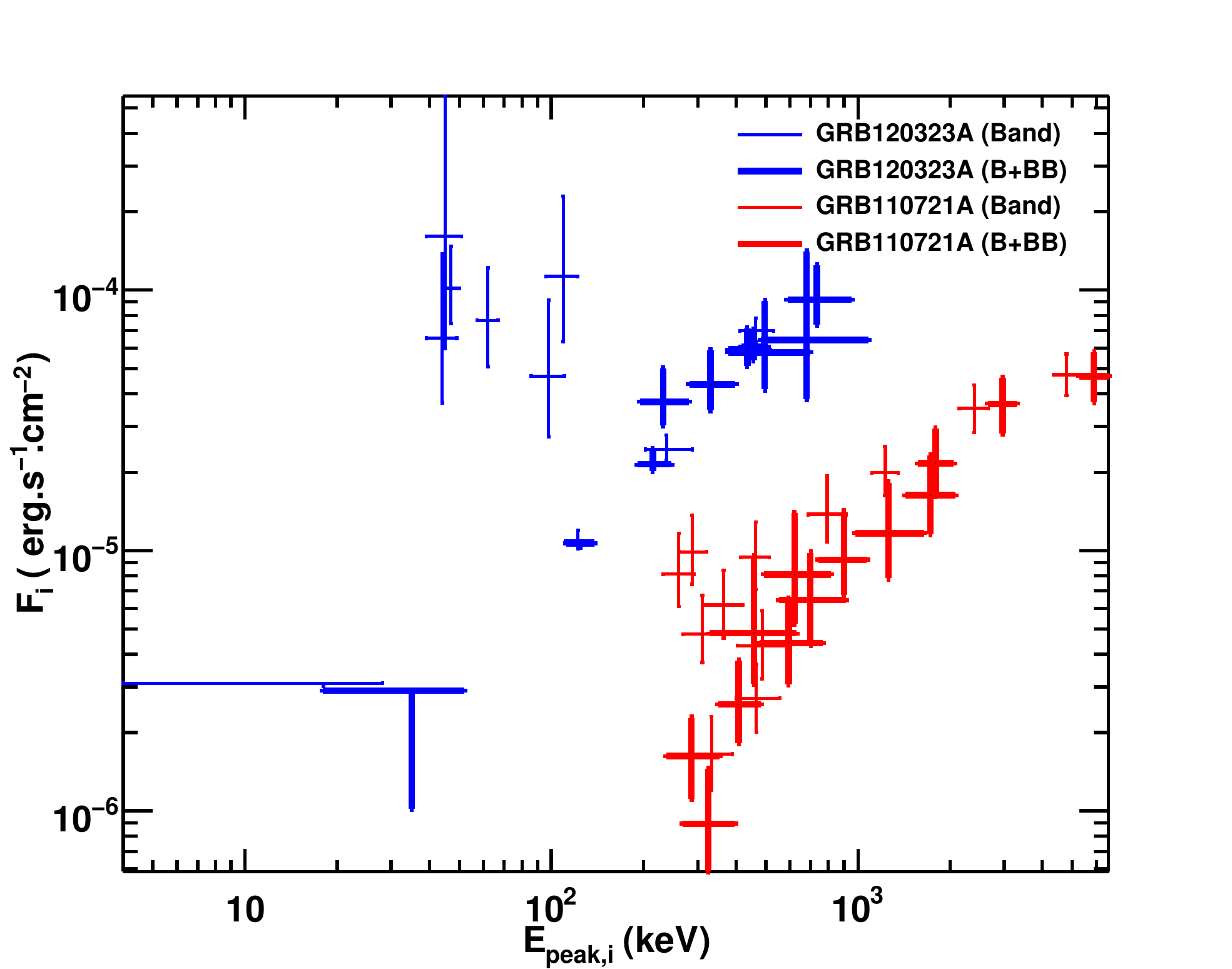}
\includegraphics[totalheight=0.286\textheight, clip,viewport=0 0 520 403]{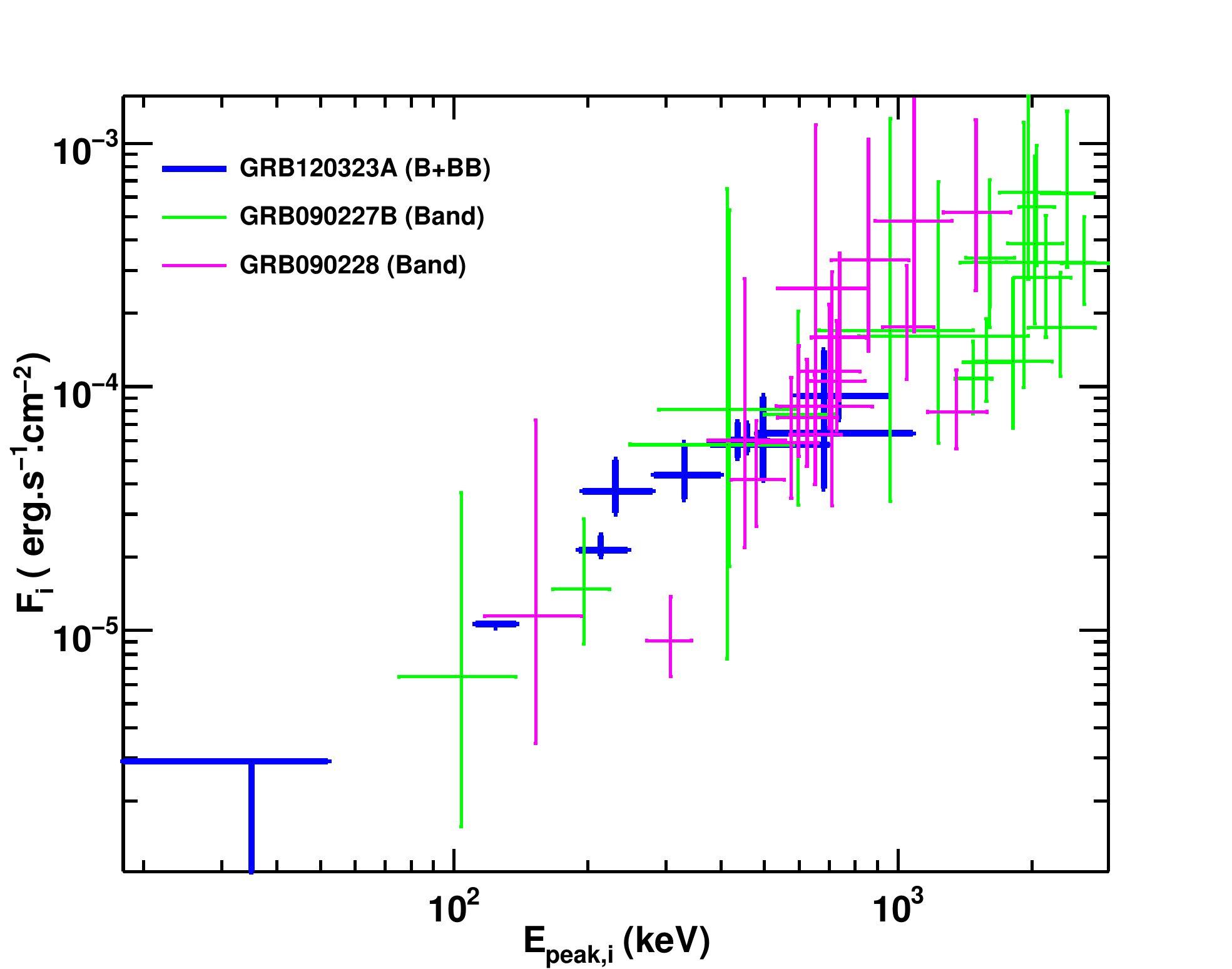}
\includegraphics[totalheight=0.286\textheight, clip,viewport=0 0 520 403]{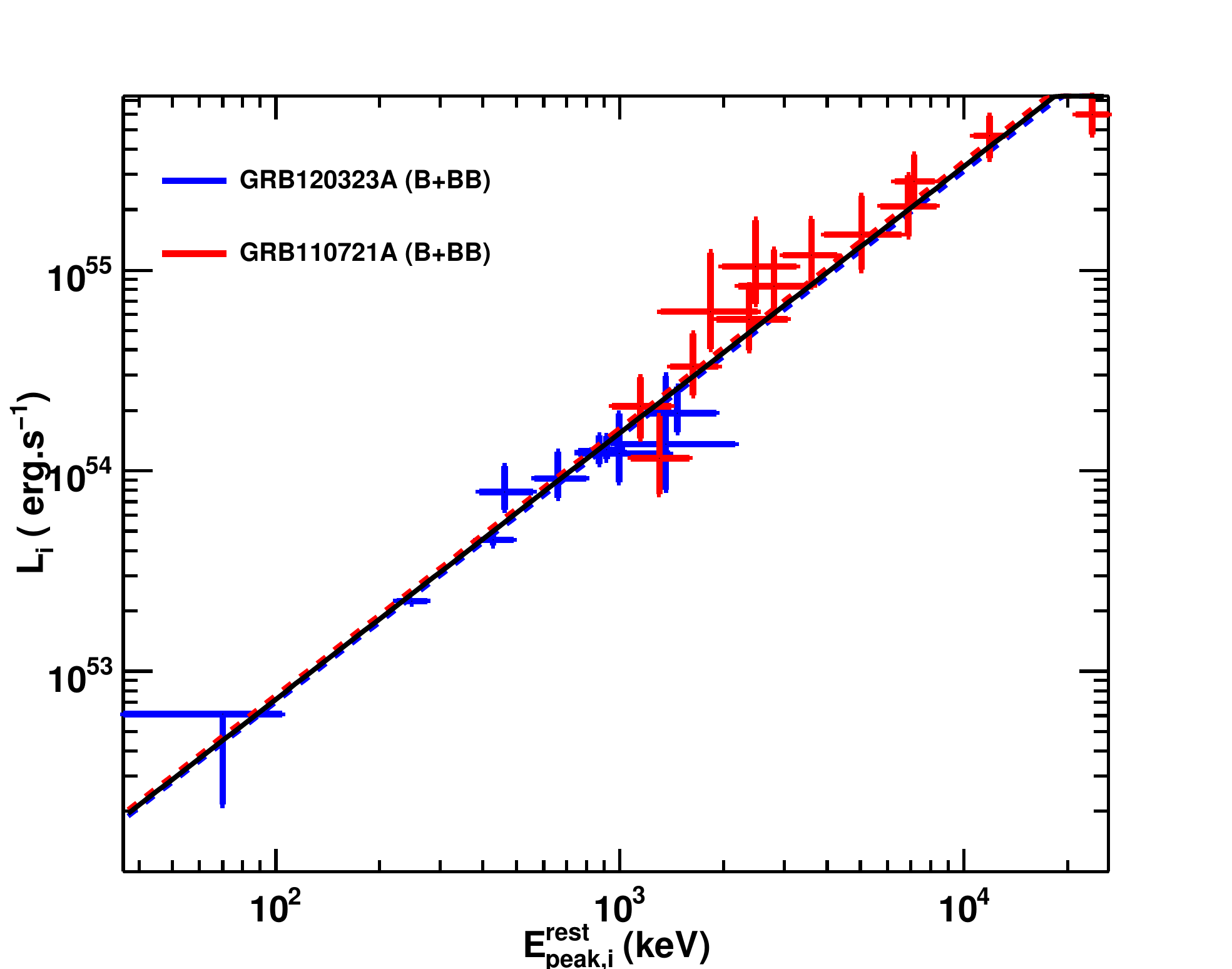}

\includegraphics[totalheight=0.286\textheight, clip,viewport=0 0 520 403]{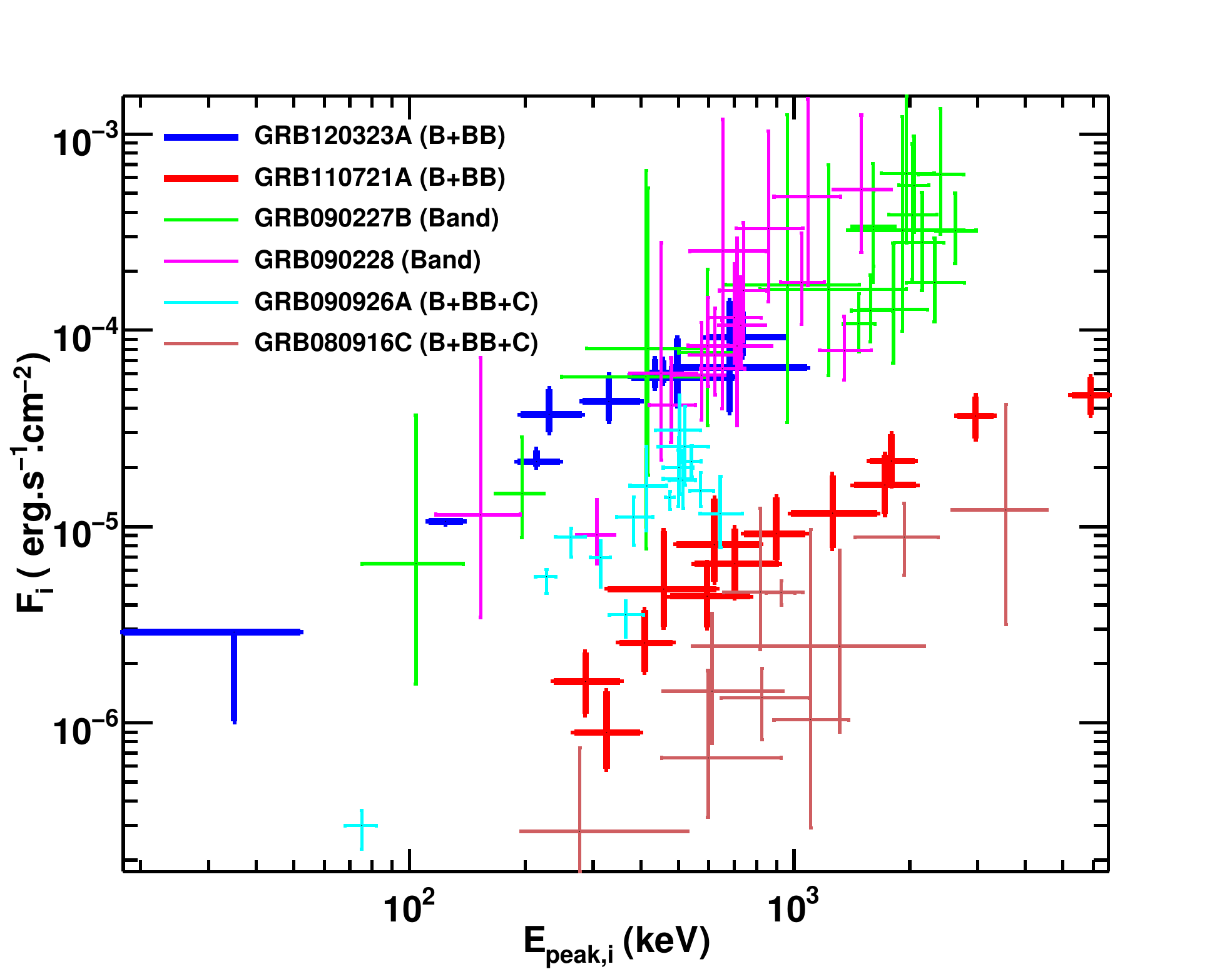}
\includegraphics[totalheight=0.286\textheight, clip,viewport=0 0 520 403]{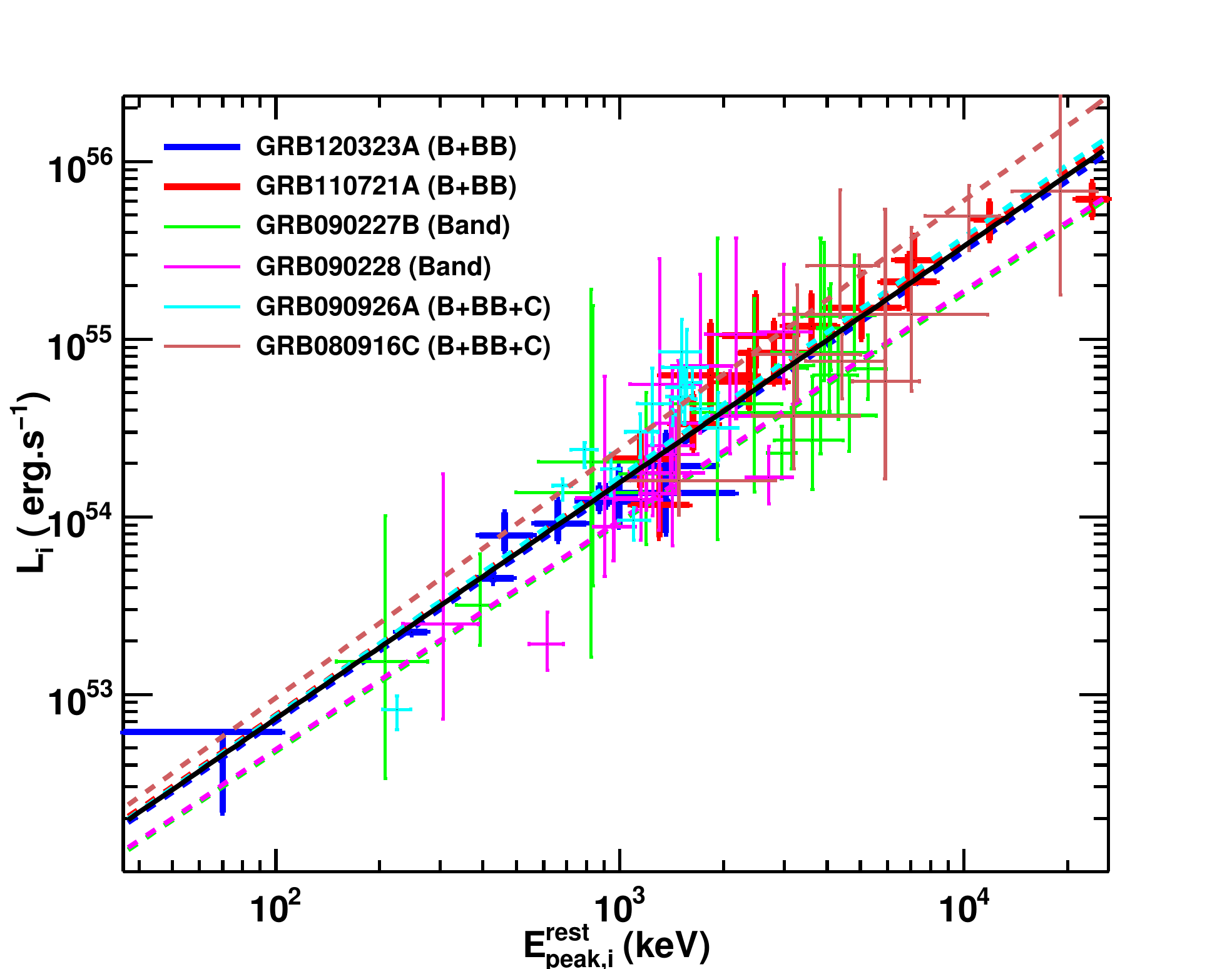}

\caption{\label{fig:GRB120323A_Flux-vs-Epeak} {Top left: Band function $\nu$F$_\nu$ spectral peak versus the flux of the Band function computed from 8 keV to 40 MeV (i.e., E$_\mathrm{peak,i}$-F$_\mathrm{i}$ relation) obtained when fitting Band (thin) and B+BB (thick) to the time-resolved data of \grb (blue) and GRB~$110721$A (red). Top right: Band function $\nu$F$_\nu$ spectral peak versus the flux of the Band function computed from 8 keV to 40 MeV (i.e., E$_\mathrm{peak,i}$-F$_\mathrm{i}$ relation) obtained when fitting B+BB (thick blue) to the time-resolved data of \grb (blue) and Band to the time-resolved spectra of two other short GRBs analyzed in \citet{Guiriec:2010}(thin). Center: E$_\mathrm{peak,i}$-F$_\mathrm{i}$ relation in the rest frame (i.e., E$_\mathrm{peak,i}^\mathrm{rest}$-L$_\mathrm{i}$ relation) obtained when fitting B+BB to the time-resolved data of \grb (blue) and GRB~$110721$A (red). The thick dash blue and red lines correspond to the fit using a power law to the data of \grb and GRB~$110721$A, respectively. The thin black line corresponds to the simultaneous fit of all data with a power law. Bottom left: Band function $\nu$F$_\nu$ spectral peak versus the flux of the Band function computed from 8 keV to 40 MeV (i.e., E$_\mathrm{peak,i}$-F$_\mathrm{i}$ relation) obtained over a sample of short and long GRBs. Bottom right: E$_\mathrm{peak,i}$-F$_\mathrm{i}$ relation in the rest frame (i.e., E$_\mathrm{peak,i}^\mathrm{rest}$-L$_\mathrm{i}$ relation) obtained over a sample of short and long GRBs.The thin black line corresponds to the simultaneous fit of all data with a power law.}}
\end{center}
\end{figure*}

In Figure~\ref{fig:GRB120323A_Flux-vs-Epeak} (top left panel), we plot the energy flux of the Band function (i.e., F$_\mathrm{i}^\mathrm{Band}$)\footnote{i being the index of a time-resolved spectrum within a burst.} between 8 keV and 40 MeV versus the Band $\nu$F$_\nu$ peak energy, E$_\mathrm{peak,i}$, of a single Band function fit (thin line) or a B+BB (thick line) to the time resolved spectra of \grb (blue) and GRB~$110721$A (red). While no correlation is observed between F$_\mathrm{i}^\mathrm{Band}$ and E$_\mathrm{peak,i}$ when fitting Band-only to the data of \grbnos, a strong correlation emerges between the flux of the Band function, F$_\mathrm{i}^\mathrm{Band}$, and E$_\mathrm{peak,i}$ when fitting a model with two humps (i.e., B+BB). We notice the large errors on F$_\mathrm{i}^\mathrm{Band}$ when fitting only Band compared to B+BB. We would expect the opposite behavior since with more parameters in the B+BB scenario we would also expect larger uncertainties, which should propagate accordingly to the errors of F$_\mathrm{i}^\mathrm{Band}$. The E$_\mathrm{peak,i}$-F$_\mathrm{i}^\mathrm{Band}$ relation obtained using the B+BB model is similar to the previously reported results using Band alone~\citep{Ryde:2001,Ghirlanda:2010,Ghirlanda:2011a,Ghirlanda:2011b,Lu:2012}.  We conclude that this behavior also favors the existence of a hump in the low-energy power law of the Band function.

The difference in the results between Band-only and B+BB fits is mainly the shift of E$_\mathrm{peak,i}$ to higher energies, when using the latter model as well as the decrease of the Band function contribution to the total energy flux, since another component is intense at low energies (see also section~\ref{sec:twocomps}). Figures~\ref{fig:GRB120323A_reconstructed_LC_B+BB} and~\ref{fig:GRB120323A_energy_LC} clearly exhibit the strong presence (over 60\% of the total flux) of the BB in the first pulse between $8-150$ keV, while this contribution becomes 40\% and less during the second pulse. These results are reflected in Figure~\ref{fig:GRB120323A_Flux-vs-Epeak} (top left panel): the Band fit data points compatible with the relation obtained with B+BB correspond to the second peak of the light curve, where the BB is statistically the most significant but at the same time the least intense component, thus affecting the Band E$_\mathrm{peak,i}$ in the least. In contrast, the points of the Band-only fit that lie off the straight line relation correspond to the first peak of the light curve when the BB is statistically less significant but most intense and thus would affect E$_\mathrm{peak,i}$ the most. This again reinforces the two component scenario (B+BB) in the first peak of the light curve. Although less extreme, similar results are obtained with GRB~$110721A$ (red lines) (see Figure~\ref{fig:GRB120323A_Flux-vs-Epeak}) for which an intense BB emission was also identified~\citep{Axelsson:2012:GRB110721A}. The relation between F$_\mathrm{i}^\mathrm{Band}$ and E$_\mathrm{peak,i}$ seems to be especially strong during the decay phase of individual pulses. Detailed analysis of several very intense bursts is necessary to assess if E$_\mathrm{peak,i}$ tracks the energy flux during the rising phase of a pulse or if a hard to soft evolution of E$_\mathrm{peak,i}$ is really observed during this phase. In the latter case, the E$_\mathrm{peak,i}$-F$_\mathrm{i}^\mathrm{Band}$ relation would only exist during the decay phase of individual pulses.

In Figure~\ref{fig:GRB120323A_Flux-vs-Epeak} (top right panel), we plot the E$_\mathrm{peak,i}$-F$_\mathrm{i}^\mathrm{Band}$ relation of two short GRBs, GRB~$090227$B and~$090228$, analyzed in~\citet{Guiriec:2010} (thin line) together with \grb (thick line). In these two short GRBs, a weak BB could also be present, but it has very little effect on both the measured Band function flux as well as on E$_\mathrm{peak}$. Band-only is then a good enough model for these two GRBs. The three GRBs lie along the same E$_\mathrm{peak,i}$-F$_\mathrm{i}^\mathrm{Band}$ relation. Short GRBs are nearby events with a narrow redshift distribution: $\sim$80\% of them have $z<$1 \citep{Racusin:2011}.
Therefore, the E$_\mathrm{peak,i}$-F$_\mathrm{i}^\mathrm{Band}$ relation is expected to lead to a similar correlation between the luminosity of the Band function, L$_\mathrm{i}^\mathrm{Band}$, and E$_\mathrm{peak,i}^\mathrm{rest}$ in the rest frame of the central engine suggesting a possible universal L$_\mathrm{i}^\mathrm{Band}$-E$_\mathrm{peak,i}^\mathrm{rest}$ relation for short GRBs. The average short GRB distances would then be consistent with the similar relation for the three bursts in Figure~\ref{fig:GRB120323A_Flux-vs-Epeak}.

We now extend this analysis including several long GRBs. Figure~\ref{fig:GRB120323A_Flux-vs-Epeak} (center) shows the E$_\mathrm{peak,i}^\mathrm{rest}$-L$_\mathrm{i}^\mathrm{Band}$ relations for both \grb and GRB~$110721$A.
We used a typical short GRB redshift of 1 for \grbnos. \citet{Berger:2011} reported tentative spectroscopic redshifts for GRB~$110721$A at either 3.512 or 0.382 based on absorption line observations with GMOS on the Gemini-South 8-m telescope; the former value is consistent with the possible redshift of $\sim$3.2 reported by \citet{Greiner:2011}. However, doubt remains on the real identification of the afterglow for GRB~$110721$A. The data for the two GRBs line up perfectly showing a very strong correlation between E$_\mathrm{peak,i}^\mathrm{rest}$ and L$_\mathrm{i}^\mathrm{Band}$ only when using a redshift of $\sim$3.2 for GRB~$110721$A. This suggests that the  long GRB~$110721$A is harder and more intense than \grb in the rest frame.
The blue and red dashed lines correspond to the fits to E$_\mathrm{peak,i}^\mathrm{rest}$-L$_\mathrm{i}^\mathrm{Band}$ data with a power law for \grb and GRB~$110721$A, respectively.
The best parameters of these fits with their 1-$\sigma$ uncertainties are:
\[
\mathrm{L_{120323A,i}^{Band}}=\mathrm{(1.57\pm1.26)10^{50}~(E_{peak,i}^{rest})^{1.32\pm0.13} erg~s^{-1}}
\]
\[
\mathrm{L_{110721A,i}^{Band}}=\mathrm{(1.65\pm2.64)10^{50}~(E_{peak,i}^{rest})^{1.34\pm0.19} erg~s^{-1}}
\]
The simultaneous fit of the two data sets with a power law corresponds to the solid black line given by:
\begin{equation}
\mathrm{L_i^{Band}}=\mathrm{(1.60\pm0.65)10^{50}~(E_{peak,i}^{rest})^{1.33\pm0.06} erg~s^{-1}}
\label{eq:L-Epeak_both}
\end{equation}
The very good consistency between these three relations supports the possible universal behavior of the E$_\mathrm{peak,i}^\mathrm{rest}$-L$_\mathrm{i}^\mathrm{Band}$ relation.

In Figure~\ref{fig:GRB120323A_Flux-vs-Epeak} (bottom left), we added the data for long GRBs~$080916$C and $090926$A from \citet{Guiriec:inPrep} to the sample of the GRBs presented above. These two additional GRBs have measured redshifts estimated at 4.35$\pm$0.15 \citep{Greiner:2009} and at $\sim$2.1062 \citep{Malesani:2009}, respectively, and are fitted with a combination of three components, a Band function, a BB component and an addition power law \citep{Guiriec:2011b}. We notice that the E$_\mathrm{peak,i}$-F$_\mathrm{i}^\mathrm{Band}$ relations for the various GRBs are shifted, with the highest $z$ GRBs being also the dimmest. Figure~\ref{fig:GRB120323A_Flux-vs-Epeak} (bottom right) shows the E$_\mathrm{peak,i}^\mathrm{rest}$ vs L$_\mathrm{i}^\mathrm{Band}$ for the same sample of GRBs, assuming a redshift of 1.0 for all short GRBs. All data points now line up; a power law fit (solid black line) gives:
\begin{equation}
\mathrm{L_i^{Band}}=\mathrm{(1.59\pm0.84)10^{50}~(E_{peak,i}^{rest})^{1.33\pm0.07} erg~s^{-1}}
\label{eq:L-Epeak_all}
\end{equation}

The color dashed lines correspond to the individual fit to the data of each burst with a power law, and the results are :
\[
\mathrm{L_{090227B,i}^{Band}}=\mathrm{(1.26\pm1.65)10^{50}~(E_{peak,i}^{rest})^{1.29\pm0.17} erg~s^{-1}}
\]
\[
\mathrm{L_{090228,i~}^{Band}}=\mathrm{(1.28\pm3.22)10^{50}~(E_{peak,i}^{rest})^{1.29\pm0.36} erg~s^{-1}}
\]
\[
\mathrm{L_{080916C,i}^{Band}}=\mathrm{(1.50\pm4.02)10^{50}~(E_{peak,i}^{rest})^{1.41\pm0.30} erg~s^{-1}}
\]
\[
\mathrm{L_{090926A,i}^{Band}}=\mathrm{(1.50\pm0.28)10^{50}~(E_{peak,i}^{rest})^{1.34\pm0.03} erg~s^{-1}}
\]

The relationships described by equations \ref{eq:L-Epeak_both} and \ref{eq:L-Epeak_all} are identical within errors and reinforce the possible universality of the E$_\mathrm{peak,i}^\mathrm{rest}$-L$_\mathrm{i}^\mathrm{Band}$ relation across short and long GRBs. We note here that this GRB sample \citep{Guiriec:2010,Guiriec:inPrep} is limited; an analysis using a larger sample is the subject of a followup study. Our GRB data sample is also too limited to quantify dispersion effects which could be due to multiple physical parameters like for instance the bulk Lorentz factor and the jet opening angle, or possible selection effects which could prevent the detection of possible outliers to this relation.

The presence of the E$_\mathrm{peak,i}$-F$_\mathrm{i}^\mathrm{Band}$ and E$_\mathrm{peak,i}^\mathrm{rest}$-L$_\mathrm{i}^\mathrm{Band}$ relations in these six GRBs leads to multiple conclusions:
\begin{itemize}
\item The E$_\mathrm{peak,i}$-F$_\mathrm{i}^\mathrm{Band}$ relation seems to be {\it intrinsic} across the time-resolved spectra of single bursts, and the E$_\mathrm{peak,i}^\mathrm{rest}$-L$_\mathrm{i}^\mathrm{Band}$ seems to be similar from burst to burst on a large sample of events \citep{Ghirlanda:2011a,Ghirlanda:2011b} although the slope slightly differs from what has been previously reported. It can, therefore, not be only attributed to instrumental selection effects as often suggested to explain the so called Amati \citep{Amati:2002} and Ghirlanda \citep{Ghirlanda:2004} relations~\citep{Ghirlanda:2008,Kocevski:2012}, nor to effects of the redshift correction simultaneously impacting L$_\mathrm{i}^\mathrm{Band}$ and E$_\mathrm{peak,i}^\mathrm{rest}$, resulting in an artificial correlation.
\item It has been suggested that the correlation between the parameters of the Band function and its $\nu$F$_\nu$ peak energy is driven by the intrinsic correlation between the parameters of the Band function itself \citep{Massaro:2008,Goldstein:2012b}. With \grb and GRB~$110721$A, we have counter examples showing that fits to the data with Band-only do not lead to any correlation between E$_\mathrm{peak,i}$ and F$_\mathrm{i}^\mathrm{Band}$, while B+BB fits do. If the correlation were mostly a model artifact, instead of physically driven, it should also be present when fitting Band-only to the data.
\item Contrary to the Amati-like relations, the E$_\mathrm{peak,i}^\mathrm{rest}$-L$_\mathrm{i}^\mathrm{Band}$ relation does not lead to a universal scenario for the central engine, but instead to the similarity of the relativistic jet evolution and radiation mechanisms dissipating the energy released by the central engine.
It is in fact an extension of the so called Yonetoku relation~\citep{Yonetoku:2004}, which  correlates for a sample of bursts, the peak flux of each burst (integrated over 1 s) with its corresponding E$_\mathrm{peak}$. This relationship is similar to the E$_\mathrm{peak,i}^\mathrm{rest}$-L$_\mathrm{i}^\mathrm{Band}$ relation albeit with only one data point per GRB.
\item For \grb the E$_\mathrm{peak,i}$-F$_\mathrm{i}^\mathrm{Band}$ relation holds only when B+BB is fit to the data. Any physical interpretation must reproduce this relation to be viable, thus making the E$_\mathrm{peak,i}^\mathrm{rest}$-F$_\mathrm{i}^\mathrm{Band}$ relation {\it a tool} that discriminates between theoretical scenarios trying to explain GRB prompt emission. In addition, GRBs deviating from this relation when fit with the Band function only, such as for \grb and GRB~$110721$A, may include evidence for a strong additional component such as BB. We note that the outlier GRBs from the E$_\mathrm{peak}$-L relation from \citet{Ghirlanda:2010}, namely GRB~$080916$C, GRB~$090510$ and GRB~$090902$B are GRBs known to exhibit strong additional spectral components to the Band function \citep{Guiriec:2010,Guiriec:2011b,Guiriec:inPrep}. Our sample also include GRB~$080916$C and we showed that with a detailed spectral analysis, GRB~$080916$C is perfectly consistent with our new E$_\mathrm{peak,i}^\mathrm{rest}$-F$_\mathrm{i}^\mathrm{Band}$ relation.
\item Yonetoku, Amati and Ghirlanda relations exhibit dispersion effects which could eventually be reduced using a similar multi-component spectral analysis as the one proposed in this paper.
\item Finally, since (redshift corrected) data of both short and long GRBs satisfy very similar E$_\mathrm{peak,i}^\mathrm{rest}$-L$_\mathrm{i}^\mathrm{Band}$ relations, a well-calibrated and dispersion corrected formula could eventually be used to estimate redshifts for GRBs in the absence of multi-wavelength follow-up observations. Such estimates would only require a relatively intense GRB to enable accurate time-resolved spectroscopy.
\end{itemize}

Beyond the previous conclusions, if the universality of this new E$_\mathrm{peak,i}^\mathrm{rest}$-L$_\mathrm{i}^\mathrm{Band}$ relation is confirmed, then it will open new perspectives for the development of future instruments. In many cases such as in the gravitational wave research field, the redshift of a GRB is a crucial quantity to measure and requires a complex chain of operations consisting of repointing telescopes at various wavelengths to the source. However, the initial localization is often not good enough to initiate the process at all. Here, the E$_\mathrm{peak,i}^\mathrm{rest}$-L$_\mathrm{i}^\mathrm{Band}$ relation would allow a redshift determination only from the study of the spectral evolution in the gamma ray emission of GRBs. Thus a large GBM-like instrument with high sensitivity would be ideal to determine GRB redshifts.

\fdaigne{In addition,~\citet{Nemmen:2012} reported striking similarities in the energetics of jets produced in GRBs and active galactic nuclei (AGNs). It would be interesting to compare the spectral properties of AGNs -- more specifically blazars -- and GRBs in order to investigate whether the physical mechanisms and radiative processes are similar in all relativistic jets. Therefore, AGNs may exhibit similar E$_\mathrm{peak,i}$-F$_\mathrm{i}^\mathrm{Band}$ and/or E$_\mathrm{peak,i}^\mathrm{rest}$-L$_\mathrm{i}^\mathrm{Band}$ relations.}

\section{Interpretation}
\label{section:interpretation1}

Gamma-ray bursts are associated with ultra-relativistic outflows ejected by a newborn compact source \citep[see e.g.,][]{Piran:2004} and their prompt emission is very likely due to internal dissipation within the ejecta \citep{Sari:1997}.
Assuming that \grb were a standard short GRB with an intense BB component, it is tempting to associate this component to the photospheric emission produced by the relativistic outflow when it becomes transparent at large distances from the central engine. Without any additional dissipation process at the photosphere, the predicted photospheric spectrum is indeed close to a BB with two main modifications : (i) the low-energy slope is affected by the complex geometry of the photosphere, which leads to a photon slope $\alpha\simeq +0.4$ instead of  $\alpha=+1$ for an exact Planck function \citep{Goodman:1986,Peer:2008,Beloborodov:2010}; (ii) the observed peak can be broadened if the temperature evolves on a timescale which is shorter than the time interval used for the spectral analysis. The second effect should be limited in \grb as its brightness allows a refined analysis with time bins shorter than the duration of the two main pulses. 

The spectral analysis presented above is not sensitive to the precise value of the low-energy spectral slope of the low energy hump (i.e., BB component). While data require a component with a positive low energy spectral slope compatible with a Planck function shape to describe the low energy hump, a modified BB component with a low energy spectral index around +0.4 does not affect the results and is perfectly compatible with the data as well (see \S\ref{sec:twocomps}). Then, by assuming that the BB component in \grb is a thermal component of photospheric origin, it is possible to put some constraints on three important physical parameters: the radius $R_0$ at the base of the flow, the Lorentz factor $\Gamma$, and the photospheric radius $R_\mathrm{ph}$ \fdaigne{(see e.g., \citealt{Daigne:2002,Peer:2007}).
\citet{Hascoet:2013} have generalized the procedure proposed by \citet{Peer:2007} to the case of magnetized outflows, under very general assumptions: (i) the flow becomes radial within an opening angle $\theta_\mathrm{j}$ above a radius $R_\mathrm{sph}$ which is smaller than the saturation radius $R_\mathrm{sat}$ where the acceleration is complete, and than the photospheric radius $R_\mathrm{ph}$; there is no significant sub-photospheric dissipation (i.e. no conversion of magnetic energy or kinetic energy into internal energy below the photosphere); (iii) acceleration is completed at the photosphere, i.e. $R_\mathrm{sat}<R_\mathrm{ph}$. Under these assumptions, $R_0$, $\Gamma$ and $R_\mathrm{ph}$ are related to observed quantities by \citep{Hascoet:2013}: } 
%\fdaignesupp{(see e.g., \citealt{Daigne:2002,Peer:2007} and \citet{Hascoet:2013} for a generalization to the case of magnetized outflows).} 
%\fdaignesupp{These are given by:}
\begin{eqnarray}
R_0 & \simeq & \left[ \frac{D_\mathrm{L} \mathcal{R}}{2(1+z)^2}\left(\frac{\phi}{1-\phi}\right)^{3/2} \right] \times \left[ \frac{f_\mathrm{NT}}{\epsilon_\mathrm{T}}\right]^{3/2}\, ,\label{eq:R0}\\
\Gamma & \simeq & \left[ \frac{\sigma_\mathrm{T}}{m_\mathrm{p}c^3}\frac{(1+z)^2 D_\mathrm{L} F_\mathrm{BB}}{\mathcal{R}} \frac{1-\phi}{\phi} \right]^\frac{1}{4} \times \left[\left(1+\sigma\right) f_\mathrm{NT}\right]^{-1/4}\, ,\label{eq:Gamma}\\
R_\mathrm{ph} & \simeq & \left[ \frac{\sigma_\mathrm{T}}{16 m_\mathrm{p}c^3}\frac{D^5_\mathrm{L} F_\mathrm{BB} \mathcal{R}^3}{(1+z)^6} \frac{1-\phi}{\phi} \right]^\frac{1}{4} \times \left[\left(1+\sigma\right) f_\mathrm{NT}\right]^{-1/4}\, ,\label{eq:Rph}
\end{eqnarray}
where $z$ and $D_\mathrm{L}$ are the redshift and the luminosity distance of the source, $F_\mathrm{BB}$ is the measured  flux of the BB component in a given time bin, $\phi=F_\mathrm{BB}/F_\mathrm{tot}$ is the ratio of the flux of the BB component over the total flux, and $\mathcal{R}$ is computed from $F_\mathrm{BB}$ and the measured temperature of the BB component by
\begin{equation}
\mathcal{R}=\left(\frac{F_\mathrm{BB}}{\sigma T_\mathrm{BB}^4}\right)^{1/2}\, .
\end{equation} 
In addition to these quantities that can be directly measured, there are two unknown parameters related to the GRB physics, the ratio $\epsilon_\mathrm{T}/f_\mathrm{NT}$ and the product $\left(1+\sigma\right) f_\mathrm{NT}$, where $\epsilon_\mathrm{T}$ is the fraction of the initial energy released by the source which is in thermal form (the initial fraction of magnetic energy
is  $1-\epsilon_\mathrm{T}$), $f_\mathrm{NT}$ is the efficiency of the dissipation mechanism responsible for the non-thermal component observed in the spectrum, and $\sigma$ is the magnetization of the relativistic outflow at the end of the acceleration process. \fdaigne{As described in \citet{Hascoet:2013}, the parameters $\epsilon_\mathrm{T}$ and $\sigma$ allow to study different classes of models for GRB outflows: (i) the standard thermally accelerated fireball model ($\epsilon_\mathrm{T}=1$; $\sigma=0$); (ii) outflows that are Poynting flux dominated close to the central engine ($\epsilon_\mathrm{T}\ll 1$) with either a good conversion of the magnetic energy into kinetic energy (low $\sigma$) or not (high $\sigma$); (iii) intermediate cases. The parameter $f_\mathrm{NT}$ allows to discuss different mechanisms for the non-thermal emission above the photosphere, such as internal shocks (low to moderate $f_\mathrm{NT}$) or magnetic reconnection (moderate to high $f_\mathrm{NT}$). Equations (\ref{eq:R0}--\ref{eq:Rph}) are valid for any acceleration law $\Gamma(R)$ for the outflow, as long as the saturation radius is below the photospheric radius. We discuss below the validity of this assumption in the case of GRB 120323A.}\\

\begin{table*}
\caption{\label{tab:model}Values of the initial radius, Lorentz factor, photospheric radius and total (isotropic equivalent) injected power derived from the results of the time-integrated and time-dependent spectral analysis, using equations~(\ref{eq:R0}--\ref{eq:Rph}).
}
\begin{center}
\small
%\tiny
\begin{tabular}{cccccc}
\multicolumn{2}{c}{Time interval} & Initial radius (cm) & Lorentz factor & Photospheric radius (cm) & Total power (erg/s)\\
T$_\mathrm{start}$ & T$_\mathrm{stop}$ & $R_0 \times \left[ {f_\mathrm{NT}}/{\epsilon_\mathrm{T}}\right]^{-3/2} $ & $\Gamma \times \left[\left(1+\sigma\right) f_\mathrm{NT}\right]^{1/4}$ & $R_\mathrm{ph}  \times \left[\left(1+\sigma\right) f_\mathrm{NT}\right]^{1/4}$ & $\dot{E}\times f_\mathrm{NT}$\\
\hline
\multicolumn{6}{c}{\textit{Time-integrated spectral analysis}}\\
\hline
$-0.016$ s  & $ 0.548$ s & $ 4.2\times 10^{       8}$ & $162.$ & $ 3.4\times 10^{      12}$ & $ 2.5\times 10^{      52}$ \\
\hline
\multicolumn{6}{c}{\textit{Time-dependent spectral analysis: 4 bins}}\\
\hline
$-0.018$ s &  $ 0.058$ s & $ 1.1\times 10^{       9}$ & $172.$ & $ 7.3\times 10^{      12}$ & $ 6.3\times 10^{      52}$ \\
$ 0.058$ s &  $ 0.100$ s & $ 2.9\times 10^{       9}$ & $176.$ & $ 1.0\times 10^{      13}$ & $ 9.5\times 10^{      52}$ \\
$ 0.100$ s &  $ 0.174$ s & $ 7.5\times 10^{       8}$ & $138.$ & $ 8.8\times 10^{      12}$ & $ 4.0\times 10^{      52}$ \\
$ 0.174$ s &  $ 0.600$ s & $ 2.2\times 10^{       8}$ & $ 85.$ & $ 5.0\times 10^{      12}$ & $ 5.3\times 10^{      51}$ \\
\hline
\multicolumn{6}{c}{\textit{Time-dependent spectral analysis: 12 bins}}\\
\hline
$-0.018$ s &  $-0.006$ s & $ 4.4\times 10^{       7}$ & $330.$ & $ 7.0\times 10^{      11}$ & $ 4.3\times 10^{      52}$ \\
$-0.006$ s &  $ 0.006$ s & $ 9.1\times 10^{       8}$ & $258.$ & $ 2.2\times 10^{      12}$ & $ 6.6\times 10^{      52}$ \\
$ 0.006$ s &  $ 0.014$ s & $ 3.7\times 10^{       9}$ & $218.$ & $ 3.5\times 10^{      12}$ & $ 6.1\times 10^{      52}$ \\
$ 0.014$ s &  $ 0.022$ s & $ 3.2\times 10^{       9}$ & $202.$ & $ 3.9\times 10^{      12}$ & $ 5.5\times 10^{      52}$ \\
$ 0.022$ s &  $ 0.044$ s & $ 1.2\times 10^{      10}$ & $161.$ & $ 5.8\times 10^{      12}$ & $ 4.1\times 10^{      52}$ \\
$ 0.044$ s &  $ 0.054$ s & $ 1.6\times 10^{      10}$ & $136.$ & $ 8.3\times 10^{      12}$ & $ 3.6\times 10^{      52}$ \\
$ 0.054$ s &  $ 0.080$ s & $ 7.2\times 10^{       9}$ & $157.$ & $ 8.5\times 10^{      12}$ & $ 5.6\times 10^{      52}$ \\
$ 0.080$ s &  $ 0.094$ s & $ 3.5\times 10^{       9}$ & $167.$ & $ 1.1\times 10^{      13}$ & $ 8.8\times 10^{      52}$ \\
$ 0.094$ s &  $ 0.132$ s & $ 2.6\times 10^{       9}$ & $145.$ & $ 1.1\times 10^{      13}$ & $ 5.8\times 10^{      52}$ \\
$ 0.132$ s &  $ 0.178$ s & $ 2.0\times 10^{       9}$ & $ 98.$ & $ 1.3\times 10^{      13}$ & $ 2.0\times 10^{      52}$ \\
$ 0.178$ s &  $ 0.250$ s & $ 2.0\times 10^{       8}$ & $ 96.$ & $ 6.6\times 10^{      12}$ & $ 1.0\times 10^{      52}$ \\
$ 0.250$ s &  $ 0.700$ s & $ 2.7\times 10^{       8}$ & $ 76.$ & $ 3.7\times 10^{      12}$ & $ 2.8\times 10^{      51}$ \\
\end{tabular}
\end{center}
\end{table*}

Unfortunately, the redshift of \grb is not known. We assume $z=0.5$, which is a typical value for a short GRB.
The tendency is that a lower redshift will reduce the constraints derived below but we checked that our conclusions are unchanged for $z=0.1$ or $z=1$.
Using the results of the spectral analysis (B+BB) presented in sections \ref{sec:Coarse time resolved} and \ref{sec:Fine time resolved}, we measure $F_\mathrm{BB}$, $\phi$ and $\mathcal{R}$ and, using equations (\ref{eq:R0}--\ref{eq:Rph}), we obtain $R_0$, $\Gamma$ and $R_\mathrm{ph}$ listed in Table~\ref{tab:model}.
The efficiency of the photospheric emission $f_\mathrm{T}=L_\mathrm{ph}/\dot{E}$, where $L_\mathrm{ph}$ is the luminosity of the photosphere and $\dot{E}$ the injected energy flux in the relativistic ejecta, can be compared to the efficiency of the dissipative mechanism responsible for the non-thermal emission $f_\mathrm{NT}=L_\mathrm{NT}/\dot{E}$, where $L_\mathrm{NT}$ is the non-thermal luminosity. Using the formulae above, we find
\begin{equation}
\frac{f_\mathrm{T}}{f_\mathrm{NT}}\simeq \frac{\phi}{1-\phi}\sim0.01-0.1\, .
\end{equation}
Since the ratio $\phi/(1-\phi)$ is within the range $0.01$--$0.1$, the non thermal dissipative process is  
dominant  in the case of \grbnos.\\

For a pure fireball, the acceleration is thermal, so that $\epsilon_\mathrm{T}=1$ and $\sigma =0$. Then, if the dissipation mechanism responsible for the non-thermal component has a very high efficiency ($f_\mathrm{NT}\simeq 1$), the values listed in Table~\ref{tab:model} give direct estimates of $R_0$, $\Gamma$ and $R_\mathrm{ph}$. They are in good agreement with typical values expected for GRBs, except for the initial radius $R_0$ which seems too large.

If the jet opening angle is $\theta_\mathrm{j}$ and the size of the initial region, where the outflow is launched, is $\ell$, then $R_0 \simeq \ell / \theta_\mathrm{j}$. If short GRBs are associated with the merger of a NS+NS binary system, the expected central engine is an accreting highly rotating black hole with a mass of $2-3\, M_\mathrm{\odot}$. 
Then, the radius of the innermost stable orbit, which can give an estimate of $\ell$, is of the order of $8-13$ km (we assume $a\simeq 0.8$ for the black hole spin). This leads to $R_0 \la 700$ km if $\theta_\mathrm{j} \ga 1^\circ$. The value of $R_0$ obtained in  Table~\ref{tab:model} is a factor $10-10^3$ larger. Even if higher black hole masses can be expected for NS+BH mergers, an initial size $R_0$ above 1000 km seems quite unrealistic for most theoretical models of short GRB central engines.
However, the assumption $f_\mathrm{NT}\simeq 1$ is quite extreme.
The main mechanism to dissipate energy above the photosphere for an unmagnetized outflow is the extraction of kinetic energy by internal shocks. This process is known to have a low efficiency $f_\mathrm{NT}\la 0.1$ \citep{Daigne:1998}. As $R_0 \propto \left(f_\mathrm{NT}/\epsilon_\mathrm{T}\right)^{3/2}$, a realistic efficiency $f_\mathrm{NT}\simeq 0.05$ also leads  to a lower value of $R_0$ (by a factor $f_\mathrm{NT}^{3/2}\simeq 10^{-2}$) which is in much better agreement with theoretical expectations for short GRBs. The impact on the other quantities is weak (a factor $f_\mathrm{NT}^{-1/4}\simeq 2$). The Lorentz factor is found in the range $160$--$660$, in good agreement again with the theoretical expectations, especially the constraints obtained from the $\gamma\gamma$ opacity argument (for recent discussions in light of Fermi-LAT results, see \citealt{Racusin:2011,Zhao:2011,Hascoet:2012}).

It is interesting to compare the  value of the photospheric radius, which is found to be in the range $1.5\times 10^{12}$--$2.7\times 10^{13}\, \mathrm{cm}$, with the radius where the internal shock phase starts, which is given by
\begin{equation}
R_\mathrm{is}\simeq 2\Gamma^2 c t_\mathrm{var} \simeq 1.5\times 10^{13}\, \mathrm{cm} \left(\frac{\Gamma}{160}\right)^2 \left(\frac{t_\mathrm{var}}{0.01\, \mathrm{s}}\right)\, ,
\end{equation}
where $t_\mathrm{var}$ is the minimum variability timescale in the initial distribution of the Lorentz factor of the outflow. Assuming $f_\mathrm{NT}=0.05$, it is found that for short timescale variability $t_\mathrm{var}=0.01$ s, the ratio $R_\mathrm{is}/R_\mathrm{ph}$ is in the range $2$--$200$ except for bin \# 10 where $R_\mathrm{is}/R_\mathrm{ph}\simeq 0.97$.
As most of the non-thermal flux seems to be associated with variability on timescales larger than 0.01 s, the observed values are therefore fully compatible with the scenario where the non-thermal emission is due to internal shocks above the photosphere.  \\

Most GRBs, however, are not compatible with the simplest scenario where $\epsilon_\mathrm{T}=1$, as is the case for \grbnos.
At least in long bursts, it seems that no thermal component can usually be detected. In GRB 100724B, where an additional BB component was found in the spectrum in a similar way as in \grbnos, this additional component was weaker \citep{Guiriec:2011a}. To reconcile most bursts with the standard GRB scenario, it is necessary to assume that $\epsilon_\mathrm{T}\la 0.1$ (see e.g., \citealt{Daigne:2002,Zhang:2009,Guiriec:2011a,Hascoet:2012a,Hascoet:2013}). In this case, most of the energy initially released by the source is in magnetic form rather than thermal and the jet acceleration can be magnetically driven \citep[see e.g. ][]{Begelman:1994,Daigne:2002b,Vlahakis:2003,Komissarov:2009,Tchekhovskoy:2010,Komissarov:2010,Granot:2011}. This leads to two possible scenarios:
\begin{itemize}
\item The magnetization of the outflow at large distances from  the central engine is still large ($\sigma\ga1$). Then internal shocks cannot form and the best candidate for the mechanism responsible for the non thermal emission is magnetic reconnection \citep[see e.g.][]{Spruit:2001,Lyutikov:2003,Zhang:2011}.
\item Most of the initial magnetic energy is converted into kinetic energy (efficient magnetic acceleration) and the magnetization at large distances is low ($\sigma\la 0.1$). Then, as in the standard fireball model, internal shocks are  the best candidate for the non-thermal mechanism.
\end{itemize}
Does \grb favor one of these two possibilities? A possible indication is given by the condition that the initial radius $R_0$ is expected to be small for short GRBs. As illustrated in Figure~\ref{fig:constraints}, the condition $R_0 \la 100\, \mathrm{km}$ leads directly, from the values listed in Table~\ref{tab:model}, to 
low values of the non thermal efficiency,  $f_\mathrm{NT}\la 0.008$--$0.3$ if $\epsilon_\mathrm{T}\simeq 1$, and even lower values of $f_\mathrm{NT}$ if $\epsilon_\mathrm{T}$ is much lower. Such efficiencies are in the range expected for internal shocks but are rather lower than the usually considered range for magnetic reconnection. A high efficiency $f_\mathrm{NT}=0.5$ for magnetic reconnection would lead to $R_0\ga 1000$ km for most bins, and even $R_0\ga 10^\mathrm{4}$ km for some bins. Such a large value of the initial radius seems quite challenging for 
most theoretical candidates
of the central engine of short GRBs, especially NS+NS mergers.
\begin{figure}[t]
\begin{center}
\includegraphics[totalheight=0.36\textheight, clip,viewport=10 20 568 562]{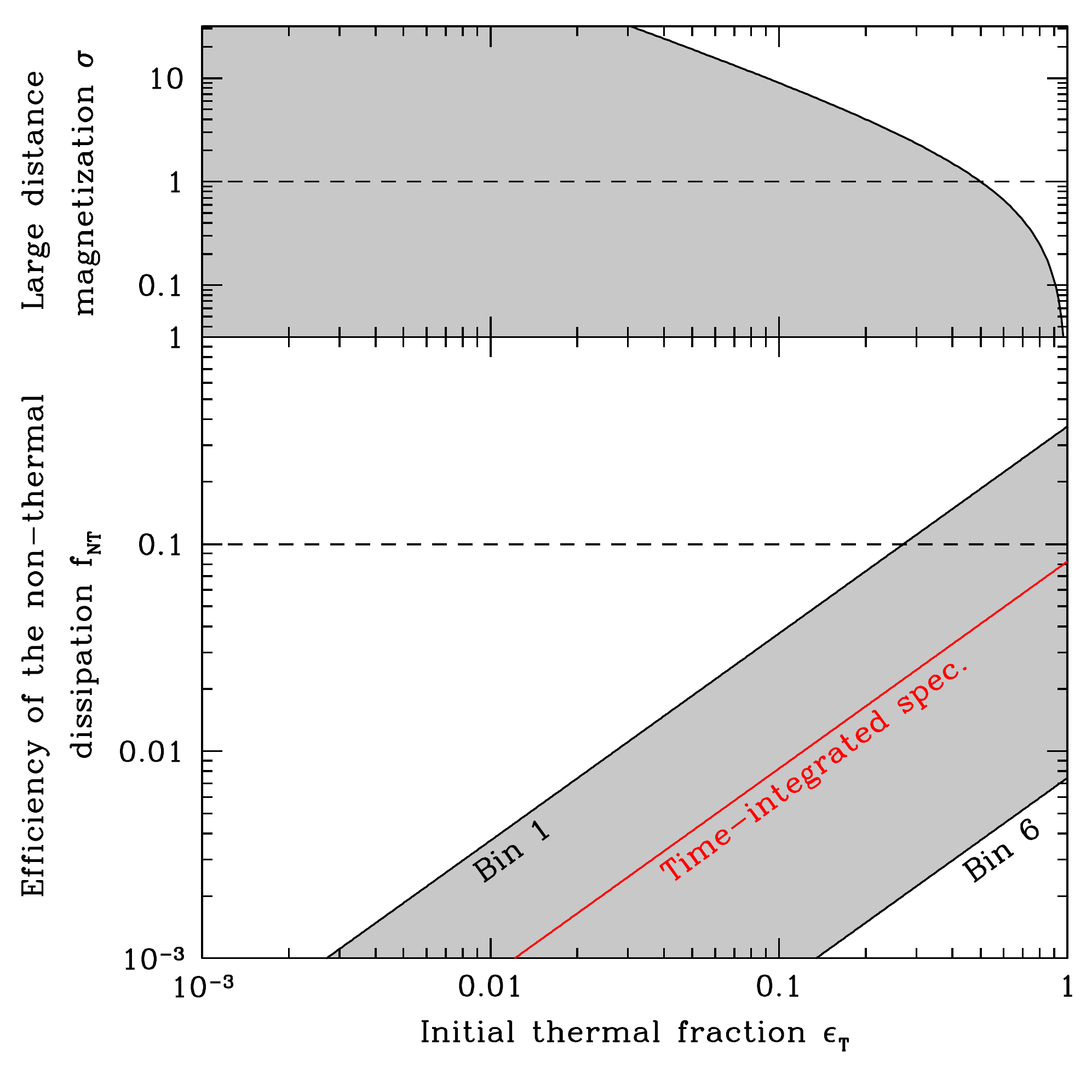}
\caption{{\textbf{Constraints on the efficiency of the non-thermal dissipation process.} \textit{Bottom:} the maximum value of $f_\mathrm{NT}$ given by the condition $R_0 \le R_{0,\mathrm{max}}=100\, \mathrm{km}$ is plotted as a function of the initial thermal fraction $\epsilon_\mathrm{T}$ during the energy release by the central engine, using the results listed in Table~\ref{tab:model} (red: time-integrated analysis; shaded area: time-resolved analysis ; the two limits correspond to bin 1 ($-0.018\to-0.006$ s) and bin 6 ($0.044$--$0.054$ s)).  If the value of $R_\mathrm{0,max}$ is increased, all the lines are shifted towards higher values of $f_\mathrm{NT}$. A dashed horizontal line indicates the location of $f_\mathrm{NT}\simeq 0.1$. \textit{Top:}  the magnetization $\sigma$ at the end of the acceleration of the outflow is plotted as a function of $\epsilon_\mathrm{T}$ (thick solid line), assuming no conversion of magnetic energy into kinetic energy, or of thermal energy into magnetic energy, and assuming a perfect conversion of thermal energy into kinetic energy (i.e., a passive magnetic field  frozen in the expanding fireball, with a final magnetization $\sigma_\mathrm{passive}=(1-\epsilon_\mathrm{T})/\epsilon_\mathrm{T}$). The shaded area with $\sigma < \sigma_\mathrm{passive}$ corresponds to
efficient magnetic acceleration models.
The non-shaded area corresponds to scenarios where the magnetic field does not contribute to the acceleration and where, in addition, part of the thermal energy is converted into magnetic energy. Such scenarios are usually not considered in GRB models. A dashed horizontal line indicates the limit $\sigma=1$ between magnetically dominated outflows ($\sigma>1$, where the best candidate for internal dissipation is magnetic reconnection, and matter dominated outflows ($\sigma<1$), where internal shocks are the best candidate for dissipation. }}
\label{fig:constraints}
\end{center}
\end{figure}
Therefore, the first scenario (large magnetization $\sigma \ga 1$ at the end of the acceleration of the flow, leading to magnetic dissipation as the dominant process) seems 
the least probable for \grbnos:
except for especially inefficient acceleration mechanisms, where most of the initial thermal energy is converted into magnetic rather than kinetic energy,
this scenario would require a low value of $\epsilon_\mathrm{T}$ to get a large final magnetization (see Figure~\ref{fig:constraints}) and then imply a very inefficient dissipative process. Even in the second scenario (efficient magnetic acceleration leading to $\sigma\ll1$ at large distance), a low initial thermal fraction $\epsilon_\mathrm{T}\la 0.1$ in \grb implies a really low efficiency for internal shocks, typically $f_\mathrm{NT}\la 0.01$ (see Figure~\ref{fig:constraints}). \grb would then represent a case where $\epsilon_\mathrm{T}$ is a little larger than in most GRBs. For instance, $\epsilon_\mathrm{T}\simeq 0.5$ implies $f_\mathrm{NT}\la 0.05$. The best candidate for the internal dissipation above the photosphere is therefore internal shocks, as the magnetization at large distance is expected to be low in this scenario.
\fdaigne{This result is obtained under the very general assumptions listed above, which were used by \citet{Hascoet:2013} to derive equations (\ref{eq:R0}--\ref{eq:Rph}). The assumption that the acceleration is completed below the photosphere, i.e. $R_\mathrm{sat}<R_\mathrm{ph}$ may not be valid for very slow acceleration mechanism for the outflow. By considering an acceleration law $\Gamma(R)\propto R^\alpha$, we have checked the validity of this assumption  in the case of GRB 120323A for the different scenarios discussed above regarding $\epsilon_\mathrm{T}$, $\sigma$, and $f_\mathrm{NT}$. We find that the saturation radius is always smaller than the photospheric radius, as long as $\alpha\ga 0.3-0.4$, which includes thermal acceleration ($\alpha=1$) and several classes of magnetic acceleration \citep[see e.g.][]{Tchekhovskoy:2010,Granot:2011} but marginally eliminates the slowest magnetic acceleration mechanism with $\alpha=1/3$. When $R_\mathrm{sat}>R_\mathrm{ph}$, modified equations (\ref{eq:R0}--\ref{eq:Rph}) can be derived \citep[see appendix in][]{Hascoet:2013}. It is found that only equations~(\ref{eq:Gamma}-\ref{eq:Rph}) are modified, but that equation~(\ref{eq:R0}) for the initial radius $R_0$ is unchanged. Then, our conclusion that a low efficiency $f_\mathrm{NT}$ for the non-thermal emission process above the photosphere is required in the case of GRB 1220323A to avoid too large initial radii is robust, as it remains valid even for a slow acceleration law}\footnote{\fdaigne{In the case of a slow acceleration with $\alpha=1/3$, the value of the Lorentz factor $\Gamma$ and the photospheric radius $R_\mathrm{ph}$ should be corrected by a factor $(R_\mathrm{ph}/R_\mathrm{sat})^{1/12}$ and $(R_\mathrm{ph}/R_\mathrm{sat})^{-1/4}$ which are very close to unity, except if $R_\mathrm{ph}\ll R_\mathrm{sat}$, which is not expected for GRBs.}}\fdaigne{.} 

From this discussion, it appears the observations of \grb are compatible with the  simplest GRB scenario where the relativistic ejecta are thermally accelerated and the non-thermal emission is produced by internal shocks. It is also compatible with the scenario where the fireball is initially magnetized and where most of the magnetic energy is converted into kinetic energy (efficient magnetic 	acceleration, $\sigma\la 0.1$) so that the dominant dissipative process remains internal shocks. \grb would  represent a case where the initial magnetization is lower than in other GRBs ($\epsilon_\mathrm{T}\simeq 0.5$--$1$ rather than $0.1$ or lower). On the other hand, it is  difficult to reconcile the data with a scenario where the outflow is magnetically dominated at large distance ($\sigma\ga 1$) and the non thermal emission is due to magnetic reconnection, unless this dissipative mechanism is much less efficient than usually considered.\\

The spectral evolution observed in \grb can now be discussed in the framework of the preferred scenario identified above.
As shown in Section~\ref{sec:Fine time resolved}, the spectral analysis based on the B+BB fits leads to a dramatic change of the low-energy slope $\alpha$ of  the Band component, compared to the Band-only analysis. Instead of very steep values, often well above $\alpha=0$, the low-energy slope $\alpha$ is found for the B+BB analysis to remain for the whole burst in the range $-1.9\to-1.1$. This is well below the synchrotron slow cooling limit ($\alpha=-2/3$), and well inside the predicted range  for the fast cooling regime. The latter regime is expected during most of the prompt phase, i.e., $-1.5\le \alpha \la -1$, where $\alpha=-1.5$ corresponds to pure fast cooling synchrotron radiation \citep{Sari:1998} and where steeper values are obtained if low-energy photons experience inverse Compton scatterings in the Klein Nishina regime \citep{Derishev:2001,Bosnjak:2009,Nakar:2009,Daigne:2011}. Therefore, the spectral analysis of \grb based on the B+BB fits agrees well with the scenario where the non-thermal prompt soft gamma-ray emission is dominated by fast cooling synchrotron radiation of shock-accelerated electrons in internal shocks. A similar result was already found by \citet{Guiriec:2011a} in GRB 100724B. Finding the same behavior in \grb offers a promising possibility to solve, at least partially, the so-called synchrotron death line problem \citep{Preece:1998,Ghisellini:2000}.

It should also be noted that in the B+BB interpretation, the spectral evolution observed for the Band component follows the E$_\mathrm{peak}$-Luminosity correlation observed in most GRBs (see Section~\ref{section:Flux-Epeak correlation}). This has been studied by several authors in the context of the internal shock model: the spectral evolution is governed by the dynamical timescale associated with the propagation of shock waves which \fdaigne{reproduces} successfully the E$_\mathrm{peak}$-Luminosity correlation \citep{Daigne:1998,Daigne:2003,Bosnjak:2009,Asano:2011,Asano:2012,Daigne:2011,Bosnjak:2012submit}. On the other hand, no similar strong correlation is found in \grb between the flux and the temperature for the BB component. As both quantities have a similar dependency with the Lorentz factor of the outflow, but a different dependency on $\dot{E}$, this may indicate that not only the Lorentz factor, but also other parameters such as the total injected power, are variable during the relativistic ejection by the central engine. \fdaigne{In addition, we found in Section~\ref{section:Flux-Epeak correlation} that the E$_\mathrm{peak}$-Luminosity correlation was very similar for 
the short GRB 120323A studied here and
a sample made of a few long GRBs detected by \textit{Fermi}. This points out towards a universal mechanism during the prompt phase, both for short and long bursts. Our detailed study of GRB 120323A suggests that this mechanism occurs above the photosphere. }

\begin{figure}
\begin{center}
\includegraphics[totalheight=0.35\textheight, clip,viewport=7 15 553 523]{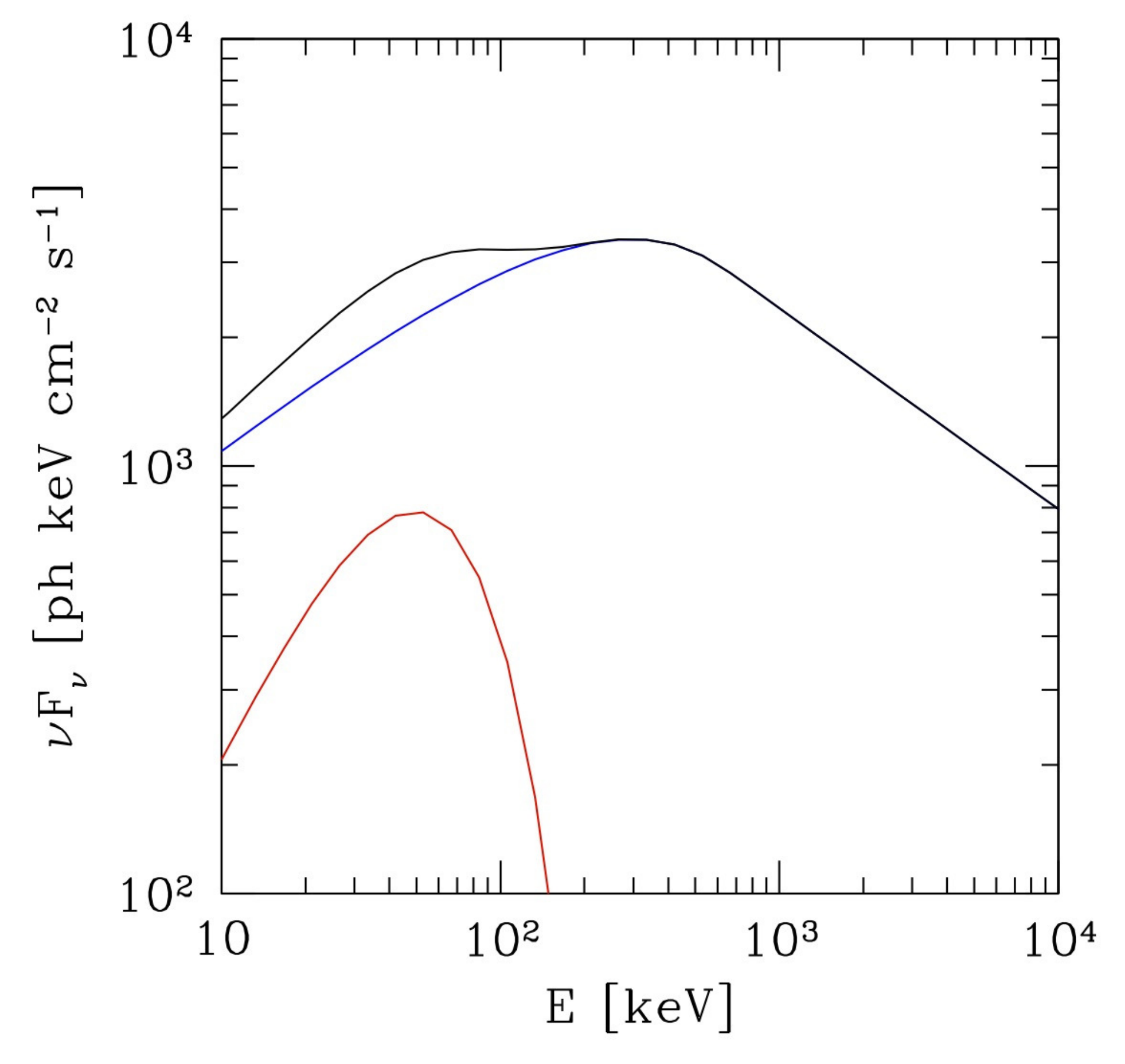}
\caption{{\textbf{The internal shock + weak thermal photospheric emission scenario.} This scenario is illustrated by a synthetic burst showing the same spectral shape as observed in the B+BB analysis of \grbnos. A redshift $z=0.5$ is assumed. The outflow is ejected at $R_0=300\, \mathrm{km}$ with an initial thermal fraction $\epsilon_\mathrm{T}=0.5$. The duration of the ejection is $0.2/(1+z)$ s. The acceleration is efficient so that $\sigma\ll 1$ at large distance. The initial Lorentz factor varies between 90 and 450 on a typical timescale $0.1/(1+z)$ to produce two main episodes of emission. The isotropic equivalent kinetic power is assumed to be constant and equals $3.5\times 10^{54}\, \mathrm{erg.s^{-1}}$. The non-thermal emission from internal shocks is computed assuming that a fraction $\epsilon_\mathrm{e}=1/3$ of the dissipated energy is injected in relativistic electrons, so that the efficiency is this case is $f_\mathrm{NT}=0.023$. The observed time-integrated spectrum (over a duration 0.2 s) is plotted in black. The fast cooling synchrotron component due to internal shocks is plotted in blue and the thermal photospheric emission in red. As explained in the text, the spectrum of the latter is not an exact Planck function. The low-energy slope of the non-thermal component does not fully agree with the observations of \grb as we have assumed pure fast cooling synchrotron ($\alpha=-1.5$) for simplicity (see text).}}
\label{fig:grb}
\end{center}
\end{figure}

Finally, it should be noted that for the low efficiency $f_\mathrm{NT}\simeq 0.01-0.05$ implied by the scenario discussed here, the isotropic equivalent total power of the outflow should be of the order of $\dot{E}\simeq 6\times 10^{52}\to 9\times 10^{54}\, \mathrm{erg/s}$ to reproduce the observed flux (from the values listed in Table~\ref{tab:model}), leading to a total isotropic equivalent energy $E_\mathrm{iso}=\left(1.5-7.5\right)\times 10^{53}\, \mathrm{erg}$, which would favor a small opening angle $\theta_\mathrm{j}\simeq 1^\mathrm{\circ}$ for the true energy $E\simeq \left(\theta_\mathrm{j}^2/2\right) E_\mathrm{iso}$ to be consistent with the energy budget discussed for instance for the popular NS+NS merger scenario  (see e.g., \citealt{Aloy:2005}). Such a value of $\theta_\mathrm{j}$ is smaller than what is usually considered for short GRBs \citep{Ruffert:1999,Rosswog:2002,Rezzolla:2011}.
Obviously, this result is affected by our choice of source redshift : the estimate of $E_\mathrm{iso}$ is reduced by a factor $\sim 30$ for $z=0.1$\fdaigne{, leading to a more acceptable constraint $\theta_\mathrm{j}\la 5^\mathrm{\circ}$. Note that the same reduction of the redshift from $0.5$ to $0.1$ only affects the value of $R_0$ by a factor $\sim 3$, so that the discussion above about the non-thermal efficiency $f_\mathrm{NT}$ is unchanged.}\\

To illustrate the scenario where the prompt emission of \grb is associated with photospheric and internal shock emission in a (initially magnetized) relativistic outflow which is matter dominated at large distance ($\sigma\ll 1$), we have simulated a synthetic burst %with 
similar to \grb during the first $\sim 0.2$ s. We assume $R_0=300\, \mathrm{km}$ and $\epsilon_\mathrm{T}=0.5$. Other parameters are found in the caption of Figure~\ref{fig:grb}.
The photospheric emission is computed using the method described in \citet{Hascoet:2013} and takes into account the modification of the spectral shape due to the complex geometry of the photosphere. The final photospheric spectrum is, therefore, more complicated than a BB but the value of the temperature and the flux are reproduced. 
The internal shock emission is computed using the multi-shell model developed by \citet{Daigne:1998} and assuming for simplicity a pure fast-cooling synchrotron spectrum (i.e., with $\alpha=-1.5$), without any correction for inverse Compton scattering in the Klein-Nishina regime. Therefore, the low-energy slope observed in the Band component of \grb is not perfectly reproduced.
The resulting spectrum (total and separated components) is plotted in Figure~\ref{fig:grb}. The overall  shape of the spectrum of \grb is reproduced. This is encouraging but this scenario should clearly be investigated in more details, especially to test if the observed spectral evolution of both components can also be reproduced.\\

The discussion above has been limited to GRB models which assume that the non thermal emission is due to internal dissipation in the relativistic outflow above the photosphere. There is however another theoretical possibility, where the whole spectrum would be of photospheric origin. 
 The spectrum originating from the photosphere can be significantly modified if there is  additional dissipation close to the photosphere \citep{Rees:2005,Peer:2005}. Such dissipation processes could for instance be due to internal, or oblique, shocks, magnetic reconnection \citep{Giannios:2008}, or collisional mechanisms \citep{Beloborodov:2010}.  If the dissipation produces a population of energetic leptons and a strong magnetic field, the original Planck spectrum can be modified  by Comptonization, causing the spectrum to extend to higher energies, and by additional low-energy synchrotron photons, causing the spectrum to extend to lower energies. Depending on the conditions at the dissipation site, foremost the optical depth, the spectrum partly  thermalizes again. These processes are capable of producing a broad spectrum much resembling a Band function \citep{Peer:2006}, but possibly showing steep low-energy slopes. The conditions that need to be met are that the energy given to the electrons should be comparable to the energy in thermal photons and that a strong magnetic field exists. The details of the spectrum formation can be found in \citet{Peer:2006,Vurm:2011} and  \citet{Ryde:2011}. This scenario seems a natural candidate to explain the results of the Band-only spectral analysis. As for the previous discussion, the capacity of this scenario to reproduce the observed spectral evolution needs to be tested in details. In particular, one puzzling fact must be investigated furthermore: the transition from the quasi-thermal (\fdaigne{$\alpha>0$}) to the non-thermal (\fdaigne{$\alpha<-1.5$}) spectrum at T$_{0}$+0.094 in \grbnos. This may be a signature of a threshold for the dissipative process to occur at the photosphere such as suggested for instance in the collisional model by~\citet{Belobodorov:2011b}. \fdaigne{
 On the other hand, the detailed data analysis presented in Section~\ref{Model description} favors the two component scenario (Band+BB). This is further strengthened by the results of Section~\ref{section:Flux-Epeak correlation}, which show that GRB 120323A recovers a standard hardness-intensity correlation in this two component scenario. 
The independent behavior of the two spectral components despite an assumed common origin is difficult to understand in dissipative photospheric models. 
Therefore, these observations strongly suggest that most of the emission in GRB 120323A  is produced in the optically thin regime, and that the photospheric emission is only sub-dominant.
 Recently, \citet{Zhang:2012} have shown that the dissipative photospheric model was also disfavored in GRB 110721A, where the peak energy reaches 15 MeV. This similar conclusion in two different GRBs, combined with the evidence for a universal hardness-intensity behavior that points out to a unique mechanism for the GRB prompt emission, leads to an 
 emerging consistent  picture where most, if not all, GRBs would be produced by non-thermal dissipation above the photosphere.}

% \citet{Zhang:2012} found in GRB 110721A that dissipative photospheric models were clearly disfavored, due to  a high peak energy in the earliest epoch ($E_\mathrm{p}\sim 15$ MeV), and therefore that the emission was mostly produced in the optically thin regime above the photosphere. In the present case of GRB 120323A, our spectral analysis suggests that two components (B+BB) are present in this burst, with different temporal evolutions. As discussed in this section, this also clearly leads to prefer the scenario where most of the emission is produced in the optically thin regime, and where the photospheric emission is only sub-dominant.}

\newpage
\section{Conclusions}
\label{section:conclusion}

We have presented here observational results and their associated theoretical interpretation of \grbnos, the most intense {\it short} GRB observed thus far with GBM. This GRB is especially bright below 150 keV. We associated the extreme intensity of the soft energy photons with the presence of a photospheric component. It is arguable whether the intensity of this event is entirely due to the existence of this component, detected for the first time due to the low energy range of GBM (starting at 8 keV) and possibly the vicinity of the source. \grb is either an unusually soft and intense short GRB or a regular short GRB but exhibiting an intense additional thermal-like component at low energy. {\it Regardless of the origin of the component, \grb is a rare event.} 

In summary, our observational analyses of the prompt $\gamma-$ray emission of \grb have led to the following conclusions:

\begin{itemize}
\item The presence of spectral evolution during a burst can create artificial features in the spectral shape. One should, therefore, be very cautious when interpreting time-integrated spectra. Time-resolved spectroscopy is required to remove these effects.

\item We determined that the spectra of \grb (time-integrated and time-resolved) are better described with a double curvature spectral shape than with the single curvature of the Band function. The spectrum can thus be interpreted as consisting of two components, one being of thermal origin, compatible with a BB or similar shapes with steep slopes (i.e., $>$0) and the other produced by a non-thermal radiation mechanism. The former component is energetically subdominant compared to the non-thermal one.

\item The simultaneous fit of a thermal and a non-thermal component to the data dramatically changes the shape of the spectra of \grbnos. Using a single Band function we find that the time evolution of all spectral parameters exhibits a pronounced discontinuity, whereas all parameters evolve very smoothly in the two components scenario. In the latter scenario, the thermal component, whose shape is compatible with the expected shape of the photospheric emission of a relativistic jet, is most intense at the beginning of the prompt emission with a constant cooling trend thereafter, which closely follows its intensity decline. The parameters of the non-thermal component are compatible with fast cooling synchrotron emission.

\item Intriguingly, no correlation is found between the E$_\mathrm{peak,i}^\mathrm{rest}$ and the luminosity of the burst, L$_\mathrm{i}$, when using a Band-only fit, while a strong correlation is obtained between L$_\mathrm{i}$ and the E$_\mathrm{peak,i}^\mathrm{rest}$ of the non- thermal component, when both thermal and non-thermal emission are fit simultaneously. In the latter case, the \grb E$_\mathrm{peak,i}^\mathrm{rest}$-L$_\mathrm{i}$ relation is perfectly consistent with those reported for a larger sample of GRBs. This result reinforces the two component scenario and supports the physical origin of this relation as well as the possibility to use it as a discriminator for the prompt emission models. This relation could also eventually be used as a possible redshift estimator for cosmology. From our (limited) sample, we estimate

$\mathrm{L_i^{Band}}=\mathrm{(1.59\pm0.84).10^{50}~(E_{peak,i}^{rest})^{1.33\pm0.07} erg.s^{-1}}$.

However, a more detailed analysis on a larger sample of GRBs is required to estimate the dispersion of this relation.

\end{itemize}

Our theoretical interpretation leads to the following conclusions:

\begin{itemize}

\item The single component spectral analysis of \grb clearly favors a photospheric origin, due to the steep low-energy slopes that are not compatible with the synchrotron radiation from shock-accelerated electrons. Additional dissipative processes at the photosphere are however necessary to reproduced the shape of the observed spectrum.   It remains to be tested if this dissipative photospheric emission scenario can reproduce the observed spectral evolution and especially the transition from a quasi-thermal (\fdaigne{$\alpha>0$}) to a non-thermal (\fdaigne{$\alpha<-1.5$}) spectrum at T$_{0}$+0.094s. \fdaigne{However, as listed above, there are several arguments to rather favor the two spectral components analysis. In this case, the dissipative photospheric model is disfavored, due to the independent behavior of the two spectral components.}

\item 
\fdaigne{In the two component analysis, the found peak-energy-luminosity correlation favors a unique mechanism for the prompt emission of short and long GRBs. In addition, the spectral analysis of GRB 120323A clearly favors models where this mechanism is non thermal emission associated to a dissipative process above the photosphere. The detailed analysis shows that}
%The two component spectral analysis of 
\grb is compatible with the standard fireball scenario where the thermal component is associated with the photospheric emission of a relativistic jet that has been thermally accelerated, and the non-thermal emission is due to synchrotron radiation from accelerated electrons in internal shocks above the photosphere. It is also compatible with an alternative scenario suggested by other bursts, such as GRB~$100724$B, where the initial magnetization of the ejecta is large and where most of the magnetic energy is converted into kinetic energy below the photosphere (efficient magnetic acceleration). 
\grb would however correspond to a case where the initial magnetization is lower and the thermal energy content larger than in previous cases, such as GRB~$100724$B.  
Indeed, our analysis shows that a large initial magnetization would lead to a very low efficiency of the non thermal emission and therefore to an energy crisis for this burst.
On the other hand, it is difficult to reconcile GRB~$100724$B with a scenario where the outflow is still highly magnetized at large radius and where the non-thermal emission is due to magnetic reconnection above the photosphere. 
This would lead to an injection radius where the outflow is launched much too large for most models of the central engine of short GRBs.

\item The physical interpretation of the two-component spectral analysis suggests that the same processes (photospheric emission + synchrotron radiation from electrons accelerated in internal shocks) may be at work in both short and long GRBs, but that the composition of the relativistic jet may slightly differ in the two classes, due to different progenitors. \grb data suggest an initial magnetization which is lower in short GRBs. The detection of a two component spectrum in other short GRBs would be necessary to confirm this result, which can shed light on the nature of the central engine and the jet acceleration mechanism in the different classes of GRBs.

\end{itemize}

\section{Acknowledgments}

SG was supported by the NASA Postdoctoral Program (NPP) at the NASA/Goddard Space Flight Center, administered by Oak Ridge Associated Universities through a contract with NASA. SG acknowledges financial support through the Cycle-4 NASA Fermi Guest Investigator program. SF acknowledges the support of the Irish Research Council for Science, Engineering and Technology, cofunded by Marie Curie Actions under FP7. We thank Valentin Pal'shin for his private communication on the propagation time between the WIND and Fermi spacecrafts as well as Adams Goldstein, Michael Briggs and Valerie Connaughton for their useful comments which helped to improve the quality of the manuscript. Finally, we thank the referee and the editor for their useful comments, which increased the quality of the article.

\newpage

\begin{appendix}
\section{Simulations}
\label{section:Simulations}

We used Monte Carlo simulations to (i) validate our ability to reconstruct properly the spectra in our shortest time intervals where the statistical fluctuations could be critical, and (ii) determine the probability of one model being better than another.

\subsection{Technique}
\label{subsection:Technique}

For each simulated set (i.e., one time interval and one input model), we generated 30,000 synthetic spectra for each relevant detector, with each spectrum covering the same duration as the real source time interval. For each synthetic spectrum we sum in each energy channel the number of background counts estimated from the real data (by fitting a polynomial function to off-source time intervals and by extrapolating it during the source active period), and the number of counts expected from the theoretical input photon model (i.e., Band or C+BB with the parameters obtained when fitting the real data in this time interval) when folded through the detector response matrix used for fitting the real data. Poisson fluctuations are then applied in each energy channel to the sum of the signal and background counts. For each data set background is also simulated based on the real background fit.

During the simultaneous fit of the data from all the detectors, a new background is simulated for each synthetic spectrum and each detector by adding Poisson fluctuations in each energy channel to the previously estimated background. The rest of the fit process is the same as with the real data fits.

We performed all simulations using the same version of Rmfit as the one used to fit the real data.

\subsection{Model comparison}
\label{subsection:Model comparisons}

We computed the significance of the improvement of a C+BB fit over a Band-only fit in the time intervals T$_\mathrm{0}$+0.080\,s to T$_\mathrm{0}$+0.094\,s and T$_\mathrm{0}$+0.094\,s to T$_\mathrm{0}$+0.132\,s, where the additional BB component improves the Band-only fit by 25 and 51 units of Cstat, respectively. We defined a set of simulations using Band as the input model (i.e., null hypothesis) with parameter values being the central values obtained by fitting the real data. We then fitted the resulting synthetic spectra with both Band and C+BB.

The resulting distributions of the Band function parameters obtained from the fits of the synthetic spectra have symmetrical and peaked shapes for E$_\mathrm{peak}$ and $\beta$, while the distribution of their amplitudes is wider. In the cases of positive values of $\alpha$ in input, the distribution of $\alpha$ is less peaked, but still with positive values. Taking into account the errors in the parameters, the input parameters are adequately reconstructed within 2$\sigma$. This indicates our ability to reconstruct a Band function in this time interval, given the parameters of the Band function, the level of the background and the signal strength.

We then fitted the synthetic spectra with C+BB ; most of the 30,000 fits gave a higher Cstat value than those using Band. Although C+BB has 1 dof more than Band, C+BB is usually worst than Band alone with Cstat values larger than 100 units, besides a few cases were C+BB improves the Band-only fit by only a few units of Cstat. For instance, for the time interval T$_\mathrm{0}$+0.080\,s to T$_\mathrm{0}$+0.094\,s, the maximal value for $\Delta$Cstat computed between Band and C+BB is $\sim$3 units from the simulation. We observe a $\Delta$Cstat of 25 units in the real data set, which indicates that the probability that C+BB is better than Band due to statistical fluctuations in the real data is very likely to be much lower than 3.3$\times$10$^\mathrm{-5}$.

We performed a similar analysis chosing C+BB as the null hypothesis. When fitted with C+BB, the parameter distributions are narrow and symmetrical showing the very good quality of the reconstruction when using this model for the two time intervals. When fit with a Band function alone, the resulting parameter distributions are perfectly compatible with what is obtained when fitting the real data with a Band function. For instance, for time interval T$_\mathrm{0}$+0.080\,s to T$_\mathrm{0}$+0.094\,s, E$_\mathrm{peak}$ is reconstructed around 50 keV with a dispersion of a few keV, $\alpha$ has values around 0 and the distribution of $\beta$ values is compact around -2. In addition, the distribution of $\Delta$Cstat measured between Band and C+BB from the synthetic spectra is narrow and peaks around the observed value (i.e., 25), which was not the case when choosing Band as a null hypothesis. This indicates that if C+BB is the correct description for the spectral shape of the real data, then we expect from the simulation that a fit of the real data with Band-only would lead to the observed results. Conversely, if Band is the real spectral shape, then based on the simulations we would expect different values for the spectral parameters when fitting C+BB to the real data than those measured.

In both time intervals, C+BB is statistically significantly better than Band alone.

\subsection{Reliability of the reconstruction}
\label{subsection:Reliability of the reconstruction}

Using the same method as in Section~\ref{subsection:Model comparisons}, we verified our ability to reconstruct the observed spectra in two time intervals, T$_\mathrm{0}$+0.022\,s to T$_\mathrm{0}$+0.044\,s and T$_\mathrm{0}$+0.054\,s to T$_\mathrm{0}$+0.080\,s, where the models Band and C+BB or C and C+BB, respectively, give similar Cstat values, but for which the spectral shape is dramatically different. In both time intervals, whatever is the input model, the input parameters are adequately recovered. We then confirm our ability to reconstruct the spectrum in such fine time intervals. We also confirm that the models are indistinguishable, and that when we fit the synthetic spectra with the model which is not the input one, we recover the parameters as obtained from the real data. This confirms that the two models are both possible options to describe the spectra of \grbnos. From a statistical point of view, the model with the lowest number of dofs would be preferred. However, since the more complex model makes more physical sense and is significantly better in some time intervals (see Section~\ref{subsection:Model comparisons}), both options are considered here.
\end{appendix}

\newpage
\begin{table*}
\caption{\label{tab:GRB120323A_resolved_spectra_large}Coarse time-resolved spectral fits of \grb using the detectors n0, n1, n3 and b0 (see section~\ref{sec:Coarse time resolved}).}
\begin{center}
{\tiny
\begin{tabular}{|l|l|l|l|l|l|l|l|l|l|l|l|l|l|}
\hline
\multicolumn{2}{|c|}{Time} &\multicolumn{1}{|c|}{Models} &\multicolumn{6}{|c|}{Standard Model} & \multicolumn{3}{|c|}{Additional Model} & \multicolumn{1}{|c|}{Cstat/dof} \\
\hline
 \multicolumn{2}{|c|}{} & \multicolumn{1}{|c|}{} &\multicolumn{6}{|c|}{Band, Compt or 2BPL} & \multicolumn{3}{|c|}{ BB, Compt or Band}  &  \multicolumn{1}{|c|}{}\\
\hline
\multicolumn{1}{|c|}{T$_\mathrm{start}$} & \multicolumn{1}{|c|}{T$_\mathrm{stop}$} & \multicolumn{1}{|c|}{Parameters} & \multicolumn{1}{|c|}{E$_{\rm peak}$} & \multicolumn{1}{|c|}{$\alpha$} & \multicolumn{1}{|c|}{$\beta$}  & \multicolumn{1}{|c|}{E$_{\rm b}$} & \multicolumn{1}{|c|}{E$_{\rm f}$} & \multicolumn{1}{|c|}{index} & \multicolumn{1}{|c|}{$\alpha$} & \multicolumn{1}{|c|}{$\beta$} & \multicolumn{1}{|c|}{kT or E$_{\rm 0}$} & \\
\hline
 \multicolumn{13}{|c|}{ }   \\  
-0.018 & +0.058 & Band & 73 & +0.10 & -2.16 & -- & -- & -- & -- & -- & -- & 506/470 \\
            &               &     & $\pm5$ & $\pm0.15$ & $\pm0.03$ & -- & -- & -- & -- & -- & -- & -- \\
             &               &  2BPL & 43 & -0.51 & -2.03 & 810 & -- & -2.83 & -- & --  & -- & 514/468 \\
             &               &             & $\pm3$ & $\pm0.10$ & $\pm0.04$ & $^\mathrm{+167}_\mathrm{-320}$ & -- & $^\mathrm{+0.41}_\mathrm{-61}$ & -- & --  & -- & -- \\   
             &               &   B+Cut & 69 & +0.19 & -2.10 & 718 & 700 & fix & -- & -- & -- & 504/468 \\
             &               &               & $\pm4$ & $^\mathrm{+0.17}_\mathrm{-0.15}$ & $\pm0.04$ & $^\mathrm{+1000}_\mathrm{-300}$ & fix & -- & -- & -- & -- & -- \\         
             &               &   B+BB & 118 & -0.14 & -2.23 & -- & -- & -- & -- & -- & 10.50 & 495/468 \\
             &               &              & $\pm14$ & $^\mathrm{+0.34}_\mathrm{-0.23}$ & $\pm0.04$ & -- & -- & -- & -- & -- & $^\mathrm{+1.11}_\mathrm{-1.07}$ & -- \\
             &               &   B+C2 & 105 & -0.40 & -2.24 & -- & -- & -- &  +7.35 & -- & 4.11 & 492/467 \\
             &               &        & $\pm11$ & $\pm0.12$ & $\pm0.04$ & -- & -- & -- &  $^\mathrm{+3.5}_\mathrm{-4.5}$  & -- & $^\mathrm{+2.49}_\mathrm{-1.83}$ & -- \\
 \multicolumn{13}{|c|}{ }   \\ 
 +0.058 & +0.100 & Band  & 46 & +0.07 & -2.01 & -- & -- & -- & --  & -- & -- & 558/470 \\
             &               &       & $\pm3$ & $\pm0.21$ & $\pm0.02$ & -- & -- & -- & -- & --  & -- & -- \\
             &               &  2BPL & 32 & -0.87 & -1.94 & 1040 & -- & -3.94 & -- & -- & -- & 534/468 \\
             &               &             & $\pm2$ & $\pm0.09$ & $\pm0.03$ & $^\mathrm{+274}_\mathrm{-194}$ & -- & $^\mathrm{+0.85}_\mathrm{-1.91}$ & --  & -- & -- & -- \\
             &               &   B+Cut & 44 & +0.10 & -1.94 & 836 & 393 & -- & --  & -- & -- & 532/468 \\
             &               &                & $\pm2$ & $^\mathrm{+0.27}_\mathrm{-0.19}$ & $\pm0.03$ & $^\mathrm{+632}_\mathrm{-252}$ & $^\mathrm{+704}_\mathrm{-269}$ & --  & -- & -- & -- & -- \\
             &               &   C+BB  & 556 & -1.35 & -- & -- & -- & --  & -- & -- & 10.94 & 511/469 \\
             &               &       &  $^\mathrm{+79}_\mathrm{-62}$ & $\pm0.05$ & -- & -- & -- & -- & -- & -- & $^\mathrm{+0.40}_\mathrm{-0.37}$ & -- \\
             &               &   B+BB & 547 & -1.35 & $<-2.35$ & -- & -- & -- & -- & -- & 10.93 & 512/469 \\
             &               &       & $^\mathrm{+78}_\mathrm{-62}$ & $\pm0.05$ & -- & -- & -- & -- & -- & -- & $^\mathrm{+0.40}_\mathrm{-0.38}$ & -- \\
             &               &   C+C2 & 570 & -0.85 & -- & -- & -- & -- & 0.02 & -- & 22.62 & 506/468 \\
             &               &       &  $^\mathrm{+64}_\mathrm{-51}$ & $^\mathrm{+0.38}_\mathrm{-0.30}$ & -- & -- & -- & -- & $^\mathrm{+0.45}_\mathrm{-0.24}$ & -- & $^\mathrm{+4.72}_\mathrm{-4.96}$ & -- \\
             &               &   C+B2 & 563 & -0.85 & -- & -- & -- & -- & 0.05 & $<-10$ & 22.12 & 506/467 \\
             &               &       & $^\mathrm{+70}_\mathrm{-44}$ & $^\mathrm{+0.39}_\mathrm{-0.27}$ & -- & -- & -- & -- & $^\mathrm{+0.42}_\mathrm{-0.28}$ & -- & $^\mathrm{+5.24}_\mathrm{-4.50}$ & -- \\
\multicolumn{13}{|c|}{ }   \\ 
 +0.100 & +0.174 & CPL & 405 & -1.55 & -- & -- & -- & -- & -- & -- & -- & 457/471 \\
             &               &      & $^\mathrm{+57}_\mathrm{-45}$ & $\pm0.03$ & -- & -- & -- & -- & -- & --  & -- & -- \\
             &               & Band & 401 & -1.55 & $<-3.40$ & -- & -- & -- & -- & -- & -- & 458/470 \\
             &               &      & $^\mathrm{+57}_\mathrm{-44}$ & $\pm0.03$ & -- & -- & -- & -- & -- & --  & -- & -- \\
             &               &  2BPL & 19.59 & -1.26 & -1.67 & 505 & -- & -3.23 & -- &  --  & -- & 447/468 \\
             &               &              & $^\mathrm{+6.27}_\mathrm{-2.90}$ & $^\mathrm{+0.23}_\mathrm{-0.18}$ & $\pm0.02$ & $^\mathrm{+100}_\mathrm{-71}$ & -- & $^\mathrm{+0.37}_\mathrm{-0.70}$ & -- &  -- & -- & -- \\
             &               &   B+Cut &  38 & -0.87 & -1.68 & 270 & 400 & -- & -- & -- & -- & 447/474 \\
             &               &                & -- & -- & -- & -- & -- & -- & -- & -- & -- & -- \\
             &               &   C+BB & 395 & -1.37 & -- & -- & -- & -- & -- & -- & 6.67 & 440/469 \\
             &               &       &  $^\mathrm{+48}_\mathrm{-38}$ & $\pm0.07$ & -- & -- & -- & -- & -- & -- & $^\mathrm{+0.59}_\mathrm{-0.51}$ & -- \\
             &               &   B+BB &  391 & -1.37 & $<-3.5$ & -- & -- & --  & -- & -- & 6.65 & 441/469 \\
             &               &        & $^\mathrm{+47}_\mathrm{-38}$ & $\pm0.07$ & -- & -- & -- & -- & -- & -- & $^\mathrm{+0.58}_\mathrm{-0.51}$ & -- \\
             &               &   C+C2 & 373 & -1.18 & -- & -- & -- & -- & -0.07 & -- & 13.06 & 439/468 \\
             &               &       &   $^\mathrm{+41}_\mathrm{-31}$ & $^\mathrm{+0.20}_\mathrm{-0.16}$ & -- & -- & -- & -- & $^\mathrm{+0.94}_\mathrm{-0.58}$ & -- & $^\mathrm{+6.16}_\mathrm{-4.30}$ & -- \\
             &               &   C+B2 & 378 & -1.24 & -- & -- & -- & -- & 0.27 & -5.92 & 11.13 & 439/467 \\
             &               &       & $^\mathrm{+43}_\mathrm{-35}$ & $^\mathrm{+0.70}_\mathrm{-0.10}$ & -- & -- & -- & -- & $^\mathrm{+0.60}_\mathrm{-1.00}$ & -- & $^\mathrm{+18.90}_\mathrm{-2.59}$ & -- \\
 \multicolumn{13}{|c|}{ }   \\    
  +0.174 & +0.600 & CPL & 61 & -1.77 & -- & -- & -- & -- & --  & -- & -- & 516/471 \\
             &               &      & $\pm5$ & $\pm0.04$ & -- & -- & -- & --  & -- & --  & -- & -- \\
             &               & Band & 62 & -1.79 & -4.42 & -- & -- & -- & -- & -- & -- & 516/470 \\
             &               &      & $\pm5$ & $\pm0.04$ & -- & -- & -- & -- & -- & --  & -- & -- \\
             &               &  2BPL & 13 & -0.87 & -1.95 & 127 & -- & -2.73 & --  & --  & -- & 512/468 \\
             &               &              & $^\mathrm{+1.14}_\mathrm{-1.81}$ & $^\mathrm{+2.92}_\mathrm{-0.43}$ & $\pm0.03$ & $^\mathrm{+31}_\mathrm{-24}$ & -- & $^\mathrm{+0.20}_\mathrm{-0.34}$ & -- & --  & -- & -- \\
             &               &   B+BB  & 72 & -1.68 & $<-2.66$ & -- & -- & --  & -- & -- & 3.463 & 514/468 \\
             &               &        & $\pm11$ & $\pm0.10$ & -- & -- & -- & --  & -- & -- & $^\mathrm{+1.60}_\mathrm{-1.06}$ & -- \\ 
 \multicolumn{13}{|c|}{ }   \\  
\hline
\end{tabular}
}
\end{center}
\end{table*}

\newpage
\begin{table*}
\caption{\label{tab:GRB120323A_resolved_spectra_fine}Fine time-resolved spectral fits of \grb using the detectors n0, n1, n3 and b0 (see section~\ref{sec:Fine time resolved}).}
\begin{center}
{\tiny
\begin{tabular}{|l|l|l|l|l|l|l|l|l|l|l|l|}
\hline
\multicolumn{2}{|c|}{Time} &\multicolumn{1}{|c|}{Models} &\multicolumn{3}{|c|}{Standard Model} & \multicolumn{2}{|c|}{Additional Model} & \multicolumn{1}{|c|}{Cstat/dof} \\
\hline
 \multicolumn{2}{|c|}{} & \multicolumn{1}{|c|}{} &\multicolumn{3}{|c|}{Band, Compt or 2BPL} & \multicolumn{2}{|c|}{ BB, Compt}  &  \multicolumn{1}{|c|}{}\\
\hline
\multicolumn{1}{|c|}{T$_\mathrm{start}$} & \multicolumn{1}{|c|}{T$_\mathrm{stop}$} & \multicolumn{1}{|c|}{Parameters} & \multicolumn{1}{|c|}{E$_{\rm peak}$} & \multicolumn{1}{|c|}{$\alpha$} & \multicolumn{1}{|c|}{$\beta$} & \multicolumn{1}{|c|}{index} & \multicolumn{1}{|c|}{kT or E$_{\rm 0}$} & \\
\hline
 \multicolumn{9}{|c|}{ }   \\  
-0.018 & -0.006 & PL & -- & -1.37$^\mathrm{+0.04}_\mathrm{-0.05}$ & -- & -- & -- & 386/472 \\
             &               & CPL & 583$^\mathrm{+183}_\mathrm{-125}$ & -0.86$^\mathrm{+0.21}_\mathrm{-0.37}$ & -- & -- & -- & 372/471 \\
             &               & Band & 249$^\mathrm{+90}_\mathrm{-116}$ & -0.47$^\mathrm{+0.43}_\mathrm{-0.27}$ & -1.78$^\mathrm{+0.14}_\mathrm{-0.18}$ & --  & -- & 364/470 \\
             &               &   P+BB & -- & -1.38$\pm0.08$ & -- & -- & 45.26$^\mathrm{+8.20}_\mathrm{-6.56}$ & 365/470 \\
             &               &   C+BB & 6277$^\mathrm{+9010}_\mathrm{-3160}$ & -1.21$\pm0.12$  & -- & --  & 40.78$^\mathrm{+8.66}_\mathrm{-7.27}$ & 362/469 \\
             &               &   B+BB & 5227$^\mathrm{+10325}_\mathrm{-4225}$ & -1.19$^\mathrm{+0.19}_\mathrm{-0.11}$ & $<-1.7$ & --  & 40.47$^\mathrm{+9.19}_\mathrm{-7.49}$ & 362/468 \\
\multicolumn{9}{|c|}{ }   \\  
-0.006 & +0.006 & Band & 123$^\mathrm{+28}_\mathrm{-21}$ & +0.31$^\mathrm{+0.44}_\mathrm{-0.34}$ & -2.03$^\mathrm{+0.09}_\mathrm{-0.13}$ & -- & -- & 417/470 \\
             &               &   C+BB & 1760$^\mathrm{+794}_\mathrm{-604}$ & -1.12$\pm0.11$ & -- & -- & 29.40$^\mathrm{+2.63}_\mathrm{-2.43}$ & 422/469 \\
\multicolumn{9}{|c|}{ }   \\  
+0.006 & +0.014 & Band &  109$\pm13$ & +0.17$^\mathrm{+0.30}_\mathrm{-0.23}$ & -2.20$\pm0.10$ &  -- & -- & 420/470 \\

             &               &   C+BB & 680$^\mathrm{+401}_\mathrm{-198}$ & -1.09$\pm0.15$ & -- & -- & 24.14$^\mathrm{+1.92}_\mathrm{-1.86}$ & 426/469 \\
             &               &   B+BB &  462$^\mathrm{+317}_\mathrm{-225}$ & -0.96$^\mathrm{+0.44}_\mathrm{-0.19}$ & $<-2.4$ & -- & 23.25$^\mathrm{+2.22}_\mathrm{-3.39}$ & 423/469 \\
\multicolumn{9}{|c|}{ }   \\  
 +0.014 & +0.022 & Band & 98$\pm12$ & -0.29$\pm0.23$ & -2.28$\pm0.11$ & -- & -- & 389/470 \\
             &               &   C+BB & 497$^\mathrm{+203}_\mathrm{-120}$ & -1.31$\pm0.11$ & -- & -- & 20.11$^\mathrm{+1.88}_\mathrm{-1.72}$ & 385/469 \\
             &               &   B+BB & 501$^\mathrm{+218}_\mathrm{-144}$ & -1.32$\pm0.13$ & $<-3$ & -- & 20.24$^\mathrm{+1.89}_\mathrm{-1.78}$ & 386/469 \\
\multicolumn{9}{|c|}{ }   \\  
 +0.022 & +0.044 & Band & 62$\pm5$ & +0.54$^\mathrm{+0.28}_\mathrm{-0.24}$ & -2.34$\pm0.06$ & -- & -- & 475/470 \\
             &               &   C+BB &  330$^\mathrm{+69}_\mathrm{-48}$ & -1.16$\pm0.11$ & -- & -- & 15.53$\pm0.73$ & 475/469 \\
             &               &   B+BB &  110$^\mathrm{+25}_\mathrm{-20}$ & -0.12$^\mathrm{+0.66}_\mathrm{-0.38}$ & -2.35$^\mathrm{+0.09}_\mathrm{-0.09}$ & -- & 11.77$\pm1.50$ & 473/468 \\
\multicolumn{9}{|c|}{ }   \\  
 +0.044 & +0.054 & Band & 44$\pm5$ & +0.75$^\mathrm{+0.56}_\mathrm{-0.49}$ & -2.29$\pm0.07$ & -- & -- & 354/470 \\
             &               &   C+BB & 231$^\mathrm{+49}_\mathrm{-36}$ & -1.12$^\mathrm{+0.22}_\mathrm{-0.17}$ & -- & -- & 10.99$^\mathrm{+0.83}_\mathrm{-0.75}$ & 352/469 \\
             &               &   B+BB & 111$^\mathrm{+56}_\mathrm{-40}$ & -0.24$\pm0.07$ & $<-2.40$ & -- & 9.07$^\mathrm{+1.65}_\mathrm{-2.06}$ & 353/468 \\
\multicolumn{9}{|c|}{ }   \\  
 +0.054 & +0.080 & Band & 47$\pm3$ & +0.29$^\mathrm{+0.28}_\mathrm{-0.25}$ & -2.12$\pm0.03$ & -- & -- & 480/470 \\
             &               &   C+BB & 435$^\mathrm{+77}_\mathrm{-58}$ & -1.36$^\mathrm{+0.08}_\mathrm{-0.07}$ & -- & -- & 11.79$^\mathrm{+0.54}_\mathrm{-0.50}$ & 471/469 \\
             &               &   B+BB &  433$^\mathrm{+84}_\mathrm{-62}$ & -1.36$\pm0.08$ & $<-3.3$ & -- & 11.81$^\mathrm{+0.55}_\mathrm{-0.52}$ & 472/469 \\
\multicolumn{9}{|c|}{ }   \\  
 +0.080 & +0.094 & Band &  45$\pm6$ & -0.08$\pm0.04$ & $-1.93$$\pm0.03$ & -- & -- & 471/470 \\
             &               & Band &  476$^\mathrm{+146}_\mathrm{-105}$ & -1.60$\pm0.04$ & $<-2.53$ &  -- & -- & 508/470 \\
             &               &   CPL         &  492$^\mathrm{+149}_\mathrm{-96}$ & -1.60$\pm0.04$ & -- &  -- & -- & 508/471 \\
             &               &   C+BB & 735$^\mathrm{+217}_\mathrm{-146}$ & -1.36$^\mathrm{+0.10}_\mathrm{-0.08}$ & -- & -- & 10.16$^\mathrm{+0.63}_\mathrm{-0.56}$ & 446/469 \\
             &               &   B+BB &  722$^\mathrm{+211}_\mathrm{-150}$ & -1.35$^\mathrm{+0.10}_\mathrm{-0.08}$ & $<-2.78$ & -- & 10.14$^\mathrm{+0.61}_\mathrm{-0.56}$ & 446/468 \\
             &               &   C+C2 &  743$^\mathrm{+238}_\mathrm{-163}$ & -1.36$^\mathrm{+0.55}_\mathrm{-0.15}$ & -- & +1.49$^\mathrm{+0.45}_\mathrm{-0.24}$ & 11.34$^\mathrm{+10.9}_\mathrm{-4.76}$ & 446/468 \\
\multicolumn{9}{|c|}{ }   \\  
 +0.094 & +0.132 & CPL & 464$^\mathrm{+67}_\mathrm{-53}$ & -1.50$\pm0.03$ & -- & -- & -- & 437/471 \\
             &               & Band & 458$^\mathrm{+66}_\mathrm{-52}$ & -1.50$\pm0.03$ & $<-3.50$ & -- & -- & 438/470 \\
             &               &   C+BB & 457$^\mathrm{+50}_\mathrm{-40}$ & -1.14$\pm0.10$ & -- & -- & 7.89$^\mathrm{+0.40}_\mathrm{-0.37}$ & 387/469 \\
             &               &   B+BB & 454$^\mathrm{+49}_\mathrm{-40}$ & -1.14$\pm0.10$ & $<-3.6$ & -- & 7.89$^\mathrm{+0.40}_\mathrm{-0.37}$ & 388/470 \\
             &               &   C+C2 & 429$^\mathrm{+37}_\mathrm{-30}$ & -0.73$^\mathrm{+0.27}_\mathrm{-0.23}$ & -- & +0.11$^\mathrm{+0.48}_\mathrm{-0.37}$ & 15.58$^\mathrm{+4.55}_\mathrm{-3.47}$ & 383/468 \\
\multicolumn{9}{|c|}{ }   \\  
   +0.132 & +0.178 & CPL & 237$^\mathrm{+52}_\mathrm{-35}$ & -1.62$\pm0.05$ &  -- & -- & -- & 461/471 \\
             &               & Band &  236$^\mathrm{+52}_\mathrm{-35}$ & -1.62$\pm0.05$ & $<-10$ & -- & -- & 461/470 \\
             &               &   C+BB & 214$^\mathrm{+32}_\mathrm{-23}$ & -1.34$\pm0.15$ & -- & -- & 4.126$^\mathrm{+0.58}_\mathrm{-0.71}$ & 456/469 \\
             &               &   B+BB & 194$^\mathrm{+30}_\mathrm{-24}$ & -1.24$^\mathrm{+0.20}_\mathrm{-0.17}$ & -2.78$^\mathrm{+0.33}_\mathrm{-0.93}$ & -- & 4.17$^\mathrm{+0.50}_\mathrm{-0.60}$ & 454/468 \\
\multicolumn{9}{|c|}{ }   \\ 
  +0.178 & +0.250 & CPL & 122$^\mathrm{+14}_\mathrm{-11}$ & -1.50$\pm0.06$ & -- & -- & -- & 372/471 \\
              &               & Band & 122$^\mathrm{+14}_\mathrm{-11}$ & -1.50$\pm0.06$ & $<-10$ &  -- & -- & 372/470 \\
              &               &   C+BB &  124$^\mathrm{+14}_\mathrm{-12}$ & -1.42$\pm0.17$ & -- & -- & 3.59$^\mathrm{+2.14}_\mathrm{-1.15}$ & 371/469 \\
              &               &   B+BB & 124$^\mathrm{+14}_\mathrm{-11}$ & -1.41$^\mathrm{+0.16}_\mathrm{-0.18}$ & $<-10$ & -- & 3.59$^\mathrm{+1.99}_\mathrm{-1.11}$ & 371/468 \\
\multicolumn{9}{|c|}{ }   \\  
  +0.250 & +0.700 & CPL & 18$^\mathrm{+10}_\mathrm{-14}$ & -1.94$\pm0.06$ & -- & -- & -- & 564/471 \\
              &               & Band &  19$^\mathrm{+9}_\mathrm{-16}$ & -1.92$\pm0.07$  & $<-2.72$ & -- & -- & 564/470 \\
              &               &   C+BB & 35$\pm17$ & -1.85$\pm0.14$ & -- & -- & 3.39$^\mathrm{+1.16}_\mathrm{-0.73}$ & 563/469 \\
 \multicolumn{9}{|c|}{ }   \\  
\hline
\end{tabular}
}
\end{center}
\end{table*}


\begin{thebibliography}{}

\bibitem[\protect\astroncite{{Abdo} et al.}{2009a}]{Abdo:2009:GRB080916C}
Abdo et al.: 2009a,
\newblock {Science}, 323, 1688

\bibitem[\protect\astroncite{{Abdo} et~al.}{2009b}]{Abdo:2009:GRB090902B}
Abdo et al.: 2009b,
\newblock {\apjl}, 706, L138

\bibitem[\protect\astroncite{{Ackermann} et al.}{2010}]{Ackermann:2010:GRB090510}
Ackermann et al.: 2010,
\newblock {\apj}, 716, 1178

\bibitem[\protect\astroncite{{Ackermann} et al.}{2011}]{Ackermann:2011:GRB090926A}
Ackermann et al.: 2011,
\newblock {\apj}, 729, 114

\bibitem[\protect\astroncite{{Ackermann} et al.}{2012}]{Ackermann:2012} 
Ackermann et al.: 2012,
\newblock {\apj}, 754, 121

\bibitem[\protect\astroncite{{Aloy} et al.}{2005}]{Aloy:2005}
Aloy, M.~A., Janka, H.-T., {\ M\&uuml}ller, E.: 2005,
\newblock{\aap}, 436, 273

\bibitem[\protect\astroncite{{Amati} et al.}{2002}]{Amati:2002}
Amati, L., Frontera, F., Tavani, M., et al.: 2002,
\newblock{\aap}, 390, 81

\bibitem[\protect\astroncite{{Amati} et al.}{2009}]{Amati:2009}
Amati, L., Frontera, F., \& Guidorzi, C.: 2009,
\newblock{\aap}, 508, 173

\bibitem[\protect\astroncite{{Asano} \& {M{\'e}sz{\'a}ros}}{2011}]{Asano:2011}
Asano, K., \& M{\'e}sz{\'a}ros, P.: 2011,
\newblock{\apj}, 739, 103

\bibitem[\protect\astroncite{{Asano} \& {M{\'e}sz{\'a}ros}}{2012}]{Asano:2012}
Asano, K., \& M{\'e}sz{\'a}ros, P.: 2012,
\newblock{arXiv:1206.0347}

\bibitem[\protect\astroncite{{Axelsson} et~al.}{2012}]{Axelsson:2012:GRB110721A}
Axelsson et al.:2012,
\newblock{arXiv:1207.6109}

\bibitem[\protect\astroncite{{Band} et al.}{1993}]{Band:1993}
Band, D., Matteson, J., Ford, L., et al.: 1993,
\newblock{\apj}, 413, 281

\bibitem[\protect\astroncite{{Bhat} \& {Guiriec}}{2011}]{Bhat:2011}
Bhat, P.~N., \& Guiriec, S.: 2011,
\newblock{Bulletin of the Astronomical Society of India}, 39, 471

\bibitem[\protect\astroncite{{Begelman} \& {Li}}{1994}]{Begelman:1994}
Begelman, M.~C., \& Li, Z.-Y.: 1994,
\newblock{\apj}, 426, 269

\bibitem[\protect\astroncite{{Beloborodov}}{2010}]{Beloborodov:2010}
Beloborodov, A.~M.: 2010, 
\newblock{\mnras}, 407, 1033

\bibitem[\protect\astroncite{{Beloborodov}}{2011a}]{Beloborodov:2011}
Beloborodov, A.~M.: 2011a, 
\newblock{\apj}, 737, 68

\bibitem[\protect\astroncite{{Beloborodov} et al.}{2011b}]{Belobodorov:2011b}
Beloborodov, A.~M., Daigne, F., Mochkovitch, R., \& Uhm, Z.~L.: 2011b,
\newblock{\mnras}, 410, 2422

\bibitem[\protect\astroncite{{Berger}}{2011}]{Berger:2011}
Berger, E.: 2011,
\newblock{GCN}, 12193, 1

\bibitem[\protect\astroncite{{Bo{\v s}njak} et al.}{2009}]{Bosnjak:2009}
Bo{\v s}njak, {\v Z}., Daigne, F., \& Dubus, G.: 2009,
\newblock{\aap}, 498, 677

\bibitem[\protect\astroncite{{Bo{\v s}njak} and {Daigne}}{2012}]{Bosnjak:2012submit}
{Bo{\v s}njak}, {\v Z}. and {Daigne}, F.: 2012,
\newblock{to be submitted to \aap}

\bibitem[\protect\astroncite{{Cavallo} \& {Rees}}{1978}]{Cavallo:1978}
Cavallo, G., \& Rees, M.~J.: 1978,
\newblock{\mnras}, 183, 359

\bibitem[\protect\astroncite{{Crider} et al.}{1997}]{Crider:1997}
Crider, A., Liang, E.~P., Smith, I.~A., et al.: 1997,
\newblock{\apjl}, 479, L39

\bibitem[\protect\astroncite{{Daigne} \& {Mochkovitch}}{1998}]{Daigne:1998}
Daigne, F., \& Mochkovitch, R.: 1998,
\newblock{\mnras}, 296, 275

\bibitem[\protect\astroncite{{Daigne} \& {Mochkovitch}}{2000}]{Daigne:2000}
Daigne, F., \& Mochkovitch, R.: 2000,
\newblock{\aap}, 358, 1157

\bibitem[\protect\astroncite{{Daigne} \& {Mochkovitch}}{2002}]{Daigne:2002}
Daigne, F., \& Mochkovitch, R.: 2002,
\newblock{\mnras}, 336, 1271

\bibitem[\protect\astroncite{{Daigne} \& {Drenkhahn}}{2002}]{Daigne:2002b}
Daigne, F., \& Drenkhahn, G.: 2002b,
\newblock{\aap}, 381, 1066

\bibitem[\protect\astroncite{{Daigne} \& {Mochkovitch}}{2003}]{Daigne:2003}
Daigne, F., \& Mochkovitch, R.: 2003,
\newblock{\mnras}, 342, 587

\bibitem[\protect\astroncite{{Daigne} et al.}{2011}]{Daigne:2011}
Daigne, F., Bo{\v s}njak, {\v Z}., \& Dubus, G.: 2011,
\newblock{\aap}, 526, A110

\bibitem[\protect\astroncite{{Derishev} et al.}{2001}]{Derishev:2001}
Derishev, E.~V., Kocharovsky, V.~V., \& Kocharovsky, V.~V.: 2001,
\newblock{\aap}, 372, 1071

\bibitem[\protect\astroncite{{Ford} et al.}{1995}]{Ford:1995}
Ford, L.~A., Band, D.~L., Matteson, J.~L., et al.: 1995,
\newblock{\apj}, 439, 307

\bibitem[\protect\astroncite{{Fryer} et al.}{1999}]{Fryer:1999}
Fryer, C.~L., Woosley, S.~E., Herant, M., \& Davies, M.~B.: 1999,
\newblock{\apj}, 520, 650

\bibitem[\protect\astroncite{{Gehrels}}{1997}]{Gehrels:1997}
Gehrels, N.: 1997,
\newblock{Nuovo Cimento B Serie}, 112, 11

\bibitem[\protect\astroncite{{Ghirlanda} et al.}{2003}]{Ghirlanda:2003}
Ghirlanda, G., Celotti, A., \& Ghisellini, G.: 2003,
\newblock{\aap}, 406, 879

\bibitem[\protect\astroncite{{Ghirlanda} et al.}{2004}]{Ghirlanda:2004}
Ghirlanda, G., Ghisellini, G., \& Lazzati, D.: 2004,
\newblock{\apj}, 616, 331 

\bibitem[\protect\astroncite{{Ghirlanda} et al.}{2007}]{Ghirlanda:2007}
Ghirlanda, G., Bosnjak, Z., Ghisellini, G., Tavecchio, F., \& Firmani, C.: 2007,
\newblock{\mnras}, 379, 73 

\bibitem[\protect\astroncite{{Ghirlanda} et al.}{2008}]{Ghirlanda:2008}
Ghirlanda, G., Nava, L., Ghisellini, G., Firmani, C., \& Cabrera, J.~I.: 2008,
\newblock{\mnras}, 387, 319

\bibitem[\protect\astroncite{{Ghirlanda} et al.}{2010}]{Ghirlanda:2010} 
Ghirlanda, G., Nava, L., \& Ghisellini, G.: 2010,
\newblock{\aap}, 511, A43

\bibitem[\protect\astroncite{{Ghirlanda} et al.}{2011a}]{Ghirlanda:2011a}
Ghirlanda, G., Ghisellini, G., Nava, L., \& Burlon, D.: 2011a,
\newblock{\mnras}, 410, L47

\bibitem[\protect\astroncite{{Ghirlanda} et al.}{2011b}]{Ghirlanda:2011b}
Ghirlanda, G., Ghisellini, G., \& Nava, L.: 2011b,
\newblock{\mnras}, 418, L109

\bibitem[\protect\astroncite{{Ghisellini} et al.}{2000}]{Ghisellini:2000}
Ghisellini, G., Celotti, A., \& Lazzati, D.: 2000,
\newblock{\mnras}, 313, L1

\bibitem[\protect\astroncite{{Giannios}}{2008}]{Giannios:2008}
Giannios, D.: 2008,
\newblock{\aap}, 480, 305

\bibitem[\protect\astroncite{{Goldstein} et~al.}{2012a}]{Goldstein:2012a}
Goldstein et al.: 2012a,
\newblock{\apjs}, 199, 19

\bibitem[\protect\astroncite{{Goldstein} et~al.}{2012b}]{Goldstein:2012b}
Goldstein et al.: 2012b,
\newblock{PoS}, GRB 2012, 082

\bibitem[\protect\astroncite{{Golenetskii} et al.}{1983}]{Golenetskii:1983}
Golenetskii, S.~V., Mazets, E.~P., Aptekar, R.~L., \& Ilinskii, V.~N.: 1983,
\newblock{\nat}, 306, 451

\bibitem[\protect\astroncite{{Golenetskii} et al.}{2012}]{Golenetskii:2012A}
Golenetskii, S., et al.: 2012,
\newblock{GCN}, 13102, 1

\bibitem[\protect\astroncite{{Golenetskii} et al.}{2012}]{Golenetskii:2012B}
Golenetskii, S., et al.: 2012,
\newblock{GCN}, 13103, 1

\bibitem[\protect\astroncite{{Goodman}}{1986}]{Goodman:1986}
Goodman, J.: 1986,
\newblock{\apjl}, 308, L47 

\bibitem[\protect\astroncite{{Gorbovskoy} et al.}{2012}]{Gorbovskoy:2012}
Gorbovskoy, E., et al.: 2012,
\newblock{GCN}, 13116, 1

\bibitem[\protect\astroncite{{Gruber} \& {Connaughton}}{2012}]{Gruber:2012}
Gruber, D., \& Connaughton, V.: 2012,
\newblock{GCN}, 13099, 1

\bibitem[\protect\astroncite{{Guiriec} et al.}{2010}]{Guiriec:2010}
Guiriec, S., et al.: 2010,
\newblock{\apj}, 725, 225 

\bibitem[\protect\astroncite{{Guiriec} et al.}{2011a}]{Guiriec:2011a}
Guiriec, S., et al.: 2011a,
\newblock{\apjl}, 727, L33

\bibitem[\protect\astroncite{{Guiriec} et al.}{2011b}]{Guiriec:2011b}
Guiriec, S., et al.: 2011b,
\newblock{AAS/HEAD}, 12, 01.04

\bibitem[\protect\astroncite{{Guiriec} et al.}{2013 in preparation}]{Guiriec:inPrep}
Guiriec, S., et al.:
\newblock{2013, in preparation}

\bibitem[\protect\astroncite{{Granot} et al.}{2011}]{Granot:2011}
Granot, J., Komissarov, S.~S., \& Spitkovsky, A.: 2011,
\newblock{\mnras}, 411, 1323

\bibitem[\protect\astroncite{{Greiner} et al.}{2009}]{Greiner:2009}
Greiner, J., Clemens, C., Kr{\"u}hler, T., et al.: 2009,
\newblock{\aap}, 498, 89

\bibitem[\protect\astroncite{{Greiner} et al.}{2011}]{Greiner:2011}
Greiner, J., Updike, A.~C., Kruehler, T., \& Sudilovsky, V.: 2011,
\newblock{GCN}, 12192, 1

\bibitem[\protect\astroncite{{Hasco{\"e}t} et al.}{2012a}]{Hascoet:2012a}
Hascoet, R., Daigne, F., \& Mochkovitch, R.: 2012a,
\newblock{Gamma-Ray Bursts 2012 Conference (GRB 2012)},

\bibitem[\protect\astroncite{{Hasco{\"e}t} et al.}{2012b}]{Hascoet:2012}
Hasco{\"e}t, R., Daigne, F., Mochkovitch, R., \& Vennin, V.: 2012b,
\newblock{\mnras}, 421, 525

\bibitem[\protect\astroncite{{Hasco{\"e}t} et al.}{2013}]{Hascoet:2013}
Hasco{\"e}t, R., Daigne, F., Mochkovitch, R.,:\fdaigne{2013},
\newblock{\fdaigne{to appear in  \aap, arXiv:1302.0235}}

\bibitem[\protect\astroncite{{Kaneko} et al.}{2006}]{Kaneko:2006}
Kaneko, Y., Preece, R.~D., Briggs, M.~S., et al.: 2006,
\newblock{\apjs}, 166, 298

\bibitem[\protect\astroncite{{Kobayashi} et al.}{1997}]{Kobayashi:1997}
Kobayashi, S., Piran, T., \& Sari, R.: 1997,
\newblock{\apj}, 490, 92

\bibitem[\protect\astroncite{{Kocevski}}{2012}]{Kocevski:2012}
Kocevski, D.: 2012,
\newblock{\apj}, 747, 146

\bibitem[\protect\astroncite{{Komissarov} et al.}{2009}]{Komissarov:2009}
Komissarov, S.~S., Vlahakis, N., K{\"o}nigl, A., \& Barkov, M.~V.: 2009,
\newblock{\mnras}, 394, 1182

\bibitem[\protect\astroncite{{Komissarov} et al.}{2010}]{Komissarov:2010}
Komissarov, S.~S., Vlahakis, N., K{\"o}nigl, A.: 2010,
\newblock{\mnras}, 407, 17

\bibitem[\protect\astroncite{{Kouveliotou} et al.}{1993}]{kouveliotou:1993}
Kouveliotou, C., et al.: 1993,
\newblock{\apjl}, 413, L101 

\bibitem[\protect\astroncite{{Liang} et al.}{2004}]{Liang:2004}
\fdaigne{Liang, E.~W., Dai, Z.~G., \& Wu, X.~F.: 2004
\newblock{\apjl}, 606, L29}

\bibitem[\protect\astroncite{{Lu} et al.}{2012}]{Lu:2012}
Lu, R.-J., Wei, J.-J., Liang, E.-W., et al.: 2012,
\newblock{arXiv:1204.0714}

\bibitem[\protect\astroncite{{Lyutikov} \& {Blandford}}{2003}]{Lyutikov:2003}
Lyutikov, M., \& Blandford, R.: 2003,
\newblock{arXiv:0312347}

\bibitem[\protect\astroncite{{MacFadyen} \& {Woosley}}{1999}]{MacFadyen:1999}
MacFadyen, A.~I., \& Woosley, S.~E.: 1999,
\newblock{\apj}, 524, 262

\bibitem[\protect\astroncite{{Malesani} et al.}{2009}]{Malesani:2009}
Malesani, D., Goldoni, P., Fynbo, J.~P.~U., et al.: 2009,
\newblock{GCN}, 9942, 1

\bibitem[\protect\astroncite{{Massaro} et al.}{2008}]{Massaro:2008}
Massaro, F., Cutini, S., Conciatore, M.~L., \& Tramacere, A.: 2008,
\newblock{AIPC Series}, 1000, 84

\bibitem[\protect\astroncite{{Meegan} et al.}{1992}]{Meegan:1992}
Meegan, C.~A., et al.: 1992,
\newblock{\nat}, 355, 143

\bibitem[\protect\astroncite{{Meegan} et al.}{2009}]{Meegan:2009}
Meegan, C., Lichti, G., Bhat, P.~N., et al.: 2009,
\newblock{\apj}, 702, 791

\bibitem[\protect\astroncite{{Meszaros} \& {Rees}}{1993}]{Meszaros:1993}
Meszaros, P., \& Rees, M.~J.: 1993,
\newblock{\apjl}, 418, L59

\bibitem[\protect\astroncite{{M{\'e}sz{\'a}ros}}{2002}]{Meszaros:2002}
M{\'e}sz{\'a}ros, P.: 2002,
\newblock{\araa}, 40, 137

\bibitem[\protect\astroncite{{Nakar} et al.}{2009}]{Nakar:2009}
Nakar, E., Ando, S., \& Sari, R.: 2009,
\newblock{\apj}, 703, 675

\bibitem[Nemmen et al.(2012)]{Nemmen:2012}
\fdaigne{Nemmen, R.~S., Georganopoulos, M., Guiriec, S., et al.: 2012,
\newblock{Science}, 338, 1445} 

\bibitem[\protect\astroncite{{Norris} et al.}{2000}]{Norris:2000}
Norris, J.~P., Marani, G.~F., \& Bonnell, J.~T.: 2000,
\newblock{\apj}, 534, 248

\bibitem[\protect\astroncite{{Norris} \& {Bonnell}}{2006}]{Norris:2006}
Norris, J.~P., \& Bonnell, J.~T.: 2006,
\newblock{\apj}, 643, 266

\bibitem[\protect\astroncite{{Paciesas} et al.}{1999}]{Paciesas:1999}
Paciesas, W.~S., et al.: 1999,
\newblock{\apjs}, 122, 465

\bibitem[\protect\astroncite{{Paczynski}}{1986}]{Paczynski:1986}
Paczynski, B.: 1986,
\newblock{\apjl}, 308, L43

\bibitem[\protect\astroncite{{Piran}}{2004}]{Piran:2004}
Piran, T.: 2004,
\newblock{Reviews of Modern Physics}, 76, 1143

\bibitem[\protect\astroncite{{Pe'er} \& {Waxman}}{2005}]{Peer:2005}
Pe'er, A., \& Waxman, E.: 2005,
\newblock{\apj}, 628, 857

\bibitem[\protect\astroncite{{Pe'er} et al.}{2006}]{Peer:2006}
Pe'er, A., M{\'e}sz{\'a}ros, P., \& Rees, M.~J.: 2006,
\newblock{\apj}, 642, 995

\bibitem[\protect\astroncite{{Pe'er} et al.}{2007}]{Peer:2007}
Pe'er, A., Ryde, F., Wijers, R.~A.~M.~J., M{\'e}sz{\'a}ros, P., \& Rees, M.~J.: 2007,
\newblock{\apjl}, 664, L1

\bibitem[\protect\astroncite{{Pe'er}}{2008}]{Peer:2008}
Pe'er, A.: 2008,
\newblock{\apj}, 682, 463

\bibitem[\protect\astroncite{{Preece} et al.}{1998}]{Preece:1998}
Preece, R.~D., Briggs, M.~S., Mallozzi, R.~S., et al.: 1998,
\newblock{\apjl}, 506, L23

\bibitem[\protect\astroncite{{Preece} et al.}{2000}]{Preece:2000}
Preece, R.~D., Briggs, M.~S., Mallozzi, R.~S., et al.: 2000,
\newblock{\apjs}, 126, 19

\bibitem[\protect\astroncite{{Racusin} et al.}{2011}]{Racusin:2011}
Racusin, J.~L., Oates, S.~R., Schady, P., et al.: 2011,
\newblock{\apj}, 738, 138

\bibitem[\protect\astroncite{{Rees} \& {Meszaros}}{1992}]{Rees:1992}
Rees, M.~J., \& Meszaros, P.: 1992,
\newblock{\mnras}, 258, 41P

\bibitem[\protect\astroncite{{Rees} \& {Meszaros}}{1994}]{Rees:1994}
Rees, M.~J., \& Meszaros, P.: 1994,
\newblock{\apjl}, 430, L93

\bibitem[\protect\astroncite{{Rees} \& {M{\'e}sz{\'a}ros}}{2005}]{Rees:2005}
Rees, M.~J., \& M{\'e}sz{\'a}ros, P.: 2005,
\newblock{\apj}, 628, 847

\bibitem[\protect\astroncite{{Rezzolla} et al.}{2011}]{Rezzolla:2011}
Rezzolla, L., Giacomazzo, B., Baiotti, L., et al.: 2011,
\newblock{\apjl}, 732, L6

\bibitem[\protect\astroncite{{Rosswog}}{2003}]{Rosswog:2003}
Rosswog, S.: 2003,
Gamma-Ray Burst and Afterglow Astronomy 2001: A Workshop Celebrating the First Year of the HETE Mission, 662, 220

\bibitem[\protect\astroncite{{Borgonovo} \& {Ryde}}{2001}]{Ryde:2001}
Borgonovo, L., \& Ryde, F.: 2001,
\newblock{\apj}, 548, 770

\bibitem[\protect\astroncite{{Rosswog} \& {Ramirez-Ruiz}}{2002}]{Rosswog:2002}
Rosswog, S., \& Ramirez-Ruiz, E.: 2002,
\newblock{\mnras}, 336, L7

\bibitem[\protect\astroncite{{Ruffert} \& {Janka}}{1999}]{Ruffert:1999}
Ruffert, M., \& Janka, H.-T.: 1999,
\newblock{\aap}, 344, 573

\bibitem[\protect\astroncite{{Ryde}}{2004}]{Ryde:2004}
Ryde, F.: 2004,
\newblock{\apj}, 614, 827

\bibitem[\protect\astroncite{{Ryde}}{2005}]{Ryde:2005}
Ryde, F.: 2005,
\newblock{\apjl}, 625, L95

\bibitem[\protect\astroncite{{Ryde} \& {Pe\' er}}{2009}]{Ryde:2009}
Ryde, F., \& Pe'er, A.: 2009,
\newblock{\apj}, 702, 1211

\bibitem[\protect\astroncite{{Ryde} et al.}{2010}]{Ryde:2010}
Ryde, F., Axelsson, M., Zhang, B.~B., et al.: 2010,
\newblock{\apjl}, 709, L172

\bibitem[\protect\astroncite{{Ryde} et al.}{2011}]{Ryde:2011}
Ryde, F., Pe'er, A., Nymark, T., et al.: 2011,
\newblock{\mnras}, 415, 3693

\bibitem[\protect\astroncite{{Sari} \& {Piran}}{1997}]{Sari:1997}
Sari, R., \& Piran, T.: 1997,
\newblock{\apj}, 485, 270

\bibitem[\protect\astroncite{{Sari} et al.}{1998}]{Sari:1998}
Sari, R., Piran, T., \& Narayan, R.: 1998,
\newblock{\apjl}, 497, L17

\bibitem[\protect\astroncite{{Scargle}}{1998}]{Scargle:1998}
Scargle, J.~D.: 1998,
\newblock{\apj}, 504, 405

\bibitem[\protect\astroncite{{Spruit} et al.}{2001}]{Spruit:2001}
Spruit, H.~C., Daigne, F., \& Drenkhahn, G.: 2001,
\newblock{\aap}, 369, 694

\bibitem[\protect\astroncite{{Tam} \& {Kong}}{2012}]{Tam:2012}
Tam, P.~H.~T., \& Kong, A.~K.~H.: 2012,
\newblock{GCN}, 13104, 1

\bibitem[\protect\astroncite{{Tchekhovskoy} et al.}{2010}]{Tchekhovskoy:2010}
Tchekhovskoy, A., Narayan, R., \& McKinney, J.~C.: 2010,
\newblock{New Astronomy}, 15, 749

\bibitem[\protect\astroncite{{van Paradijs} et al.}{1997}]{vanParadijs:1997}
van Paradijs, J., Groot, P.~J., Galama, T., et al.: 1997,
\newblock{\nat}, 386, 686

\bibitem[\protect\astroncite{{Vlahakis} \& {K\"onigl}}{2003}]{Vlahakis:2003}
Vlahakis, N., {K\"onigl}, A.: 2003,
\newblock{\apj}, 596, 1080

\bibitem[\protect\astroncite{{Vurm} et al.}{2011}]{Vurm:2011}
Vurm, I., Beloborodov, A.~M., \& Poutanen, J.: 2011,
\newblock{\apj}, 738, 77

\bibitem[\protect\astroncite{{Woosley}}{1993}]{Woosley:1993}
Woosley, S.~E.: 1993,
\newblock{\apj}, 405, 273

\bibitem[\protect\astroncite{{Woosley} \& {Heger}}{2006}]{Woosley:2006}
Woosley, S.~E., \& Heger, A.: 2006,
\newblock{\apj}, 637, 914

\bibitem[\protect\astroncite{{Yonetoku} et al.}{2004}]{Yonetoku:2004}
Yonetoku, D., Murakami, T., Nakamura, T., et al.: 2004,
\newblock{\apj}, 609, 935

\bibitem[\protect\astroncite{{Zhang} et al.}{2006}]{Zhang:2006}
Zhang, Z., Xie, G.~Z., Deng, J.~G., \& Jin, W.: 2006,
\newblock{\mnras}, 373, 729

\bibitem[\protect\astroncite{{Zhang} \& {Pe'er}}{2009}]{Zhang:2009}
Zhang, B., \& Pe'er, A.: 2009,
\newblock{\apjl}, 700, L65

\bibitem[\protect\astroncite{{Zhang} \& {Yan}}{2011}]{Zhang:2011}
Zhang, B., \& Yan, H.: 2011,
\newblock{\apj}, 726, 90

\bibitem[\protect\astroncite{{Zhang} et al.}{2012}]{Zhang:2012}
\fdaigne{Zhang, B., Lu, R.-J., Liang, E.-W., \& Wu, X.-F.:2012,  
\newblock{\apjl}, 758, L34}

\bibitem[\protect\astroncite{{Zhao} et al.}{2011}]{Zhao:2011}
Zhao, X.-H., Li, Z., \& Bai, J.-M.: 2011,
\newblock{\apj}, 726, 89

\end{thebibliography}
\end{document}